\documentclass[11pt,a4paper]{article}
\usepackage{jheppub_kim}

\usepackage{pdflscape}
\usepackage{amsmath}
\usepackage{amssymb}
\usepackage{dcolumn}
\usepackage{bm}
\usepackage{color}
\usepackage{epsfig}
\usepackage{amsfonts}
\usepackage{graphicx}
\usepackage{subfigure}
\usepackage{dcolumn}

\begin{document}

\title{  Lorentz Violating $p$-form Gauge Theories in Superspace}
\author[a]{ Sudhaker Upadhyay,} 
\author[b]{ Mushtaq B. Shah} 
\author[b]{and Prince A.  Ganai}

\affiliation[a]{Centre for Theoretical Studies, 
Indian Institute of Technology Kharagpur,  Kharagpur-721302, WB, India}
\affiliation[b]{Department of Physics, National Institute of Technology,  Srinagar, Kashmir-190006, India}

\emailAdd{sudhakerupadhyay@gmail.com}
\emailAdd{ mbstheory72@gmail.com }
\emailAdd{princeganai@nitsri.net}
 
\abstract{ 
Very special relativity (VSR) keeps the main features of special relativity but breaks rotational invariance due to an intrinsic
preferred direction. 
We study the VSR modified extended BRST and anti-BRST symmetry of the  Batalin-Vilkovisky 
(BV) actions corresponding to the  $p=1,2,3$-form  gauge theories. 
Within VSR framework, we discuss the  extended BRST invariant and extended BRST and anti-BRST invariant  superspace 
formulations for these BV actions. Here we observe that 
the VSR modified extended BRST invariant BV actions corresponding to the  $p=1,2,3$-form  gauge theories  can be written manifestly  covariant manner in a superspace with one Grassmann coordinate.
 Moreover,  two Grassmann coordinates are required to describe 
 the VSR modified extended BRST and extended anti-BRST invariant BV actions in a superspace.
 These results are consistent with the Lorentz invariant (special relativity) formulation.
}
\maketitle

\section{Overview and motivation}	
The   standard  model,  although  phenomenologically  successful, 
is unable to explain a variety of issues satisfactorily \cite{col}. 
The standard  model is assumed to be an effective description that
 works in the low-energy limit  of a more
fundamental theory (having a quantum description  of  gravitation also). 
However, the natural scale for a fundamental theory including gravity is 
governed by the Planck mass. This leads to an  interesting  question  that  whether  any
aspects of this underlying theory could be revealed through the
definite  experiment  with  present  techniques. To understand this properly, 
one possibility may be to examine proposed fundamental theories for effects that are 
qualitatively different from standard-model physics.
In this regard, one  possibility  is that  the new physics involves a violation of Lorentz symmetry.  In this connection, Cohen and Glashow proposed that the laws of physics need not be invariant under the full Lorentz group but rather under its
proper subgroup \cite{coh}.   An advantage of
this hypothesis is that while Lorentz symmetry is violated,
the theory still follows the basic postulates of special relativity,
 like the constancy of the velocity of light.
 Any scheme with  proper  Lorentz
subgroups  along
with translations   is referred to as very special relativity (VSR).
 
As an observable consequences of VSR, a novel mechanism for neutrino
masses without introducing new particles has been studied in Ref. \cite{coh1}. 
Few other  observable consequences of VSR have also been given in \cite{fan,dun}. 
In recent past, VSR has been studied in various contexts. For instance,
 the idea of VSR is implemented to de Sitter spacetime  where breaking of de Sitter invariance arises in two different
ways \cite{12}. 
Also, it has been shown that   the gauge field in quantum gauge perspective
acquire mass naturally without the conventional Higgs mechanism \cite{sudpanigrahi}.
The modifications due to VSR  have
also been analyzed  for the reducible gauge  theories 
using the  BV formulation   \cite{sud}.
Within VSR framework,  the event space underlying the
dark matter and the dark gauge fields supports the algebraic structure \cite{13}. 
 The   VSR effect to curved space-times shows that the
  $SIM(2)$ symmetry, which leaves the preferred
null direction   invariant, does not provide the complete couplings
to the gravitational background \cite{mu}. 
 The proper Lorentz subgroups together with translations are   realized in the 
 non-commutative space-time where the behavior of non-commutativity parameter $\theta^{\mu\nu}$ is  found lightlike \cite{jab,subi}.
In  VSR scenario,  $N = 1$ 
SUSY gauge theories   contain two conserved
supercharges rather than the usual four \cite{9}. The effects of quantum correction to VSR is studied to produce a curved space-time with a cosmological constant \cite{10}, where it is shown that the symmetry group $ISIM(2)$ does admit a 2-parameter family of continuous
deformations. Recently, a violation of Lorentz invariance in quantum electrodynamics induced by a very high frequency background
wave is studied, where   averaging observables over the rapid field oscillations
provides an effective theory \cite{ald}. The  
  quantum electrodynamics   and  the
massive spin-1 particle are discussed in VSR in  Refs. \cite{alv, 15}.
The   spontaneous symmetry-breaking mechanism to give a flavor-dependent VSR
mass to the gauge bosons is  studied in \cite{vic}.
 Interestingly, a quantum field theoretic structure suitable to describe the dark matter suggest   that    VSR plays  the same  role for the dark matter 
 fields as  special relativity does for the standard model fields \cite{13}.

On the other hands, the supersymmetric version \cite{wes}  of a non-Abelian gauge theory is  
 a super-geometrical theory of a constrained super 1-form \cite{wes1}. 
 It is well-known that in a superspace formulation there exists
a superconnection  which is the gauge superfield in a superspace. 
This, eventually, extends the  relation  of   
1-forms and gauge theories in ordinary spacetime  to the superspace.
As a result, the entire formalism of differential geometry is valid 
 in the superspace approach. The higher order $p$-forms theories have been studied 
 \cite{wes1}, which
 contain the gauge superfields characterized by gauge parameters of $(p-1)$-forms.
The importance of such formulation lies to the fact that
the well-known formulation of simple supergravity in
eleven dimensions \cite{cre} explicitly contains a $3$-form component gauge field and its
extension to superspace \cite{cre1} naturally includes the introduction of a super $3$-form
gauge superfield.
 A simple reduction of eleven dimensions to four-dimensions then leads to  
 3-form
gauge superfield in the $N = 8$ supergravity. 
Also, antisymmetric tensor fields describe the low energy excitations in string theories    \cite{h,i}. The study of higher-form gauge theory is also important for the classical string theories \cite{a},  
vortex motion in an irrotational, incompressible fluid \cite{b,c},   
dual formulation of the Abelian Higgs model \cite{d,e} and for  supergravity multiplets 
\cite{g}. 

To quantize the $p$-form gauge theories,  BV formulation is  one of the most general and powerful  approaches  \cite{ht,wei,bv,bv1,bv2,subm,sm}. One of the important illustrations of this formulation is that  it
provides a systematic way of accretion of the nontrivial
ghost for ghost structure for the case of reducible gauge theories. It has been observed that the antifields of the BV formulation coincide with
antighosts of certain collective fields, which ensure that Schwinger-Dyson
equations are satisfied as a consequence of the gauge symmetry algebra \cite{alf,jon}.
 Also the quantum 
corrections for anomalous gauge theories  can be
evaluated from the functional measure as long as a suitable regularization procedure is introduced \cite{tro}.
 A superspace formalism   for the Lorentz invariant
BV action of 1-form, 2-form and 3-form gauge theories have been studied \cite{ad, ba, sudb}.
The extended BRST and extended
anti-BRST invariant formulations (including some shift symmetry) of the Lorentz invariant BV action have also
been studied \cite{ad,ba,fk}, which lead  to the 
proper identification
of the antifields through equations of motion of auxiliary field variables. The importance of
  shift symmetries introduced through collective fields lies to the fact that it ensure
  that Schwinger-Dyson
equations at the level of the BRST algebra  
 can be performed within the Feynman path integral \cite{alf}.
According to the field
redefinition theorem, the particular choice of variables should have no influence on physical quantities like
  S-matrix elements, which appreciates  to formulate the quantization
prescription in a more coordinate-independent manner. In Refs. 
\cite{alf,jon}, Schwinger-Dyson equations 
is accomplished  in different field variables following  shift symmetry. 
Although VSR has been studied in various contexts, but the extended BRST and extended
anti-BRST invariant formulations with their superspace description  remain
unstudied in VSR framework. This provides us an opportunity    to bridge this gap.

We  consider a non-Abelian 1-form gauge theory in VSR context which remains invariant under
a non-local (Lorentz breaking) gauge transformation. The equations of motion 
in VSR-type Lorentz gauge leads a Proca type equation which confirms that
non-Abelian vector to have a mass. Although this describes a theory with mass 
but remains invariant under a (VSR modified) gauge transformation.
This leads to a redundancy in gauge degrees of freedom if we 
quantize it without fixing a gauge.
Therefore, utilizing Faddeev-Popov procedure, we construct an effective action
which admits a (VSR modified) BRST transformation. 
Furthermore, we study the extended  BRST symmetry which includes a shift symmetry.
This extra symmetry is then   gauge fixed
(adding new ghosts, antighosts, and auxiliary fields) in such
a way that the original action is recovered after the extra fields
are integrated out.  In order to recover original theory (or to compensate these additional fields), we further introduce 
 anti-ghosts with exactly opposite ghost numbers. Within formulation,
these anti-ghosts coincide with the antifields of the BV formulation
analogous to Lorentz invariant case. 
We   further provide a superspace description of VSR modified 
non-Abelian 1-form gauge theory possessing extended BRST symmetry with the help of
coordinates $(x_\mu,\theta)$. The superspace description of this theory having 
extended anti-BRST invariance only needs another fermionic variable $\bar\theta$
together with $x_\mu$. However, we found that even in VSR modified theory possessing both the extended  BRST   and extended anti-BRST invariance, one
needs a superspace with two grassmann parameters $\theta,\bar\theta$
together with $x_\mu$ to provide a superfields description. 
We generalize the results of non-Abelian 1-form gauge theory to
the higher-form (for instance $2,3$-form) gauge theories also to
show consistency of results.

The paper is organized as following. 
In section II, we outline non-Abelian 1-form gauge theory in VSR.
Here, we study the VSR modified 
extended BRST and extended anti-BRST transformations (which include a shift symmetry)
 for the BV action of theory.   Within this section, we demonstrate a superspace description 
 for the 1-form gauge theory having extended BRST invariance and extended 
 anti-BRST invariance as the separate cases. Section III is devoted to 
 the generalization of results for the non-Abelian 1-form gauge theory
 to the Abelian 2-form gauge theory. We further generalize these results 
 to the Abelian 3-form gauge theory case in section IV. The 
 paper is summarize with future remarks in the last section.
 The lengthy calculations are reported in Appendix sections.

\section{Non-Abelian 1-form gauge theory: VSR modified BV action in superspace}
In this section, we describe the VSR modified BV action for non-Abelian  1-form gauge (Yang-Mills) theory in superspace.
Let us review first the VSR description of  non-Abelian  1-form gauge theory following 
Ref. \cite{vic}.
We start by defining the  classical Lagrangian density in VSR as follows \cite{vic}
\begin{equation}
{\cal L}_0=  -\frac{1}{4}\mbox{Tr}\left[\tilde{F}^{a\mu\nu}\tilde{F}_{\mu\nu}^a\right], 
\end{equation}
where field-strength tensor $\tilde{F}_{\mu\nu}$ is given by
\begin{eqnarray}
\tilde F_{\mu\nu}= F_{\mu\nu} -\frac{1}{2}m^2\left(n_\nu\frac{1}{(n\cdot D)^2}n^\alpha
F_{\mu\alpha} - n_\mu\frac{1}{(n\cdot D)^2}n^\alpha
F_{\nu\alpha}\right),
\end{eqnarray}
with
\begin{eqnarray}
F_{\mu\nu}&=&\partial_\mu A_\nu-\partial_\nu A_\mu -i[A_\mu, A_\nu]+\frac{1}{2}m^2n_\nu\left( 
\frac{1}{(n\cdot\partial)^2}\partial_\mu(n\cdot A)\right)\nonumber\\
&-& \frac{1}{2}m^2n_\mu\left( 
\frac{1}{(n\cdot\partial)^2}\partial_\nu(n\cdot A)\right)-\frac{i}{2}m^2\left[\frac{1}{(n\cdot\partial)^2}n\cdot A, (n_\mu A_\nu-n_\nu A_\mu) \right].
\end{eqnarray}
The vector  $n_\mu$ is a constant  null vector (i.e. $n^2=0$),
which transforms  under a VSR transformation
so that any term containing ratios involving $n_\mu$ are
invariant. The covariant derivative is  determined by imposing the proper transformation
property for the covariant derivative  as follows: $ D_\mu =   \partial
_\mu-i[A_\mu, \cdot ]- \frac{i}{2}m^2n_\mu \left[\left(\frac{1}{(n\cdot\partial)^2}n\cdot A
\right), \cdot \right]$.
We use  following definition to handle the non-local terms \cite{vic}
\begin{eqnarray}
\frac{1}{n\cdot D}=\int_0^\infty db\ e^{-bn\cdot D}.
\end{eqnarray}
The equations of motion  for  the vector field  
 satisfying VSR type Lorentz gauge condition 
 leads to a Proca equation which suggest that vector field has mass \cite{vic}.
 
 This Lagrangian is also invariant under a VSR modified (non-local) gauge transformation
 \begin{eqnarray}
 \delta A_\mu &=&\partial_\mu \lambda -i[A_\mu,\lambda] +\frac{i}{2}m^2n_\mu\left(\lambda, 
\frac{1}{(n\cdot\partial)^2}  n\cdot A \right)\nonumber\\
&-& \frac{1}{2}m^2n_\mu\left( 
\frac{1}{(n\cdot\partial)}\lambda \right)+\frac{i}{2}m^2n_\mu\left(\frac{1}{(n\cdot\partial)^2} n\cdot [A,\lambda] \right),
 \end{eqnarray}
where $\lambda$ is an infinitesimal parameter. 
Being a (VSR modified) gauge invariant, the non-Abelian 1-form gauge theory   contains   redundant degrees of freedom.
To quantize the theory correctly we need to choose a  gauge appropriately.
In this context, the gauge-fixed Lagrangian density together with the ghost term is given
by
\begin{eqnarray}
 {{\cal L}}_{gf}&=& \mbox{Tr} \left[   B 
 {\partial}^\mu {A_\mu }- B  \frac{m^2}{n\cdot \partial}n^\mu {A_\mu }  +i\tilde C(\square
 -m^2)C +\tilde C\partial_\mu[A^\mu, C] +
 \frac{m^2}{2}\tilde C\left(\frac{1}{n\cdot\partial}[n\cdot A,C]\right)\right.\nonumber\\
 & -&\left.\frac{m^2}{2}\tilde C
 \left(n\cdot\partial \left[C, \frac{n\cdot A}{(n\cdot\partial)^2}\right]\right)\right], 
 \label{gf}
\end{eqnarray}
where $B^a$ is an auxiliary field. Here  we note that, in comparison to VSR modified covariant gauges, the
gauge-fixed Lagrangian density  takes  simplest form  \cite{sud} in VSR modified light-cone gauge, $\eta\cdot A=0$, where $\eta_\mu$  is an arbitrary constant vector that defines a preferred axis in space.
 Now, the effective quantum Lagrangian density for Yang-Mills theory in VSR is given by
\begin{equation}
{\cal L}=  {\cal L}_0+ {{\cal L}}_{gf}. 
\label{sf}
\end{equation}
This Lagrangian density leads to following vector field and ghost propagators respectively:
\begin{eqnarray}
\Delta^{\mu\nu} &=&\frac{1}{p^2+m^2}\left[\eta^{\mu\nu}-\left(\frac{\alpha-1}{2\alpha-1}\right)\frac{1}{p^2+m^2}\left(p^\mu p^\nu +\frac{1}{2}m^2(n^\mu p^\nu +n^\nu p^\mu )\frac{1}{n\cdot p}+\frac{1}{4}m^4 \frac{n^\mu n^\nu}{(n\cdot p)^2}\right) \right],\nonumber\\
\Delta_{gh}&=&-\frac{1}{p^2+m^2}.
\end{eqnarray}
This implies clearly that both the gauge field and ghost field have same mass $m$, consequently,  this mass generation is different from Higgs mechanism.
This is matter of calculation only to show that this  effective Lagrangian density  (\ref{sf}) is   invariant  under the following BRST transformations:
\begin{eqnarray}
s_b A_\mu &= &     \partial _\mu C -  \frac{m^2}{n\cdot \partial}n_\mu C -i[A_\mu, C]
+i\frac{m^2}{2}\left[C,\frac{n\cdot A}{(n\cdot\partial)^2}\right]
+ i\frac{m^2}{2}n_\mu\left(\frac{1}{(n\cdot\partial)^2}n\cdot[A,C]\right),\nonumber \\
&=: & {\cal D}_\mu   C,\nonumber\\
s_b C &=&   iC^2, \ \
s_b \tilde{C} = iB  ,\ \
s_b B =0.
\label{brst}
\end{eqnarray}
This BRST transformations are nilpotent in nature, i.e., $s_b^2 =0$.
Since the gauge fixing and ghost part of the  effective Lagrangian density is BRST-exact,
so these can also be expressed in terms 
of BRST variation of gauge-fixing fermion ($\Psi$). Thus,
the effective Lagrangian density can also be expressed as
\begin{equation}
 {{\cal L}} =  {\cal L}_0 +\mbox{Tr}( s_b {\Psi}), 
\label{sf1}
\end{equation}
where the explicit form of the  gauge-fixing fermion $\Psi$  is 
\begin{equation}
 {\Psi}   =-i \tilde{C}\left(\partial^\mu A_\mu -\frac{1}{2}\frac{m^2}{n\cdot \partial}n^\mu A_\mu\right  ).\label{tot1}
\end{equation}
One could also check that the effective Lagrangian density  (\ref{sf}) is also  invariant under the another nilpotent transformation where the role of ghost and anti-ghost fields 
are interchanged. These  so-called  anti-BRST transformations are
\begin{eqnarray}
s_{ab} A _\mu &= &     \partial _\mu \tilde C -  \frac{m^2}{n\cdot \partial}n_\mu \tilde C -i[A_\mu, \tilde C]
+i\frac{m^2}{2}\left[\tilde C,\frac{n\cdot A}{(n\cdot\partial)^2}\right]
+ i\frac{m^2}{2}n_\mu\left(\frac{1}{(n\cdot\partial)^2}n\cdot[A,\tilde C]\right),\nonumber \\
&= & {\cal D}_\mu \tilde C,\nonumber\\
s_{ab}\tilde C  &=&   i\tilde C^2, \ \
s_{ab}  {C} =  -iB  +i C  \tilde C,\ \
s_{ab} B =-i[  B, \tilde C].
\end{eqnarray}
The gauge-fixing and ghost parts of the Lagrangian density  are anit-BRST exact also 
and can also be described  in terms of anti-BRST variation of some another gauge-fixing 
fermion. Here, we would like to  state   that  the VSR gauge fields are massive  with 
a common mass, however,  in nature gauge fields may have
different masses due to the spontaneous symmetry breaking in VSR with
non-Abelian gauge symmetry  Ref. \cite{vic}. In such a way, the  fields
can have the usual mass due to spontaneous symmetry
breaking in addition to a flavor-dependent VSR mass.

\subsection{VSR modified extended BRST invariant BV action }  
Now, there exists  an interesting question that if the gauge field in VSR is being displaced as
$ A_\mu \rightarrow A_\mu -\bar A_\mu$: does the
gauge symmetry still remain and moreover, how does this shift symmetry affect
the underlying BRST structure.  
 In this context, we  shift  all the fields (within VSR framework)
from their original value as follows
\begin{equation}
A_\mu  \longrightarrow  A _\mu- \bar A _\mu, \quad
C \longrightarrow  C  - \bar C,  \quad
\tilde C  \longrightarrow \tilde C  - \bar {\tilde C},  \quad
B  \longrightarrow B  - \bar B .
\end{equation}
Under these shifts, the Lagrangian density (\ref{sf}) in VSR  is modified by
\begin{eqnarray}
\bar{\cal L }&=& {\cal L}(A _\mu - \bar A _\mu, C  - \bar C , \tilde C  - \bar{\tilde C} , B  - \bar 
B ).\label{til} 
\end{eqnarray}
This shifted version of Lagrangian density 
remains  invariant under the BRST transformation  (\ref{brst}) with respect to the 
shifted fields $A _\mu - \bar A _\mu, C  - \bar C , \tilde C  - {\bar{\tilde C}} , B  - \bar 
B $. In addition, this Lagrangian density is also invariant under the (local) shift
symmetry 
$s_b \phi ={\cal R}(x),\ s_b \bar\phi ={\cal R}(x)$,
where collective fields $\phi$ and $\bar\phi$ are $ (A _\mu, C, \tilde C, B)$ and 
$ (\bar A _\mu, \bar C, \bar {\tilde C}, \bar B)$, respectively. ${\cal R}(x)$
is the generic notation for the Slavnov variations  of fields $\phi$ and $\bar{\phi}$.
This  deserves further being gauge fixed and, in turn, leads to an additional BRST symmetry.
The BRST symmetry together with the shift symmetry is known as the extended BRST symmetry.
This  extended BRST 
symmetry transformations corresponding to Lagrangian density (\ref{til}) in VSR  read
\begin{eqnarray}
s_b A _\mu &=&\psi_\mu  ,\ \
s_b \bar A_\mu  =\psi_\mu  - {\cal D}^{(A-\bar A)}_\mu  (C- \bar C), \nonumber\\
s_b C  &=& \epsilon   ,\ \
s_b \bar C  =\epsilon  -i(C-\bar C)^2, \ \
s_b \tilde C  =\tilde \epsilon , \nonumber\\
s_b \bar {\tilde C}  &=& \tilde \epsilon  -i (B-\bar B) ,\ \
s_b B = \rho  ,\ \
s_b \bar B  =\rho ,\label{ex}
\end{eqnarray}
where 
\begin{eqnarray}
{\cal D}^{(A-\bar A)}_\mu  (C- \bar C)&=&  \partial _\mu (C- \bar C) -  \frac{m^2}{n\cdot \partial}n_\mu (C- \bar C) \nonumber\\
&-& i[A-\bar A, C- \bar C]+i\frac{m^2}{2}\left[C- \bar C,\frac{n\cdot (A-\bar A)}{(n\cdot\partial)^2}\right]\nonumber\\
&+& i\frac{m^2}{2}n_\mu\left(\frac{1}{(n\cdot\partial)^2}n\cdot[A-\bar A, C- \bar C]\right).
\end{eqnarray}
 Here, extra fields $\psi_\mu, \epsilon, \tilde \epsilon$ and $\rho$ denote the ghost fields 
related to shift  symmetry for $A_\mu , C ,\tilde C$ and $B$, respectively.
Due to  nilpotency property of extended BRST symmetry (\ref{ex}), it is evident that 
the variation of these ghost fields  $\psi_\mu, \epsilon, \tilde \epsilon$ and $\rho$    under
 extended BRST transformation vanishes.  Now, in order to make theory
  unchanged, we need to remove the contribution of these ghosts from the physical states.
  Thus, we   introduce the antifields (anti-ghosts) $A^\star_\mu, C^\star, \tilde 
C^\star$ and $B^\star$ corresponding to each ghost field which compensates the net
contribution of these ghosts. 
The BRST variation of these antifields   are defined by
\begin{eqnarray}
s_b A^{\star}_\mu &=& -\zeta_\mu , \ \
s_b C^{\star} = - \sigma, \ \
s_b \tilde C^{\star}  = -\tilde \sigma , \ \
s_b B^{\star} = -\tilde \upsilon,\label{anbrs}
\end{eqnarray} 
where $\zeta_\mu, \sigma, \tilde \sigma $ and $\tilde \upsilon$ are  
auxiliary  fields 
corresponding to shifted fields $\bar A_\mu, \bar C, \bar {\tilde C}$ and 
$\bar B$  and do not change under
 BRST transformation.

 In order to fix the gauge for shift symmetry, we add following gauge fixing term 
 to the VSR quantum action (\ref{sf}) (and we call the resulting Lagrangian density as BV
 action):
  \begin{eqnarray}
\bar{\cal L}_{gf} &=&  \mbox{Tr}\left[ -\zeta^{\mu} \bar A_\mu - A^{\mu\star} \left[\psi_\mu - {\cal D}^{(A-\bar A)}_\mu  (C- \bar C)\right] + \sigma  \bar {\tilde C }
- C^{\star}[\tilde \epsilon -i (B-\bar B)] \right.\nonumber\\
& -& \left.\tilde \sigma  \bar C + \tilde C^{\star}\left[ \epsilon -i(C-\bar C)^2  \right] 
+ \tilde \upsilon 
\bar B + B^{\star} \rho \right].\label{i}
\end{eqnarray}
In this way,  all the tilde fields will vanish and we  then recover  our original theory.
We note that this gauge-fixed  Lagrangian density, $\bar{\cal L}_{gf}$, is also invariant under the extended BRST  symmetry transformations  (\ref{ex}).

Now, by integrating out the auxiliary fields  $\zeta_\mu, \sigma, \tilde \sigma $ and $\tilde \upsilon$, this reads
\begin{eqnarray}
\bar{\cal L}_{gf} &=&  \mbox{Tr}\left[  - A_\mu^{\star} (\psi^{\mu} - {\cal D}^\mu  C)  - C^{\star}(\tilde \epsilon^a -i B) +  \tilde C^{\star}\left(\epsilon -i C^2\right)  + B^{\star} \rho\right].
\label{h}
\end{eqnarray}
Since  the gauge-fixed Lagrangian density (\ref{gf}) is BRST-exact, therefore  one can express    in terms of a general gauge-fixing fermion 
$\Psi(A_\mu,  \tilde C, C, B)$ as 
\begin{eqnarray}
 {\cal L}_{gf}&=&  s_b \Psi =  - \frac{\delta \Psi}{\delta A_\mu}\psi _\mu + \frac{\delta \Psi}{\delta C} 
\epsilon
+ \frac{\delta \Psi}{\delta \tilde C}\tilde\epsilon - \frac{\delta \Psi}{\delta B} \rho,
\label{g}
\end{eqnarray}
In the last term, the BRST transformations (\ref{ex}) are utilized.
Now, we utilize the equations of motion for the auxiliary fields which set all the fluctuated  fields  to zero. Thus, we left with the following BV action:
\begin{eqnarray}
{\cal L}_{eff} & = &  {\cal L}_0 + {\cal L}_{gf} 
+\bar{\cal L}_{gf},\nonumber\\
  &=&   {\cal L} +  \mbox{Tr}\left[ \left(- A_\mu^{\star} - \frac{\delta \Psi}{\delta A^{\mu}}
  \right)\psi ^{\mu} +
  \left(\tilde C^{\star} + \frac{\delta \Psi}{\delta C}\right) \epsilon -
  \left(C^{\star} - \frac{\delta \Psi}{\delta \tilde C}\right)\tilde\epsilon\right. \nonumber\\
 &+&\left.  \left(  B^{\star} - \frac{\delta \Psi}{\delta B}\right) \rho
  + A_\mu^{\star}   {\cal D}_\mu  C  +i C^{\star}  B -i\tilde C^{\star}C^2 \right].\label{eff}
\end{eqnarray}
    In order to get identifications of the antifields
    in VSR, it is sufficient to integrate out the
ghost fields associated with the shift symmetry
\begin{eqnarray}
A_\mu ^{\star} &=& -\frac{\delta \Psi}{\delta A^{\mu}} , \ \
\tilde C^{\star} = - \frac{\delta \Psi}{\delta C}  , \ \
C^{\star} =   \frac{\delta \Psi}{\delta \tilde C}  , \ \
B^{\star}=\frac{\delta \Psi}{\delta B}.
\end{eqnarray}
 However, for the VSR modified gauge-fixing fermion   given in (\ref{tot1}),  we determine
 the  anti-ghost fields   as following:
\begin{eqnarray}
 \bar {A}_\mu ^{a\star} &=&  -i\partial_\mu \tilde C + \frac{i}{2}\frac{m^2}{n\cdot \partial}n_\mu \tilde C, \ \
\tilde C^{a\star} =0, \ \
\bar {C}^{a\star} =  -i\partial_\mu A^{a\mu} +\frac{i}{2}\frac{m^2}{n\cdot \partial}n_\mu  A^{a\mu}, \nonumber\\
B^{a\star} &=& 0.
\end{eqnarray}
Here, we observe that, analogous to Lorentz invariant case,   antifields get identification naturally.
This clarifies the geometric interpretation of the antifields on the line
of Maurer-Cartan 1-forms. 
Plugging these anti-ghost fields  in (\ref{eff}), we can 
recover the original Lagrangian density of YM theory in VSR.

\subsection{VSR modified extended BRST invariant superspace formulation}
We know that superspace formulations for gauge theories can be built up
 in such a manner that the BRST transformations
are realized as translations along the Grassmannian coordinate \cite{fer}.
In order to describe the VSR modified BRST invariant BV action in superspace,
we need an extra (Grassmannian) coordinate $\theta$ together with $x^\mu$.
Superspace formulations for the VSR modified BRST transformation 
are obtained by associating with each field   a 
superfield of the form
\begin{eqnarray}
\mathfrak{A}_\mu(x,\theta)&=&A_\mu(x) +\theta {\cal D}_{\mu} C,\nonumber\\
{\mathfrak{C}} (x,\theta)&=&C (x) + i\theta  C^2,\nonumber\\
\tilde{\mathfrak{C}}(x,\theta)&=&\tilde C(x) +i \theta B. 
\end{eqnarray}
The shifted superfields in VSR will be consistent only if these can be written  by
\begin{eqnarray}
\mathbb{A}_\mu(x,\theta)&=&\mathfrak{A}_\mu(x,\theta)-\bar {\mathfrak{A}}_\mu(x,\theta)=(A_\mu-\bar
 A_\mu)
 +\theta {\cal D}^{(A-\bar A)}_{\mu} (C-\bar C),\nonumber\\
\mathbb{C}(x,\theta)&=&{\mathfrak{C}}(x,\theta)-\bar{{\mathfrak{C}}}(x,\theta)=(C -\bar{C}) +i\theta 
(C-\bar C)^2,\nonumber\\
\tilde{\mathbb{C}}(x,\theta)&=&\tilde{\mathfrak{C}}(x,\theta)-\bar{\tilde{\mathfrak{C}}}(x,\theta)=\tilde C(x)-\bar{\tilde C}(x) +i \theta (B-\tilde{B}).
\end{eqnarray}
From the above one can see the arbitrariness in the extended BRST symmetries. Therefore, one can not determine the individual superfields  uniquely.
So, to be consistent with the above analysis, 
we can define the original superfields and shift superfields  
with the help of extended BRST transformation 
as follows,
\begin{eqnarray}
\mathfrak{A}_\mu(x,\theta) &=&A_\mu +\theta \psi_\mu, \ \ \   \bar {\mathfrak{A}}_\mu(x,\theta)= \bar
 A_\mu 
+\theta(\psi_\mu - {\cal D}^{(A-\bar A)}_{\mu} (C-\bar C)),\nonumber\\
{\mathfrak{C}}(x,\theta) &=& C +\theta\epsilon,\ \   \bar{{\mathfrak{C}}}(x,\theta)=  \bar{C}  
+\theta \left(\epsilon -i(C-\bar C)^2\right),\nonumber\\
\tilde{\mathfrak{C}}(x,\theta)  &=&\tilde C   +\theta\bar\epsilon,\ \ \bar{\tilde{\mathfrak{C}}}(x,\theta)= \bar{\tilde C}  - \theta (\tilde\epsilon -i B+i\tilde{B}),\nonumber\\
\mathfrak{B}(x,\theta)  &=&B+\theta\rho.\label{11}
\end{eqnarray}
Exploiting BRST transformations (\ref{anbrs}), we introduce the super antifields
with one grassmannian coordinate in VSR as
\begin{eqnarray}
\bar{\mathfrak{A}}^{\star}_\mu(x,\theta) &=& A_\mu^\star -\theta\zeta_\mu,\nonumber\\
\bar{\mathfrak{C}}^{\star}(x,\theta)&=& {{C}}^{\star}  -\theta\sigma,\nonumber\\
\bar{\tilde{\mathfrak{C}}}^{\star}(x,\theta)&=& \tilde{{C}}^{\star} -\theta\tilde\sigma,\nonumber\\
\mathfrak{B}^{\star}(x,\theta)&=& B^{\star}-\theta\tilde v.\label{22}
\end{eqnarray}
We find that the appropriate combinations of superfields of (\ref{11}) and (\ref{22}), 
leads to the  gauge-fixed Lagrangian  density  corresponding to shift symmetry in VSR  (\ref{i}) as following:
\begin{eqnarray}
{\bar{\cal L}}_{gf}=\mbox{Tr}\left[\frac{\partial}{\partial\theta} \left(\bar{\mathfrak{A}}_\mu^{\star}\mathfrak{A}^{\mu }
+\bar{\tilde{\mathfrak{C}}}^{\star}\bar{{\mathfrak{C}}}-\bar{\tilde{\mathfrak{C}}}\bar{\mathfrak{C}}^{\star}-
\mathfrak{B}^{\star}\mathfrak{B} \right)\right].
\end{eqnarray}
The  VSR modified  gauge-fixed fermion (\ref{tot1}) 
in extended BRST superspace formulation  can be written as 
\begin{eqnarray}
\Omega(x,\theta) 
&=&-i\tilde C \left(\partial_\mu- \frac{1}{2}\frac{m^2}{n\cdot \partial}n_\mu\right)  A^{\mu } +
i\theta\left[\tilde C \left(\partial_\mu -\frac{1}{2}\frac{m^2}{n\cdot \partial}n_\mu\right)
\psi^{\mu a}\right.\nonumber\\
&-&\left.\tilde{\epsilon}
\left(\partial_\mu -\frac{1}{2}\frac{m^2}{n\cdot \partial}n_\mu \right)A^{\mu} 
\right].
\end{eqnarray}
Here, it is evident that $\theta$ component of the above expression 
gives the gauge fixing Lagrangian density corresponding to
the  original  BRST symmetry (\ref{gf}), i.e., 
\begin{equation}
{\cal L}_{gf}=\mbox{Tr}\left[\frac{\partial}{\partial\theta}\Omega(x,\theta)\right],
\end{equation}
Being the $\theta$ component of a
super gauge-fixed fermion, it is obvious that  ${\cal L}_{gf}$ is invariant  under the extended BRST transformations. 

\subsection{VSR modified extended anti-BRST symmetry} 

In this subsection, we construct the VSR-modified extended anti-BRST transformation.
The importance of anti-BRST transformation lies in the fact that, while the anti-BRST invariance does not lead to any
additional information in comparison to BRST invariance,
 it is extremely important in order to put the theory
 in geometrical setting.
The extended anti-BRST transformations which leaves the
BV action invariant are,
\begin{eqnarray}
s_{ab} A _\mu &= &  A_\mu^{ \star}+  {\cal D}^{(A-\bar A) }_\mu  (\tilde C- \bar {\tilde C})  ,\ \
s_{ab} \bar A _\mu = A^{\star}_\mu ,\nonumber \\
s_{ab}\tilde C  &=& \tilde C^{ \star} +i(\tilde C-\bar{\tilde C})^2 , \ \
s_{ab}\bar{\tilde C}  = \tilde C^{\star}, \nonumber \\
s_{ab}  {C}  &=&   C^{\star}-i B +i\bar B +i(  C-\bar{ C})(\tilde C-\bar{\tilde C}),\ \
s_{ab}  \bar{C}  =  C^{\star},\nonumber \\
s_{ab} B  &= &  B^{\star}- i[(B-\bar B),  \tilde C-\bar{\tilde C}],\ \
s_{ab} \bar B  =  B^{\star},\nonumber \\
s_{ab} \psi_\mu  &= & \zeta_\mu  +{\cal D}^{(A-\bar A)}_\mu(B-\bar B) -  [{\cal D}^{(A-\bar A)}_\mu(C-\bar C)](\tilde C -\bar {\tilde C }),\nonumber \\
s_{ab} \epsilon  &= &  \sigma -[ B-\bar B, C-\bar{C} ] + (\tilde C-\bar{\tilde C}) (C-\bar{C})^2,
 \nonumber \\
s_{ab} \tilde\epsilon  &= &\tilde \sigma - [B-\bar B, \tilde C-\bar{\tilde C}], \ \
s_{ab} \rho  =\tilde v.
\end{eqnarray}
Rest fields, whose anti-BRST transformations are not written here   do not change under the extended anti-BRST transformation.
 To describe  the superspace formulation of Yang-Mills theory in VSR
having both the extended BRST and extended anti-BRST invariance,  we need
two additional Grassmannian coordinates $\theta, \bar\theta$.
Now, it is   straightforward to write the superfields in this formulation where the BRST
and anti-BRST transformations merely correspond  to translations in
the $\theta$ and $\bar{\theta}$ coordinates respectively.
Thus, we see that the results of superspace description of Lorentz invariant 1-form theory 
\cite{ba} also hold in the case of Lorentz breaking theory.

\section{2-form gauge theory: VSR modified BV action in superspace}
The study of Abelian 2-form gauge theory is important because it plays a crucial role in
studying the theory for classical strings \cite{a},  vortex motion in an irrotational, incompressible
fluid \cite{b,c} and the dual formulation of the Abelian Higgs model \cite{d}.
In this section, we discuss the VSR modified 
extended BRST and extended anti-BRST transformations (which include a shift symmetry)
 for the BV action of 2-form gauge theory.   We further demonstrate a superspace description 
 for the 2-form gauge theory having extended BRST invariance and extended 
 anti-BRST invariance.   
To do so, we  start with the classical Lagrangian density for  Abelian   
rank-2 antisymmetric tensor field ($B_{\mu\nu}$) theory  in VSR as \cite{sud,sudpanigrahi}  
 \begin{equation}
 {{\cal L}_0}=\frac{1}{12}\tilde{F}_{\mu \nu \rho}\tilde{F}^{\mu \nu \rho},\label{kin}
\end{equation}
where $\tilde{F}_{\mu \nu \rho}$ is the VSR-modified field-strength tensor   defined as $\tilde{F}_{\mu \nu \rho}\equiv   {\partial}_\mu 
B_{\nu\rho}+ {\partial}_\nu B_{\rho\mu}+ {\partial}_\rho
B_{\mu\nu}- \frac{1}{2}\frac{m^2}{n\cdot \partial}n_\mu 
B_{\nu\rho} -\frac{1}{2}\frac{m^2}{n\cdot \partial}n_\nu 
B_{ \rho\mu}-\frac{1}{2}\frac{m^2}{n\cdot \partial}n_\rho 
B_{\mu\nu}.$  Here  $n_\mu$ is a fixed null vector and transforms multiplicatively, as before, under a VSR transformation
to ensure the invariance of  non-local terms.

 This field-strength tensor and, consequently,  Lagrangian density  is not invariant under the Lorentz invariant gauge transformation 
 $ \delta B_{\mu\nu}= \partial_\mu \zeta_\nu - \partial_\nu\zeta_\mu$, where  $\zeta_{\mu}(x)$ is an arbitrary vector  
field.
Rather, this  is invariant under the following VSR-modified gauge transformation 
\begin{eqnarray}
\delta 
{B_{\mu\nu}}&=& {\partial}_{\mu}\zeta_{\nu} - {\partial}_{\nu}\zeta_{\mu} -\frac{1}{2}\frac{m^2}{n\cdot \partial}n_{\mu}\zeta_{\nu}+\frac{1}{2}\frac{m^2}{n\cdot \partial}n_{\nu}\zeta_{\mu}.
\end{eqnarray}
Since the Lagrangian density is invariant under above non-local transformation, hence,
to quantize this theory following BRST technique, it is necessary to introduce two
anticommuting vector fields $\rho_{\mu}$ and $\tilde\rho_{\mu}$, 
a commuting vector field $\beta_{\mu}$, two anticommuting scalar fields $\chi$ and $\tilde\chi$, 
and the commuting scalar fields $\sigma, \varphi$ and $ \tilde\sigma $ \cite{ht}.
Involving all  these fields, the  gauge breaking term together with the ghosts  is given as \cite{sudpanigrahi}
\begin{eqnarray}
 {{\cal L}}_{gf} 
&=& i\tilde\rho_\nu \left(\partial_\mu\partial^\mu \rho^\nu -\partial_\mu\partial^\nu
\rho^\mu -m^2\rho^\nu +\frac{1}{2}	\frac{m^2}{n\cdot \partial}n^\nu\partial\cdot\rho
+ \frac{1}{2}	\frac{m^2}{n\cdot \partial} \partial^\nu n\cdot\rho -  \frac{1}{4}\frac{m^4}{(n\cdot\partial)^2}n^\nu n\cdot\rho\right)\nonumber\\
& -&\tilde{\sigma}
(\partial_\mu\partial^\mu -m^2)\sigma +\beta_\nu\partial_\mu B^{
\mu\nu} -\frac{1}{2}m^2\beta_\nu\frac{1}{n\cdot\partial}n_\mu B^{\mu\nu}+\lambda_1\beta_\nu\beta^\nu  \nonumber\\
&-& \beta_\nu\partial^\nu\varphi -i\tilde\chi \partial_\mu\rho^\mu +\frac{i}{2}m^2\tilde\chi\frac{1}{n\cdot\partial}n_\mu \rho^\mu-i\lambda_2\tilde\chi\chi -i\tilde\rho^\mu\partial_\mu\chi-\frac{i}{2}\frac{m^2}{n\cdot\partial}\tilde\rho^\mu n_\mu\chi, \label{gfix}
\end{eqnarray}
 $k_1$ and $k_2$ are arbitrary gauge parameters.
 The ghost  and ghost of ghost  propagators in momentum space are given,  respectively,  by
 \cite{sud}
 \begin{eqnarray}
&&D_{\mu\nu}^{gh}(k) =-\frac{1}{k^2+m^2}\left[g_{\mu\nu}+\frac{k_\mu k_\nu}{m^2}\right],\nonumber\\
&&D^{ggh}(p) =-\frac{1}{p^2+m^2}.
\end{eqnarray}
These expressions suggest that ghost and ghost of ghost have same mass $m$. Also, these 
propagators follow a large momentum behavior similar to the
Lorentz-invariant case.  Therefore, the 2-form theory in VSR is a renormalizable theory.

By incorporating the gauge breaking term (\ref{gfix}), the Lagrangian density in VSR reads effectively
\begin{equation} 
 {{\cal L}}  = {{\cal L}}_0 +  {{\cal L}}_{gf}, \label{act}
\end{equation}
which is invariant under the following nilpotent BRST  
transformation:
\begin{eqnarray}
s_b B_{\mu\nu} &=& 
( {\partial}_\mu\rho_\nu - {\partial}_\nu\rho_\mu -\frac{1}{2}\frac{m^2}{n\cdot \partial}n_\mu\rho_\nu + \frac{1}{2}\frac{m^2}{n\cdot \partial}n_\nu\rho_\mu),  \nonumber \\  
s_b\rho_\mu & =&   -i {\partial}_\mu\sigma +\frac{i}{2}\frac{m^2}{n\cdot \partial}n_\mu\sigma
,     \ s_b\sigma 
= 0,\ s_b\tilde\rho_\mu  = i\beta_\mu, \nonumber\\
s_b\beta_\mu &=& 0, \ \ s_b\tilde\sigma  = -\tilde\chi,  \ \ \ 
s_b\tilde\chi =0, \ \
s_b\varphi  =  \chi,  \ \ s_b\chi =0.\label{sym}
\end{eqnarray} 
Since the gauge-fixing and ghost part of the effective Lagrangian density   
is  BRST-exact and therefore can
be expressed in terms of BRST variation of some gauge-fixed fermion $\Psi$.
 Therefore,   
\begin{equation}
 {\cal L}_{gf}=s_b \Psi,
\end{equation}
where    $\Psi$ has the following form:
\begin{eqnarray}
\Psi  
& =&-i \left[\tilde\rho_\nu {\partial}_\mu  {B^{\mu\nu}}+\tilde\sigma{\partial}_\mu\rho^\mu+\varphi\partial_\mu\tilde\rho^\mu -\tilde\rho_\nu k_1 \beta^\nu-\varphi  k_2 \tilde\chi -\frac{\tilde\rho_\nu}{2}\frac{m^2}{n\cdot \partial}n_{\mu}{B}^{\mu\nu}\right.\nonumber\\
&-&\left. \frac{\tilde\sigma}{2}\frac{m^2}{n\cdot \partial}n_{\mu}\rho^\mu -\frac{\varphi }{2}\frac{m^2}{n\cdot \partial}n_{\mu}\tilde\rho^\mu \right].  \label{gff}
\end{eqnarray}
This gauge-fixed fermion is very important to identify the antifields of 
BV action.

\subsection{VSR modified extended BRST invariant BV action }
The extended BRST and extended
anti-BRST invariant formulations  of the Lorentz invariant BV action lead  to the 
proper identification
of the antifields through equations of motion of auxiliary field variables  \cite{ad,ba,fk}. The  study of extended (including shift) symmetries introduced through collective fields 
is important because it ensure
  that Schwinger-Dyson
equations at the level of the BRST algebra  
 can be performed within the Feynman path integral \cite{alf}.
In this subsection, we study the VSR modified extended BRST invariant BV action. To do so,
 we first  deviate all the fields from their original values.
This enlarges in a trivial way the symmetry content of  the theory, adding extra  shift symmetries.
To study the  extended BRST structure of the Abelian rank-2 tensor field theory
in VSR, we shift all the fields of theory from their original values as follows 
$ {B_{\mu\nu}}- {\bar B_{\mu\nu}},\rho_\mu 
-\bar\rho_\mu, \tilde\rho_{\mu}-\bar{
\tilde\rho}_{\mu}, \sigma_\mu - \bar{\sigma}_\mu, \tilde\sigma_{\mu}-\bar{\tilde\sigma}_{\mu
}, \beta_\mu - \bar\beta_\mu, \chi-\bar{\chi}, \tilde\chi - \bar{\tilde\chi},
\varphi -\bar\varphi$.
This leads to
the following shifted  Lagrangian density:
\begin{eqnarray}
 {\bar{\cal L}}= {{\cal L}}_0  ( {B_{\mu\nu}}- {\bar B_{\mu\nu}})+ {{\cal L}}_{gf}  ( {B_{\mu\nu}}- {\bar B_{\mu\nu}},\Xi-\bar{\Xi} ), 
\end{eqnarray}
where $\Xi-\bar{\Xi} = \rho_\mu 
-\bar\rho_\mu, \tilde\rho_{\mu}-\bar{
\tilde\rho}_{\mu}, \sigma_\mu - \bar{\sigma}_\mu, \tilde\sigma_{\mu}-\bar{\tilde\sigma}_{\mu
}, \beta_\mu - \bar\beta_\mu, \chi-\bar{\chi}, \tilde\chi - \bar{\tilde\chi},
\varphi -\bar\varphi$.

The explicit form of ${{\cal L}}_{gf}  ( {B_{\mu\nu}}- {\bar B_{\mu\nu}},\Xi-\bar{\Xi})$
is given by
\begin{eqnarray}
{{\cal L}}_{gf}  &=& -i\left[\partial_\mu\tilde \rho_\nu \partial^\mu \rho^\nu + m^2\tilde\rho_\nu\rho^\nu -\partial_\mu\tilde \rho_\nu \partial^\mu \bar{\rho^\nu} - m^2\tilde\rho_\nu\bar{\rho}^\nu- \partial_\mu\tilde \rho_\nu \partial^\nu \rho^\mu+\frac{1}{2}\partial_\mu\tilde \rho_\nu\frac{m^2}{n\cdot \partial}n^\nu\rho^\mu \right.\nonumber\\
&+&\left.\partial^\nu\rho^\mu\frac{1}{2}\frac{m^2}{n\cdot \partial} n_\mu\tilde{\rho}_\nu-\frac{1}{4}\left(\frac{m^2}{n\cdot \partial}\right)^2 n_\mu\tilde\rho_\nu n^\nu\rho^\mu+ \partial_\mu\tilde \rho_\nu \partial^\nu \bar{\rho}^\mu -\frac{1}{2}\partial_\mu\tilde \rho_\nu\frac{m^2}{n\cdot \partial}n^\nu\bar{\rho}^\mu \right.\nonumber\\
&-&\left. \frac{1}{2}\partial^\nu\bar{\rho}^\mu\frac{m^2}{n\cdot \partial}n_\mu\tilde\rho_\nu+\frac{1}{4}\left(\frac{m^2}{n\cdot \partial}\right)^2 n_\mu\tilde\rho_\nu n^\nu\bar{\rho}^\mu-\partial_\mu\bar{\tilde\rho}_\mu \partial^\mu\rho^\nu  -m^2 \bar{\tilde\rho}_\nu \rho^\nu + \partial_\mu\bar{\tilde\rho}_\nu \partial^\mu\bar{\rho}^\nu \right. \nonumber\\
&+&\left. m^2\bar{\tilde\rho}_\nu \bar{\rho}^\nu + \partial_\mu\bar{\tilde\rho}_\nu \partial^\nu\rho^\mu - \frac{1}{2}\partial^\mu\bar{\tilde\rho}^\mu\frac{m^2}{n\cdot \partial}n^\nu\tilde\rho^\mu -\frac{1}{2}\partial^\nu\rho^\mu \frac{m^2}{n\cdot \partial}n_\mu\bar{\tilde\rho}^\nu + \frac{1}{4}\left(\frac{m^2}{n\cdot \partial}\right)^2 n_\mu\bar{\tilde\rho}_\nu n^\nu\rho^\mu \right.\nonumber\\
&-&\left. \partial_\mu\bar{\tilde\rho}_\nu \partial^\nu\bar{\rho}^\mu +\frac{1}{2}\partial_\mu\bar{\tilde\rho}_\nu \frac{m^2}{n\cdot \partial}n^\nu\bar{\rho}^\mu + \frac{1}{2}\partial^\nu\bar{\rho}^\mu \frac{m^2}{n\cdot \partial}n_\mu\bar{\tilde\rho}_\mu - \frac{1}{4}\left(\frac{m^2}{n\cdot \partial}\right)^2 n_\mu\bar{\tilde\rho}_\nu n^\nu\bar{\rho}^\mu   + \partial_\mu\tilde{\sigma}\partial^\mu \sigma \right. \nonumber\\
&-&\left.
\partial_\mu\tilde{\sigma}\partial^\mu \bar{\sigma} -\partial_\mu\bar{\tilde{\sigma}}\partial^\mu {\sigma} + \partial_\mu\bar{\tilde{\sigma}}\partial^\mu\bar{\sigma} + m^2\tilde{\sigma}\sigma - m^2\tilde{\sigma}\bar{\sigma} - m^2\bar{\tilde\sigma}\sigma +m^2\bar{\tilde\sigma}\bar{\sigma}\right.\nonumber\\
&+&\left.
\beta_\nu\partial_\mu B^{\mu\nu} - \beta_\nu\partial_\mu \bar B^{\mu\nu}+\beta_\nu\partial^\nu\bar{\varphi}-\bar{\beta}_\nu\partial_\mu B^{\mu\nu} - \bar{\beta}_\nu\partial_\mu \bar{B}^{\mu\nu}+\bar{\beta}_\nu\partial^\nu\varphi+\bar{\beta}\partial^\nu\bar\varphi - \beta_\nu\partial^\nu\varphi \right.\nonumber\\
&-&\left. \frac{1}{2}\beta_\nu\frac{m^2}{n\cdot \partial}n_\mu B^{\mu\nu} + \frac{1}{2}
\beta_\nu\frac{m^2}{n\cdot \partial}n_\mu \bar B^{{\mu\nu}}-\frac{\beta_\nu}{2}\frac{m^2}{n\cdot \partial}n^\nu\bar{\varphi} + \frac{\bar{\beta_\nu}}{2}\frac{m^2}{n\cdot \partial}n_\mu B^{\mu\nu}+\frac{\bar{\beta_\nu}}{2}\frac{m^2}{n\cdot \partial}n_\mu\bar{ B^{\mu\nu}} \right.\nonumber\\
&-&\left. \frac{\beta_\nu}{2}\frac{m^2}{n\cdot \partial}n^\nu\varphi - \frac{\bar{\beta}_\nu}{2}
\frac{m^2}{n\cdot \partial}n^\nu\varphi +\frac{\bar{\beta}_\nu}{2}\frac{m^2}{n\cdot \partial}
n^\nu \bar{\varphi}+k_1\beta_\nu  \beta^\nu - k_1\beta_\nu\bar{\beta}^\nu -k_1\bar{\beta}_\nu 
\beta^\nu +k_1\bar{\beta}_\nu \bar{\beta}^\nu \right.\nonumber\\
&-& \left.i\left(\tilde\chi\partial_\mu\rho^\mu - \tilde\chi \partial_\mu \bar{\rho}^\mu-
\bar{\tilde\chi}\partial_\mu\rho^\mu +\bar{\tilde\chi}\partial_\mu\bar{\rho}^\mu - 
\frac{\bar{\chi}}{2}\frac{m^2}{n\cdot \partial}n\cdot\rho+\frac{\bar{\chi}}{2}\frac{m^2}{n
\cdot \partial}n\cdot\bar{\rho}+ \frac{\bar{\chi}}{2}\frac{m^2}{n\cdot \partial}n\cdot\rho
\right.\right.\nonumber\\
&-&\left.\left.\frac{\bar{\chi}}{2}\frac{m^2}{n\cdot \partial}n\cdot\bar{\rho} \right)- i\left(\chi
\partial_\mu\tilde\rho^\mu - \chi\partial_\mu\bar{\tilde\rho}^\mu - \bar{\chi}\partial_\mu
\tilde\rho^\mu + \bar{\chi}\partial_\mu \bar{\tilde\rho}^\mu-\frac{\chi}{2}\frac{m^2}{n\cdot 
\partial}n\cdot\tilde{\rho} +\frac{\chi}{2}\frac{m^2}{n\cdot \partial}n\cdot
\bar{\tilde{\rho}}\right.\right.\nonumber\\
&+&\left.\left. \frac{\bar{\chi}}{2}\frac{m^2}{n\cdot \partial}n\cdot\tilde{\rho}- \frac{\chi}{2}
\frac{m^2}{n\cdot \partial}n\cdot\bar{{\tilde{\rho}}}\right) -k_2\chi \tilde{\chi} + k_2\chi
\bar{\tilde\chi} +k_2\bar{\chi} \tilde{\chi} -k_2\bar{\chi}\bar{\tilde\chi}\right].
\end{eqnarray} 
This Lagrangian density coincides with  ${\cal L}_{gf}$ of  (\ref{gfix}) when   bar fields vanish. 
It is evident that this Lagrangian density is invariant under the BRST transformation 
 (\ref{sym}) for the shifted   fields.  
In addition, there exists the following shift symmetry also:
\begin{eqnarray}
s_b \Phi (x)= \alpha (x),\ \ s_b \bar\Phi (x)&=& \alpha (x),
\end{eqnarray}
which leaves this Lagrangian density   invariant. Here $\Phi $ and $\bar\Phi  $ are generic notation for all
fields and shifted fields respectively. 
  The  form the extended BRST symmetry.
  The extended BRST transformation, which is comprised by the BRST symmetry along with the above shift symmetry, is then  given by
\begin{eqnarray}
s_b \Phi (x)= \alpha (x),\ \ s_b \bar\Phi (x)&=& \alpha (x)-\beta (x),
\end{eqnarray}
where $\beta (x)$ refers the original BRST transformation collectively,
whereas $\alpha (x)$ refers  the shift transformation collectively.
In order to quantize theory collectively, we need to fix the gauge for all the local 
symmetry.  Therefore,  corresponding to this local shift symmetry, one 
needs the theory to be gauge-fixed and this 
 leads to an additional BRST symmetry \cite{ba}.      
The extended BRST symmetry transformation 
for all the fields  are given by
\begin{eqnarray}
s_b B_{\mu\nu}&=&\psi_{\mu\nu},\ \ s_b {\bar B}_{\mu\nu} =\psi_{\mu\nu}-(\partial_\mu\rho_\nu -  \partial_\mu{\bar \rho}_\nu  -\partial_\nu\rho_\mu + \partial_\nu{\bar\rho}_\mu
\\
&-& \frac{1}{2}\frac{m^2}{n\cdot \partial}n_\mu \rho_\nu +\frac{1}{2}\frac{m^2}{n\cdot \partial}n_\mu \bar{\rho}_\nu + \frac{1}{2}\frac{m^2}{n\cdot \partial}n_\mu \bar{\rho}_\mu
 -\frac{1}{2}\frac{m^2}{n\cdot \partial}n_\mu \bar{\rho}_\mu) , \nonumber\\
 s_b\bar{\rho}_\mu &=&  \epsilon_\mu +i\partial_\mu\sigma -i
\partial_\mu{\bar\sigma} +\frac{i}{2}\frac{m^2}{n\cdot \partial}n_\mu\sigma +\frac{i}{2}\frac{m^2}{n\cdot \partial}n_\mu {\bar\sigma}  ,\nonumber\\
s_b\tilde\rho_\mu & =& \xi_\mu,\  s_b\bar{\tilde\rho}_\mu =\xi_\mu -i\beta_\mu +i\bar{\beta}_\mu,\ \ s_b\bar\sigma = \varepsilon,\nonumber\\ 
 s_b\sigma  &=&  \varepsilon,\
s_b\beta_\mu  =  \eta_\mu,\ \ s_b\bar{\beta}_\mu = \eta_\mu, \
s_b\tilde\sigma  = \psi,\ \ s_b\tilde\chi =\eta,\nonumber\\ 
 s_b\bar{\tilde\sigma} &=& \psi +\tilde\chi -\bar{\tilde\chi},\ \ s_b\bar{\tilde\chi} = \eta,\ \ s_b\bar\varphi =
\phi- \chi +\bar\chi,\ 
 s_b\chi  = \Sigma,\nonumber\\
s_b\varphi &=& \phi,\ \ s_b\bar\chi =\Sigma,\
 s_b \xi_i=0,\ \xi_i\equiv [\psi_{\mu\nu}, \epsilon_\mu, \xi_\mu, \varepsilon, \eta_\mu, \psi, \eta, \phi, \Sigma ]. 
\label{brs1} 
\end{eqnarray}
The fields $\tilde{\psi}_{\mu\nu},$  $\epsilon_\mu,$ $\xi_\mu,$ $\varepsilon,$ $\eta_\mu,$ $\psi,$ $\eta,$ 
$\phi$ and $ \Sigma$ are introduced as ghost fields associated with the shift symmetry corresponding to the fields 
$B_{\mu\nu},$ $\rho_\mu,$ $\tilde\rho_\mu,$ $\sigma,$ $\beta_\mu,$ $\tilde\sigma,$ $\tilde\chi,$ $\varphi$ and $\chi$ respectively.
Further, we  add following antighost  fields $B_{\mu\nu}^\star,$ $\rho_{\mu}^\star,$ 
$\tilde{\rho}_{\mu}^\star,$ $\sigma^\star,$ $\tilde{\sigma}^\star,$ $\beta_\mu^\star,$ $\chi^\star,$
$\tilde{\chi}^\star$ and $ \varphi^\star $ corresponding to the fields 
$B_{\mu\nu},$ $\rho_\mu,$ $\tilde\rho_\mu,$ $\sigma,$ $\beta_\mu,$ $\tilde\sigma,$ $\tilde\chi,$ $\varphi$ and $\chi$ respectively with opposite statistics. These 
antighost fields transform under  BRST transformations as following:
\begin{eqnarray}
s_b B_{\mu\nu}^\star &=&L_{\mu\nu},\ \ 
s_b \rho_{\mu}^\star = M_\mu,\ \
s_b \tilde{\rho}_{\mu}^\star  =  \bar{M}_\mu,\ \
s_b \sigma^\star = N, \nonumber\\
s_b \tilde{\sigma}^\star &=& \bar N, \ \
s_b \beta_\mu^\star = S_\mu, \ \
s_b \chi^\star  =  O,\ \
s_b \tilde{\chi}^\star = \bar O, \ \
 s_b \varphi^\star  = T, \label{brs2}  
\end{eqnarray}
where $L_{\mu\nu},  M_\mu , \bar{M}_\mu, N, \bar N, S_\mu, O, \bar O, T$  are  the Nakanishi-Lautrup type auxiliary fields and do not change under BRST transformation which ensure the nilpotency of BRST symmetry.

Now,  we  fix the gauge for shift symmetry  in VSR  by choosing the following gauge-fixed Lagrangian density:
\begin{eqnarray}
{\bar{\cal L}}_{gf}  
&=& L_{\mu\nu}{\bar B}^{\mu\nu} -B_{\mu\nu}^\star (\psi^{\mu\nu}-
\partial^\mu\rho^\nu + 
\partial^\mu{\bar\rho}^\nu +\partial^\nu\rho^\mu
-\partial^\nu{\bar\rho}^\mu  +\frac{1}{2}\frac{m^2}{n\cdot \partial}n^\mu \rho^\nu-\frac{1}{2}\frac{m^2}{n\cdot \partial}n^\mu \bar{\rho}^\nu     \nonumber\\
&-&  \frac{1}{2}\frac{m^2}{n\cdot \partial}n^\nu \rho^\mu+ \frac{1}{2}\frac{m^2}{n\cdot \partial}n^\nu \bar\rho^\mu)+{\bar M}_\mu{\bar\rho}^\mu +{\tilde \rho}_\mu^\star (\epsilon^\mu +i\partial^\mu\sigma -i\partial^\mu\bar\sigma -i \frac{1}{2}\frac{m^2}{n\cdot \partial}n^\mu\sigma \nonumber\\
&+&  i \frac{1}{2}\frac{m^2}{n\cdot \partial}n^\mu \bar\sigma) +M_
\mu\bar{\tilde \rho}^\mu +\rho_\mu^\star (\xi^\mu-i\beta^\mu +i{\bar \beta}^\mu ) + N\bar\sigma -\sigma^\star \varepsilon  + \bar N\bar{\tilde\sigma} -{\tilde\sigma}^\star (
\psi-\tilde\chi +\bar{\tilde\chi}) \nonumber\\
&+& \bar O\bar\chi +{\tilde\chi}^\star \Sigma +O\bar{\tilde\chi} +\chi^\star \eta+
T\bar\varphi -\varphi^\star (\phi -\chi+\bar\chi)
+S_\mu{\bar \beta}^\mu -\beta_\mu^\star \eta^\mu,
\label{la}
\end{eqnarray}
which sets all the bar fields to zero and thus we recover the original 
theory. This gauge-fixing term is also invariant 
under the extended BRST symmetry transformations given in Eqs. (\ref{brs1}) and (\ref{brs2}).
 The gauge-fixed  extended Lagrangian density, $
{\bar{\cal L}}_{gf}$, after exploiting the equations of motion of auxiliary fields  has 
the following form:
\begin{eqnarray}
\bar{\cal L}_{gf}  
&=&       -B_{\mu\nu}^\star (\psi^{\mu\nu}-
\partial^\mu\rho^\nu +\partial^\nu\rho^\mu+\frac{1}{2}\frac{m^2}{n\cdot \partial}n^\mu \rho^\nu- \frac{1}{2}\frac{m^2}{n\cdot \partial}n^\nu \rho^\mu)
+{\tilde \rho}_\mu^\star (\epsilon^\mu +i\partial^\mu\sigma \nonumber\\
&-&  \frac{i}{2}\frac{m^2}{n\cdot \partial}n^\mu \sigma)   +\rho_\mu^\star (\xi^\mu-i\beta^\mu )- \sigma^\star \epsilon   -{\tilde\sigma}^\star (\psi-\tilde\chi ) 
+{\tilde\chi}^\star \Sigma   
 + \chi^\star \eta 
-\varphi^\star (\phi -\chi )\nonumber\\
&-&\beta_\mu^\star \eta^\mu.        \label{per}
\end{eqnarray}
As the gauge-fixed  fermion $\Psi$ for Abelian rank-2 antisymmetric tensor field 
in VSR depends only on the original fields, then the
most general gauge-fixed Lagrangian density  is given by
\begin{eqnarray}
{\cal L}_{gf} &= &s_b \Psi  
= \psi_{\mu\nu}\frac{\delta\Psi}{\delta B_{\mu\nu}}+
\epsilon _\mu \frac{\delta\Psi}{\delta\rho_\mu}+ \xi_\mu \frac{\delta\Psi}
{\delta\tilde{\rho}_\mu} +
\varepsilon \frac{\delta\Psi}{\delta\sigma}\nonumber\\
&+&\psi  \frac{\delta\Psi}{\delta
\tilde\sigma} +\eta_\mu \frac{\delta\Psi}{\delta\beta_\mu}+
\Sigma  \frac{\delta\Psi}{\delta\chi}+\eta \frac{\delta\Psi}{\delta\tilde\chi}
+\phi \frac{\delta\Psi}{\delta\varphi}.
\end{eqnarray}
Utilizing the fermionic and bosonic behavior of fields, this can  further be written as  
\begin{eqnarray}
{\cal L}_{gf} &= & -\frac{\delta\Psi}{\delta B_{\mu\nu}}\psi_{\mu\nu}+
 \frac{\delta\Psi}{\delta\rho_\mu}\epsilon _\mu + \frac{\delta\Psi}
{\delta\tilde{\rho}_\mu} \xi_\mu -
 \frac{\delta\Psi}{\delta\sigma}\varepsilon\nonumber\\
&-& \frac{\delta\Psi}{\delta
\tilde\sigma}\psi -\frac{\delta\Psi}{\delta\beta_\mu}\eta_\mu +
\frac{\delta\Psi}{\delta\chi}\Sigma + \frac{\delta\Psi}{\delta\tilde\chi}\eta
-\frac{\delta\Psi}{\delta\varphi}\phi. \label{psi} 
\end{eqnarray}
Now, we are able to write the effective Lagrangian density in VSR, ${\cal L}_{eff} =  { {\cal L}_0}+
{\cal L}_{gf}+ {\bar{\cal L}}_{gf}$,  as follows
\begin{eqnarray}
{\cal L}_{eff} 
&= & \frac{1}{12}\tilde{F}_{\mu \nu \rho}\tilde{F}^{\mu \nu \rho} + B_{\mu\nu}^\star \left(
\partial^\mu\rho^\nu -\partial^\nu\rho^\mu- \frac{1}{2}\frac{m^2}{n\cdot \partial}n^\mu \rho^\nu    + \frac{1}{2}\frac{m^2}{n\cdot \partial}n^\nu \rho^\mu \right) \nonumber\\
&+& i{\tilde \rho}_\mu^\star\left( \partial^\mu \sigma- \frac{1}{2}\frac{m^2}{n\cdot \partial}n^\mu\sigma \right)
-i\rho_\mu^\star \beta^\mu 
 + {\tilde\sigma}^\star \tilde\chi
+\varphi^\star \chi -\left(B_{\mu\nu}^\star +\frac{\delta\Psi}{\delta B^{\mu\nu}}\right)\psi^{\mu\nu} \nonumber\\
&+&\left({ \rho}_\mu^\star +\frac{\delta\Psi}{\delta\tilde{\rho}^\mu}\right)\xi^\mu  
 +\left({\tilde \rho}_\mu^\star +\frac{\delta\Psi}{\delta\rho^\mu}\right)\epsilon^\mu 
-\left(\sigma^\star +\frac{\delta\Psi}{\delta\sigma}\right)\varepsilon - \left (\tilde{\sigma}^\star +\frac{\delta\Psi}{\delta\tilde\sigma} \right )\psi\nonumber\\
&+ &
\left (\tilde{\chi}^\star +\frac{\delta\Psi}{\delta\chi} \right )\Sigma 
+ \left ({\chi}^\star +\frac{\delta\Psi}{\delta\tilde\chi} \right )\eta 
-\left (\varphi^\star +\frac{\delta\Psi}{\delta\varphi} \right )\phi
- \left (\beta_\mu^\star +\frac{\delta\Psi}{\delta\beta^\mu} \right )\eta^\mu,
\end{eqnarray}
here  expressions (\ref{kin}), (\ref{per}) and (\ref{psi}) are utilized.
Exploiting the equations of motion of
ghost fields  associated with the shift symmetry, we get  the following identifications
of antighost fields: 
\begin{eqnarray}
B_{\mu\nu}^\star &=&-\frac{\delta\Psi}{\delta B^{\mu\nu}},\ \
{\tilde \rho}_\mu^\star =-\frac{\delta\Psi}{\delta\rho^\mu},\ \
{ \rho}_\mu^\star  = -\frac{\delta\Psi}{\delta\tilde{\rho}^\mu},\nonumber\\
\sigma^\star &=&-\frac{\delta\Psi}{\delta\sigma},\ \ 
\tilde{\sigma}^\star  = -\frac{\delta\Psi}{\delta\tilde\sigma},\ \ 
\tilde{\chi}^\star =-\frac{\delta\Psi}{\delta\chi},\nonumber\\
{\chi}^\star &=&-\frac{\delta\Psi}{\delta\tilde\chi},\ \
\beta_\mu^\star =-\frac{\delta\Psi}{\delta\beta^\mu}, \ \ 
\varphi^\star  = -\frac{\delta\Psi}{\delta\varphi}.
\end{eqnarray}
With a particular expression of  gauge-fixed fermion $\Psi$ as given in   (\ref{gff}),
 these leads to 
\begin{eqnarray}
&&{B}_{\mu\nu}^\star  = -i\left(\partial_\mu - \frac{1}{2}\frac{m^2}{n\cdot \partial}n_\mu\right)\tilde{\rho}_\nu,\ \
{\tilde \rho}_\mu^\star  = -i\left(\partial_\mu - \frac{1}{2}\frac{m^2}{n\cdot \partial}n_\mu 
\right)\tilde{\sigma} ,\\
&&{ \rho}_\mu^\star   = i\left(\partial^\nu B_{\nu\mu}- \frac{1}{2}\frac{m^2}{n\cdot \partial}n^\nu  B_{\nu\mu}        -k_1\beta_\mu
\right) ,\ \ \sigma^\star =0,\nonumber \\
&&\tilde{\sigma}^\star = i\left(\partial_\mu -\frac{1}{2}\frac{m^2}{n\cdot \partial}n_\mu
\right)\rho^\mu,\ \
\tilde{\chi}^\star =0,\  
{\chi}^\star  = -i k_2\varphi, \nonumber \\ 
&&\ 
\beta_\mu^\star =i k_1\tilde{\rho}_\mu,\ \ \tilde{\varphi^\star}=  i\left( \partial_\mu\tilde{\rho}
^\mu -\frac{1}{2}\frac{m^2}{n\cdot \partial}n_\mu \tilde{\rho}^\mu -k_2\tilde\chi\right). \label{antifield}
\end{eqnarray}
Here we see that these antighost  fields coincide with  the antifields 
of the theory.  We note that these antifields are nonlocal which describe
the features of VSR. Thus, analogous to Lorentz invariant theory, the antifields are obtained naturally in VSR also.
One can see that  these identifications lead  the effective Lagrangian density   to 
their original 
form as given in Eq. (\ref{act}). 
Now, we can describe the  gauge-fixing part of the effective Lagrangian density in terms of
the BRST variation of a generalized gauge-fixed fermion as follows 
\begin{eqnarray}
 {\cal L}_{eff}&=&  {\cal L}_0(B_{\mu\nu}-\bar B_{\mu\nu}) + s_b \left(B_{\mu\nu}^\star \bar B^{\mu\nu} +
\rho_{\mu}^\star \bar {\tilde\rho}^{\mu} +
\tilde\rho_{\mu}^\star \bar{ \rho}^{\mu} +\sigma^\star\bar\sigma +
\tilde\sigma^\star\bar{\tilde\sigma}+\beta_\mu^\star\bar\beta^\mu  \right.\nonumber\\
&+&\left. 
\chi^\star\bar{\tilde\chi} +\tilde\chi^\star\bar{\chi}
+\varphi^\star\bar\varphi
\right),\nonumber\\
 &\equiv & {{\cal L}_0}(B_{\mu\nu}-\bar B_{\mu\nu})+s_b \Phi^\star\bar\Phi.\label{comple}
\end{eqnarray}
Here  $\Phi^\star$ and $\bar\Phi$ are the generic notation
for antifields and (corresponding) shifted fields, respectively. The ghost number of   $ \Phi^\star\bar\Phi$   is $-1$. We thus recover the BV action for
Abelian 2-form gauge theory in VSR with the identification of antifields. 
 
\subsection{ VSR modified extended BRST invariant superspace formulation}   
In this section we discuss  a superspace
formalism of the VSR modified 2-form theory having extended BRST invariance. In this 
regard,  we extend the space to a superspace   $(x^\mu, \theta)$ by introducing 
a   fermionic coordinate  $\theta$. 
In this superspace, the ``superconnection" $2$-form is defined by
\begin{equation}
\omega^{(2)} =\frac{1}{2 !}{\cal B}_{\mu\nu}(x, \theta ) (dx^\mu\wedge dx^\nu ) +
 {\cal M}_{\mu}(x, \theta ) (dx^\mu\wedge d\theta ) + {\cal N} (x, \theta ) (d\theta\wedge d\theta ), 
\end{equation}
where   $d= dx^\mu\left(\partial_\mu - \frac{i}{2}\frac{m^2}{n\cdot \partial}n_\mu \right) +d\theta  \left(\partial_\theta - \frac{i}{2}\frac{m^2}{n\cdot \partial}n_\theta \right)$  is the exterior derivative.
By requiring  the field
strength,   $F^{(3)}= d\omega^{(2)}$,  to vanish
along the $\theta$ direction, we get
the    following form of the component of the superfields in VSR  
\begin{eqnarray}
{\cal B}_{\mu\nu}(x, \theta ) &=& B_{\mu\nu} (x) +\theta  (s_bB_{\mu\nu}),\nonumber\\
{\cal M}_{\mu}(x, \theta ) &=& \rho_\mu (x) +\theta (s_b\rho_\mu),\nonumber\\
 {\cal N} (x, \theta ) &=& \sigma (x) +\theta (s_b \sigma). 
\end{eqnarray}
In the similar fashion, 
we are able to write all the  fields  involved in extended BV action in superspace as 
\begin{eqnarray} 
{\cal B}_{\mu\nu}(x, \theta ) &=& B_{\mu\nu} (x) +\theta \psi_{\mu\nu},\ \ {\cal M}_{\mu}(x, \theta )  =  \rho_{\mu} 
(x) +\theta \epsilon_\mu,\nonumber\\
\bar{{\cal B}}_{\mu\nu}(x, \theta )  
&=&    \bar{B}_{\mu\nu} (x) +\theta (\psi_{\mu\nu}- \partial_\mu\rho_\nu+                                                      \partial_\mu\bar{\rho}_\nu+ \partial_\nu\rho_\mu- \partial_\nu\bar{\rho}_\mu + \frac{1}{2}\frac{m^2}{n\cdot \partial}n_\mu\rho_\nu   \nonumber\\                                                                                                                                                                                                             
&-&   \frac{1}{2}\frac{m^2}{n\cdot \partial}n_\mu\bar{\rho}_\nu - \frac{1}{2}\frac{m^2}{n\cdot \partial}n_\nu \rho_\mu +\frac{1}{2}\frac{m^2}{n\cdot \partial}n_\nu\bar{\rho}_\mu),                                                                                                                     
 \nonumber\\                                                                                                                                                                                                              
\bar{{\cal{M}}}_{\mu}(x, \theta )  
 &=& \bar{\rho}_{\mu} (x) +\theta (
\epsilon_\mu -i\partial_\mu\sigma +i\partial_\mu\bar\sigma+  \frac{i}{2}\frac{m^2}{n\cdot \partial}n_\mu \sigma - \frac{i}{2}\frac{m^2}{n\cdot \partial}n_\mu \bar\sigma ),\nonumber\\
{\cal{N}}(x, \theta ) &=& \sigma (x) +\theta 
\varepsilon,\ \
\bar{{\cal{N}}}(x, \theta )  =  \bar{\sigma}(x) +\theta \varepsilon,  \nonumber\\
\tilde{\cal M}_{\mu}(x, \theta ) &=& \tilde\rho_{\mu} (x) +\theta \xi_\mu,\ \
{\bar{\tilde{\cal M}}}_\mu  (x, \theta )  = { \bar{\tilde\rho}}_ \mu  (x) +\theta (
\xi_\mu -i\beta_\mu +i{\bar{\beta}}_\mu ), \nonumber\\
{\cal S}_{\mu}(x, \theta ) &=& \beta_{\mu} (x) +\theta \eta_\mu,\ \ 
\bar{\cal S}_{\mu}(x, \theta )  =  \bar\beta_{\mu} (x) +\theta \eta_\mu, \nonumber\\
\tilde{\cal N}(x, \theta ) &=& \tilde\sigma (x) +\theta \psi,\ \
{\bar{\tilde{\cal N}}}(x, \theta )  = {\bar{\tilde\sigma}} (x) +\theta (\psi -\tilde
\chi +{\bar{\tilde\psi}}), \nonumber\\
{\cal{O}}(x, \theta ) &=& \chi (x) +\theta 
\Sigma,\ \
\bar{\cal{O}}(x, \theta )  =  \bar\chi (x) +\theta 
\Sigma, \nonumber\\
\tilde{\cal{O}}(x, \theta ) &=& \tilde\chi (x) +\theta 
\eta, \ \
{\bar{\tilde{\cal{O}}}}(x, \theta )  =  \bar{\tilde\chi} (x) +\theta 
\eta, \nonumber\\
{\cal{T}}(x, \theta ) &=& \varphi (x) +\theta 
\phi, \ \
\bar{\cal{T}}(x, \theta ) =  \bar\varphi (x) +\theta 
(\phi -\chi +\bar\chi ).\label{sufi}
\end{eqnarray} 
The components of antifields of the theory in superspace  is written by
\begin{eqnarray}
\bar{\cal B}_{\mu\nu}^\star &=& B_{\mu\nu}^\star  +\theta L_{\mu\nu}, \  
\bar{{\cal M}_{\mu}}^\star = \rho_{\mu}^\star  +\theta M_{\mu},\
\bar{\tilde{\cal M}}_{\mu}^\star  =  \tilde\rho_{\mu}^\star  +\theta \bar M_{\mu},\nonumber\\
\bar{{\cal S}_{\mu}}^\star &=& \beta_{\mu}^\star  +\theta S_{\mu}, \ \
\bar{{\cal N}}^\star  =  \sigma^\star  +\theta N,\ \
\bar{\tilde{\cal N}}^\star = \tilde\sigma^\star  +\theta \bar N,\nonumber\\
\bar{{\cal O}}^\star &=& \chi^\star  +\theta O,\ \
\bar{{\cal T}}^\star = \varphi^\star  +\theta T, \ \
\bar{\tilde{\cal O}}^\star  = \tilde\chi^\star  +\theta \bar O.\label{antisufi}
\end{eqnarray}
Exploiting expressions (\ref{sufi}) and (\ref{antisufi}),
we derive   
\begin{eqnarray}
\frac{\delta}{\delta\theta}\bar{\cal B}_{\mu\nu}^\star \bar{\cal B}^{\mu\nu}&=&   L_{\mu\nu}\bar {B}^{\mu\nu} -B_{\mu\nu}^\star  \left(\psi^{\mu\nu} -\partial^\mu\rho^\nu +\partial^\mu\bar\rho^\nu
+\partial^\nu\rho^\mu -\partial^\nu\bar{\rho}^\mu +\frac{1}{2}\frac{m^2}{n\cdot \partial}n^\mu \rho^\nu \right. \nonumber\\
&-& \left. \frac{1}{2}\frac{m^2}{n\cdot \partial}n^\mu\bar{\rho}^\nu -\frac{1}{2}\frac{m^2}{n\cdot \partial}n^\nu\rho^\mu+\frac{1}{2}\frac{m^2}{n\cdot \partial}n^\nu\bar{\rho}^\mu \right), \nonumber\\
 \frac{\delta}{\delta\theta}\bar{\tilde{\cal M}}_{\mu}^\star  \bar{\cal M}^{\mu}
 &=& \bar M_{\mu}\bar{\rho}^\mu +\tilde\rho_{\mu}^\star  \left(\epsilon ^{\mu}+
i\partial^\mu\sigma -i\partial^\mu\bar\sigma - \frac{i}{2}\frac{m^2}{n\cdot \partial}n^\mu\sigma +  \frac{i}{2}\frac{m^2}{n\cdot \partial}n^\mu\bar{\sigma} \right), \nonumber\\ 
\frac{\delta}{\delta\theta}\bar{\tilde{\cal M}}_{\mu} \bar{\cal M}^{\mu\star}
&=& M_{\mu}\bar{\tilde\rho}^\mu +\rho_{\mu}^\star  (\xi^{\mu}-
i\beta^\mu +i\bar\beta^\mu ),
\nonumber\\
\frac{\delta}{\delta\theta}\bar{\cal N}^\star  \bar{\cal N}
&=& N\bar{\sigma} -\sigma^\star  \varepsilon, 
\ \
\frac{\delta}{\delta\theta}\bar{\tilde{\cal N}}^\star  \bar{\tilde{\cal N}}
 =  \bar N\bar{\tilde\sigma} -\tilde\sigma^\star  (\psi-\tilde\chi +\bar{\tilde\chi}),
\nonumber\\
\frac{\delta}{\delta\theta}\bar{\tilde{\cal O}}^\star  \bar{{\cal O}}
&=& \bar O\bar{\chi} +\tilde\chi^\star  \Sigma,
 \ \
\frac{\delta}{\delta\theta}\bar{\tilde{\cal O}}  \bar{{\cal O}}^\star
 =  O\bar{\tilde\chi} +\chi^\star  \eta,
\nonumber\\
\frac{\delta}{\delta\theta}\bar{\cal T}^\star  \bar{\cal T}
&=&  T\bar{\varphi} -\varphi^\star ( \phi -\chi +\bar\chi ),
\ \
\frac{\delta}{\delta\theta}\bar{\cal S}_\mu^\star  \bar{\cal S}^\mu
 =   S_\mu\bar{\beta}^\mu -\beta_\mu^\star  \eta^\mu.
\end{eqnarray} 
Here, we note that the RHS of the  sum of above expressions  coincides with the gauge-fixed   
Lagrangian density corresponding to the shift symmetry  (\ref{la}).
Thus, the gauge-fixed   
Lagrangian density in superspace can be written as  
\begin{eqnarray}
\bar{\cal L}_{gf} &=& \frac{\delta}{\delta\theta}\left[\bar{\cal B}_{\mu\nu}^\star \bar{\cal 
B}^{\mu\nu} +\bar{\tilde{\cal M}}_{\mu}^\star  \bar{\cal M}^{\mu}+\bar{\tilde{\cal 
M}}_{\mu} \bar{\cal M}^{\mu\star}+\bar{\cal N}^\star  \bar{\cal N}+
\bar{\tilde{\cal N}}^\star  \bar{\tilde{\cal N}}+
\bar{\tilde{\cal O}}^\star  \bar{{\cal O}}+\bar{\tilde{\cal O}}  \bar{{\cal O}}^
\star \right.\nonumber\\
&+&\left.\bar{\cal T}^\star  \bar{\cal T}+\bar{\cal S}_\mu^\star  \bar{\cal 
S}^\mu\right].\label{super}
\end{eqnarray}
Similar to the Lorentz invariant theory, the invariance of  $\bar{\cal L}_{gf} $  under the
extended BRST transformation  is evident from the above expression as   it belongs to the $\theta$ component of superfields.
As the gauge-fixing fermion depends on the original fields,  the component form
of fermionic superfield  $\Gamma (x,\theta) $ in superspace is defined  as
\begin{eqnarray}
{\Gamma  }(x,\theta)  &=&\Psi (x)  + \theta \left[-\frac{\delta\Psi}{\delta B_{\mu\nu}}\psi_{\mu\nu}+
 \frac{\delta\Psi}{\delta\rho_\mu}\epsilon _\mu + \frac{\delta\Psi}
{\delta\tilde{\rho}_\mu} \xi_\mu -
 \frac{\delta\Psi}{\delta\sigma}\varepsilon- \frac{\delta\Psi}{\delta
\tilde\sigma}\psi -\frac{\delta\Psi}{\delta\beta_\mu}\eta_\mu +
\frac{\delta\Psi}{\delta\chi}\Sigma\right.\nonumber\\
& +&\left. \frac{\delta\Psi}{\delta\tilde\chi}\eta
-\frac{\delta\Psi}{\delta\varphi}\phi\right].
\end{eqnarray}
From the above expression, 
 the most general VSR-modified gauge-fixed Lagrangian density ${\cal L}_{gf}$ can be 
 described in the superspace
  by
\begin{equation}
{\cal L}_{gf} =\frac{\delta\Gamma(x,\theta) }{\delta\theta}.\label{lagr}
\end{equation}
Being the $\theta$ component of fermionic superfield, it is evident that
  ${\cal L}_{gf}$ is invariant under the extended BRST transformation. 
 Thus, the   VSR-modified 
effective Lagrangian density
 in this superspace formalism is identified as
\begin{eqnarray}
 {\cal L}_{eff} &=& {\cal L}_0(B_{\mu\nu}-\bar B_{\mu\nu}) + \frac{\delta}{\delta\theta}\left[\bar{\cal B}_{\mu\nu}^\star \bar{\cal 
B}^{\mu\nu} +{\bar{\tilde{\cal M}}}_{\mu}^\star  \bar{\cal M}^{\mu}+{\bar{\tilde{\cal 
M}}}_{\mu} \bar{\cal M}^{\mu\star}+\bar{\cal N}^\star  \bar{\cal N}+
{\bar{\tilde{\cal N}}}^\star  \bar{\tilde{\cal N}}+
{\bar{\tilde{\cal O}}}^\star  \bar{{\cal O}}\right.\nonumber\\
&+&\left.  {\bar{\tilde{\cal O}}}  \bar{{\cal O}}^
\star +\bar{\cal T}^\star  \bar{\cal T}+\bar{\cal S}_\mu^\star  \bar{\cal 
S}^\mu + \Gamma (x,\theta) \right].                                        
\end{eqnarray}
In VSR framework also, we observe that using equations of motion of auxiliary fields and ghost fields 
of the shift symmetry, this effective Lagrangian density reduces to the 
original BRST invariant Lagrangian density.

\subsection{VSR modified extended anti-BRST invariant BV action} 
In this subsection, we discuss the VSR modified extended anti-BRST invariant BV action
for  the Abelian rank-2 antisymmetric tensor field.  First of all, let us write the   
VSR modified anti-BRST symmetry transformation ($s_{ab}$),    which leaves the
Lagrangian density  (\ref{act}) for the 2-form gauge theory   invariant, as 
follows,
\begin{eqnarray}
s_{ab} B_{\mu\nu} &=&   \partial_\mu\tilde\rho_\nu
 -\partial_\nu\tilde\rho_\mu - \frac{1}{2}\frac{m^2}{n\cdot \partial}n_\mu\tilde{\rho}_\nu
  +\frac{1}{2}\frac{m^2}{n\cdot \partial}n_\nu\tilde{\rho}_\mu  ,\nonumber\\  
s_{ab}\tilde\rho_\mu &=& -i\left(\partial_\mu -  \frac{i}{2}\frac{m^2}{n\cdot \partial}n_\mu
\right)\tilde\sigma, \ s_{ab}
\tilde\sigma = 0, s_{ab}\rho_\mu  = 
-i\beta_\mu,\nonumber\\
s_{ab}\beta_\mu &=&0,\ 
s_{ab}\sigma  = \chi,   \
s_{ab}\chi =0, \
s_{ab}\varphi  =  -\tilde\chi,  \ \ s_{ab}\tilde\chi =0.
\label{anbr}
\end{eqnarray}
This VSR modified anti-BRST transformation is nilpotent and plays an important role
in defining physical unitarity. However, this  transformation does not anticommute   with the 
BRST transformation  (\ref{sym}) in absolute fashion, i.e. $\{s_b, s_{ab}\}\not =0$
for some fields. One should not bother for this in real sense as
the absolutely anticommutativity can be achieved on ground of Curci-Ferrari (CF) type
restriction. This is emphasized in the next  section (in case 
of Abelian 3-form gauge theory) with more details. 
 
The VSR modified gauge-fixed fermion corresponding to the anti-BRST transformation,  $\bar\Psi$, is defined by
\begin{eqnarray}
\bar\Psi  
&=&i\left[\rho_\nu (\partial_\mu B^{\mu\nu} - \frac{1}{2}\frac{m^2}{n\cdot \partial}n_\mu B^{\mu\nu}  +k_1 \beta^\nu )
-\sigma
\partial_\mu \tilde \rho^\mu+\sigma \frac{1}{2}\frac{m^2}{n\cdot \partial}n_\mu \tilde \rho^\mu \right.\nonumber\\
&+&\left.
\varphi (\partial_\mu \rho^\mu -\frac{1}{2}\frac{m^2}{n\cdot \partial}n_\mu \rho^\mu + k_2 \chi)\right].  
\end{eqnarray}
As the the gauge-fixing part of the Lagrangian density is anti-BRST exact and, thus, can be 
written   in terms of anti-BRST variation of $\bar\Psi$ as following:
\begin{eqnarray}
{\cal L}_{gf}=s_{ab} \bar\Psi.
\end{eqnarray}
To discuss the extended anti-BRST symmetry within VSR framework, we do follow
  the same procedure as in the case of BRST transformation.
Thus, here we demand that extended anti-BRST operation on $(\Phi-\bar\Phi)$ 
should have same structure of  the original anti-BRST transformations with shifted fields. 
This requirement leads to the following VSR modified extended anti-BRST transformations:
\begin{eqnarray}
s_{ab} \bar {B}_{\mu\nu}&=& B_{\mu\nu}^\star,\ \ 
s_{ab} {B}_{\mu\nu}= B_{\mu\nu}^\star +
(\partial_\mu\tilde\rho_\nu -\partial_\mu\bar{ \tilde\rho}_\nu-\partial_\nu\tilde\rho_\mu+\partial_\nu\bar{\tilde\rho}_\mu - \frac{i}{2}\frac{m^2}{n\cdot \partial}n_\mu\tilde\rho_\nu
\nonumber\\
&+& \frac{i}{2}\frac{m^2}{n\cdot \partial}n_\mu \bar{ \tilde\rho}_\nu
+\frac{i}{2}\frac{m^2}{n\cdot \partial}n_\nu\tilde\rho_\mu 
 -\frac{i}{2}\frac{m^2}{n\cdot \partial}n_\nu \bar{\tilde\rho}_\mu ), \ \ \ s_{ab}\bar\rho_\mu =\rho_\mu^\star,\nonumber\\
s_{ab}\bar{\tilde{\rho}}_\mu &=& \tilde\rho_\mu^\star,  \ \
 s_{ab}\tilde{\rho}_\mu = \tilde\rho_\mu^\star 
-i\partial_\mu\tilde\sigma+i
\partial_\mu\bar{\tilde\sigma} + \frac{i}{2}\frac{m^2}{n\cdot \partial}n_\mu\sigma  -\frac{i}{2}\frac{m^2}{n\cdot \partial}n_\mu\bar{\tilde\sigma} ,\nonumber\\
s_{ab}\rho_\mu &=&\rho_\mu^\star  -i\beta_\mu +i\bar{\beta}_\mu, 
\ \
s_{ab}\bar{\tilde\sigma}  =  \tilde\sigma^\star,\   s_{ab}\tilde\sigma =\tilde\sigma^\star,
\ s_{ab}\bar\beta_\mu =  \beta_\mu^\star,\nonumber\\
 s_{ab}{\beta}_\mu &=& \beta_\mu^\star,\
s_{ab}\bar\sigma = \sigma^\star,\ \ s_{ab}{\sigma} =\sigma^\star -\chi +\bar{\chi}, \ s_{ab}\bar\chi =\chi^\star,
\nonumber\\
 s_{ab}{\chi}&=&\chi^\star, \
s_{ab}\bar\varphi  =  \varphi^\star,\ \ s_{ab}\varphi =\varphi^\star- \tilde\chi +\bar{
\tilde\chi},\  s_{ab}\bar{\tilde\chi} =\tilde\chi^\star, \  
s_{ab}\tilde\chi =\tilde\chi^\star.\label{ab}
\end{eqnarray}
The antifields $B_{\mu\nu}^\star, \tilde\rho_\mu^\star, \rho_\mu^\star, \tilde\sigma^\star, 
\beta_\mu^\star, \psi,
\sigma^\star, \chi^\star, \varphi^\star$ and $ \tilde\chi^\star$ do not vary under
extended
anti-BRST transformations as the transformations are nilpotent in nature.
Moreover, the ghost fields of the shift symmetry transform under
VSR modified  extended
anti-BRST transformations as follows,
\begin{eqnarray}
s_{ab}\psi_{\mu\nu}&=& L_{\mu\nu},\ \ s_{ab}\epsilon_{\mu} = M_{\mu},\ \
s_{ab}\xi_{\mu} =\bar M_{\mu},\ \
s_{ab}\varepsilon = N,\nonumber\\
 s_{ab}\psi  &=& \bar N,\ 
s_{ab}\eta_{\mu} = S_{\mu},\ 
s_{ab}\Sigma  =  O,\   s_{ab}\eta = \bar O,\  
s_{ab}\phi =T,\nonumber\\
s_{ab}\bar M_{\mu} &=&O,\ \ s_{ab} L_{\mu\nu}   =0, \ \ s_{ab}  M_{\mu}   =0,\ \ s_{ab} N   =0,\ \ s_{ab} \bar N   =0,\nonumber\\
 s_{ab} S_\mu   &=&0,\ \ s_{ab} O   =0,\ \ s_{ab} \bar O   =0,
\ \ s_{ab} \bar T.\label{vv}
\end{eqnarray}
The transformations (\ref{ab}) and (\ref{vv}) together leads to complete   
extended anti-BRST transformations in VSR framework, which leave
the shifted effective Lagrangian density  invariant.
With the help of these set of extended anti-BRST transformation, it is straightforward  to construct the superspace having extra fermion coordinate
$\bar{\theta}$ along with $x_\mu$. 
\subsection{VSR modified extended BRST and anti-BRST invariant superspace formulation}
In this subsection, we develop a superspace formulation for VSR modified 
2-form gauge theory which is   manifestly invariant  under the both extended
BRST  and   extended anti-BRST transformations.
To define a superspace for such theory, we need  
 two Grassmannian coordinates, $\theta$ and $\bar\theta$,
 together with $x_\mu$. Therefore, the superfields here
 depend on superspace  $(x_\mu, \theta, \bar\theta)$. 
 Within VSR framework, the 
``super connection" $2$-form ($\omega^{(2)}$) 
and the field strength ($ F^{(3)}$), respectively, are 
\begin{eqnarray}
\omega^{(2)} &=&\frac{1}{2 !}{\cal B}_{\mu\nu}(x, \theta, \bar\theta ) (dx^\mu\wedge dx^\nu ) +
 {\cal M}_{\mu}(x, \theta, \bar\theta ) (dx^\mu\wedge d\theta ) + {\cal N} (x, \theta, \bar\theta ) 
(d\theta\wedge d\theta )\nonumber\\
&+& \tilde{\cal M}_{\mu}(x, \bar\theta, \bar\theta ) (dx^\mu\wedge d\bar\theta ) + \tilde{\cal N}
 (x, \bar\theta, \bar\theta ) (d 
\bar\theta\wedge d\bar\theta ) +{\cal T} (x, \bar\theta, \bar\theta )(d \theta\wedge d\bar\theta ),\\
 F^{(3)} &=&d \omega^{(2)}.
\end{eqnarray} 
Here, the exterior derivative $d$ has the following form:
\begin{eqnarray}
d&=& 
 dx^\mu\left(\partial_\mu-\frac{1}{2}\frac{m^2}{n\cdot \partial}n_\mu\right) +d\theta\left(\partial_\theta-\frac{1}{2}\frac{m^2}{n\cdot \partial}n_\theta\right)+d\bar\theta\left( \partial_{\bar\theta}-\frac{1}{2}\frac{m^2}{n\cdot \partial}n_{\bar\theta}\right).   
\end{eqnarray} 
The components of the superfields can be computed by requiring  
  the field strength  to vanish along the directions of
$\theta$ and $\bar\theta$. 
The explicit expressions for these superfields are calculated  in (\ref{a1}).

Exploiting the expressions of superfields given in (\ref{a1}), we compute
the following expressions:  
\begin{eqnarray}
\frac{1}{2}\frac{\delta}{\delta\bar\theta}\frac{\delta}{\delta\theta}\bar{\cal B}_{\mu\nu}
\bar{\cal B}^{\mu\nu}
  &=&  L_{\mu\nu}\bar {B}^{\mu\nu} -B_{\mu\nu}^\star  \left(\psi^{\mu\nu} \partial^\mu\rho^\nu +
\partial^\mu\rho^\nu
+\partial^\nu\rho^\mu -\partial^\nu\bar{\rho}^\mu -\psi^{\mu\nu}\frac{1}{2}\frac{m^2}{n\cdot \partial}n^\mu\rho^\nu \right.\nonumber\\
&-& \left. \frac{1}{2}\frac{m^2}{n\cdot \partial}n^\mu\rho^\nu -\frac{1}{2}\frac{m^2}{n\cdot \partial}n^\nu\rho^\mu +\frac{1}{2}\frac{m^2}{n\cdot \partial}n^\nu \bar{\rho}^\mu \right),
\nonumber\\
\frac{\delta}{\delta\bar\theta}\frac{\delta}{\delta\theta}\bar{\tilde{\cal M }}_{\mu} 
\bar{\cal M}^{\mu} 
& =& \bar M_{\mu}\bar{\rho}^\mu +\tilde\rho_{\mu}^\star  \left(\epsilon ^{\mu}+
i\partial^\mu\sigma -i\partial^\mu\bar\sigma -\frac{i}{2}\frac{m^2}{n\cdot \partial}n^\mu\sigma +\frac{i}{2}\frac{m^2}{n\cdot \partial}n^\mu \bar\sigma \right)
+ M_{\mu}\bar{\tilde\rho}^\mu  \nonumber\\
&+&   \rho_{\mu}^\star  (\xi^{\mu}-
i\beta^\mu +i\bar\beta^\mu  ),
\nonumber\\
\frac{1}{2}\frac{\delta}{\delta\bar\theta}\frac{\delta}{\delta\theta}\bar{\cal N}
 \bar{\cal N}
&=&  N\bar{\sigma} -\sigma^\star  \varepsilon, 
\ \ \ \frac{1}{2}\frac{\delta}{\delta\bar\theta}\frac{\delta}{\delta\theta}\bar{\tilde{\cal N
}} \bar{\tilde{\cal N}}
= \bar N\bar{\tilde\sigma} -\tilde\sigma^\star  (\psi-\tilde\chi +\bar{\tilde\chi}),
\nonumber\\
\frac{\delta}{\delta\bar\theta}\frac{\delta}{\delta\theta}\bar{\tilde{\cal O
}}  \bar{{\cal O}}
&=& \bar O\bar{\chi} +\tilde\chi^\star  \Sigma
+ O\bar{\tilde\chi} +\chi^\star  \eta,
\ \frac{1}{2}\frac{\delta}{\delta\bar\theta}\frac{\delta}{\delta\theta}\bar{\cal T}\bar{\cal T}
= T\bar{\varphi} -\varphi^\star ( \phi -\chi +\bar\chi ),
\nonumber\\
\frac{1}{2}\frac{\delta}{\delta\bar\theta}\frac{\delta}{\delta\theta}\bar{\cal S}_
\mu  \bar{\cal S}^\mu
&=& S_\mu\bar{\beta}^\mu -\beta_\mu^\star  \eta^\mu. 
\end{eqnarray} 
By adding all the equations of above expression side by side,
we get, remarkably, that this is nothing but the expression of Lagrangian density, $\bar {\cal L}_{gf}$, given in Eq. (\ref{la}).
Thus, we can write 
\begin{eqnarray}
  \bar{\cal L}_{gf} 
=\frac{1}{2}\frac{\delta}{\delta\bar\theta}\frac{\delta}{\delta\theta}\left[
\bar {\cal B}_{\mu\nu} \bar {\cal B}^{\mu\nu} +2{\bar{\tilde {\cal M}}}_\mu{\bar {\cal M}}^\mu +\bar {\cal N}\bar {\cal N}
+{\bar{\tilde {\cal N}}}{\bar{\tilde {\cal N}}} + 2{\cal{\bar{ \tilde O}}} {\bar{\cal O}} +\bar{\cal T}\bar{\cal T}
+\bar{\cal S}_\mu\bar{\cal S}^\mu\right].\label{barl}
\end{eqnarray}
Eventually, we see that  $\bar {\cal L}_{gf} $ is 
nothing but the  $\theta\bar\theta$ component of the composite superfields.
Therefore, this certifies the invariance of  the $\bar {\cal L}_{gf} $ 
under both the extended BRST and
extended anti-BRST transformations.
The component form of super gauge-fixing fermion  in this superspace is given by
\begin{equation}
\Gamma  (x, \theta, \bar\theta )=\Psi +\theta s_b\Psi +\bar\theta s_{ab}\Psi +\theta
\bar\theta s_b s_{ab}\Psi,
\end{equation} 
to express the ${\cal L}_{gf}$ as $\frac{\delta}{\delta\theta}\left[\delta(\bar\theta)\Gamma  (x, \theta, 
\bar\theta) \right]$. 
The  $\theta\bar\theta$ component of $\Gamma  (x, \theta, \bar\theta )$  vanishes due to equations of motion
in the theories having both BRST and anti-BRST invariance. 
 
Therefore, the effective Lagrangian density (\ref{comple}), which is invariant under both the extended BRST and the extended anti-BRST 
transformations,  can  be expressed in superspace by  
\begin{eqnarray} 
{\cal L}_{eff} 
&=&{\cal L}_0(B_{\mu\nu}-\bar B_{\mu\nu})+\frac{1}{2}\frac{\delta}{\delta\bar\theta}\frac{\delta}{\delta\theta}\left[
\bar {\cal B}_{\mu\nu} \bar {\cal B}^{\mu\nu} +2{\bar{\tilde {\cal M}}}_\mu{\bar {\cal M}}^\mu +\bar {\cal N}\bar {\cal N}
+{\bar{\tilde {\cal N}}}{\bar{\tilde {\cal N}}}+2{\bar{\tilde{\cal O}}}\bar{\cal O} +\bar{\cal T}\bar{\cal T}
+\bar{\cal S}_\mu\bar{\cal S}^\mu\right]\nonumber\\
&+&\frac{\delta}{\delta\theta}\left[\delta(\bar\theta)
\Gamma (x, \theta, \bar\theta)\right]. 
\end{eqnarray}
Here, we found that the gauge fixing parts of the effective Lagrangian density
is the $\theta$ components of the certain functional.
 Thus we observe that the VSR modified  2-form gauge theory
 in superspace described by two fermionic parameters also follow the
 same structure as the Lorentz invariant case.
 \section{3-form gauge theory: VSR modified BV action in superspace}
 VSR generalization to the tensor field (reducible gauge) theories has also been done
  using a BV formulation \cite{sud,sudpanigrahi}. A rigorous construction of quantum field theory in VSR framwork is also studied  \cite{lii}. In this connection, we would like 
  study the BV action tensor field of rank-3  in superspace.
 In particular, we generalize our previous results for the 
 case of Abelian 3-form gauge theory  which 
is relevant as it plays a crucial role to study the excitations of
the quantized versions of strings, superstrings and related extended objects. The    classical Lagrangian density for the Abelian 3-form gauge theory
in VSR is given by \cite{sud,sudpanigrahi},
  as
\begin{equation}
 {{\cal L}_0} =\frac{1}{24}\tilde{F}_{\mu\nu\eta\xi}\tilde{F}^{\mu\nu\eta\xi},\label{ll}
\end{equation}
where the 4-form field strength  tensor ($\tilde{F}_{\mu\nu\eta\xi}$) has the following 
form:
\begin{eqnarray}
\tilde{F}_{\mu\nu\eta\xi} 
 &=& \partial_\mu B_{\nu\eta\xi}-\partial_\nu B_{ \eta\xi\mu} + \partial_\eta B_{\xi\mu\nu} 
-\partial_\xi B_{\mu\nu\eta} \  \
-\frac{1}{2}\frac{m^2}{n\cdot \partial}n_\mu B_{\nu\eta\xi} +\frac{1}{2}\frac{m^2}{n\cdot \partial}n_\nu B_{ \eta\xi\mu} \nonumber\\
&-&\frac{1}{2}\frac{m^2}{n\cdot \partial}n_\eta B_{\xi\mu\nu}  +\frac{1}{2}\frac{m^2}{n\cdot \partial}n_\xi B_{\mu\nu\eta}.
\end{eqnarray}
Here $B_{\mu\nu\eta}$ is totally antisymmetric rank-3 tensor gauge  field
and  $ n_\mu$ is a constant null vector.
The Lagrangian density (\ref{ll}) is not invariant under 
standard gauge transformation as the Lorentz invariance is broken by choosing
an specific direction. However, this Lagrangian density 
is invariant under the following VSR modified   gauge transformation  
\begin{equation}
\delta{B}_{\mu\nu\eta}  
 = \partial_\mu \lambda_{\nu\eta} +\partial_\nu\lambda_{\eta\mu}+\partial_\eta\lambda_{\mu\nu}-                                 \frac{1}{2}\frac{m^2}{n\cdot \partial}n_\mu\lambda_{\nu\eta} -\frac{1}{2}\frac{m^2}{n\cdot \partial}n_\nu\lambda_{\eta\mu} -  \frac{1}{2}\frac{m^2}{n\cdot \partial}n_\eta\lambda_{\mu\nu},\label{gg}
\end{equation}
where $\lambda_{\mu\nu}$ is an arbitrary antisymmetric parameter.
As this is a (VSR modified) gauge invariant theory,
 it contains redundant degrees of freedom. From the expression of (\ref{gg}),
 it is evident that the theory is reducible. Therefore, to quantize this theory correctly
we need to fix the gauge appropriately. 
The gauge-fixed Lagrangian density in VSR is calculated by
\begin{eqnarray}
{\cal L}^B_{gf} 
&=& \partial_\mu B^{\mu\nu\eta}B_{\nu\eta} -\frac{1}{2}\frac{m^2}{n\cdot \partial}n^\mu B^{\mu\nu\eta}B_{\nu\eta} +\frac{1}{2}B_{\mu\nu}\tilde B^{\mu\nu}+ \partial_\mu \tilde c_{\nu\eta}\partial ^\mu 
c^{\nu\eta} + m^2\tilde c_{\nu\eta}c^{\nu\eta}\nonumber\\
&+& \partial_\nu \tilde c_{\eta\mu}\partial ^\mu 
c^{\nu\eta} -\partial_\nu \tilde c_{\eta\mu}\frac{1}{2}\frac{m^2}{n\cdot \partial}n^\mu c^{\nu\eta} -  \partial^\mu c^{\nu\eta}\frac{1}{2}\frac{m^2}{n\cdot \partial}n_\nu \tilde c^{\eta\mu}+ n_\nu \tilde c^{\eta\mu}\frac{1}{4}\frac{m^2}{n\cdot \partial} n^\mu c^{\nu\eta}\nonumber\\
&+& \partial_\mu \tilde c_{\mu\nu}\partial^\mu c^{\nu\eta}-\partial_\mu \tilde c_{\mu\nu}\frac{1}{2}\frac{m^2}{n\cdot \partial}n^\mu c^{\nu\eta}- \partial^\mu c^{\nu\eta}\frac{1}{2}\frac{m^2}{n\cdot \partial}n_\eta \tilde c_{\mu\nu} +n^\eta \tilde c^{\mu\nu}\frac{1}{4}\frac{m^2}{n\cdot \partial}n_\eta \tilde c_{\mu\nu}\nonumber\\
&-&\partial_\mu\tilde\beta_\nu \partial^\mu\beta^\nu - m^2\tilde\beta_\nu \beta^\nu +\partial_\nu\tilde\beta_\mu \partial^\mu\beta^\nu - \partial_\nu\tilde\beta_\mu\frac{1}{2}\frac{m^2}{n\cdot \partial}n^\mu \beta^\nu -\partial^\mu\beta^\nu\frac{1}{2}\frac{m^2}{n\cdot \partial}n_\nu \tilde\beta_\mu  \nonumber\\
&+& \frac{1}{4}\frac{m^2}{n\cdot \partial}n_\nu \tilde\beta_\mu n^\mu \beta^\nu -  BB_2 -\frac{1}{2} B_1^2 +
\partial_\mu \tilde c^{\mu\nu}f_\nu - \partial_\mu c^{\mu\nu}\tilde F_\nu - \frac{1}{2}\frac{m^2}{n\cdot \partial}n_\mu\tilde c^{\mu\nu}f_\nu   \nonumber\\
&+&   \frac{1}{2}\frac{m^2}{n\cdot \partial}n_\mu c^{\mu\nu}\tilde{F}_\nu +\partial_\mu \tilde c_2 \partial^\mu c_2 + m^2\tilde c_2 c_2+\tilde{f}_\mu f_\mu -\tilde F_\mu F^\mu+\partial_\mu \beta^\mu B_2 \nonumber\\
&-&  \frac{1}{2}\frac{m^2}{n\cdot \partial}n_\mu \beta^\mu B_2+ \partial_\mu \phi^\mu B_1 -\partial_\mu\tilde\beta^\mu B-\frac{1}{2}\frac{m^2}{n\cdot \partial}n_\mu \phi^\mu B_1 + \frac{1}{2}\frac{m^2}{n\cdot \partial}n_\mu
\tilde\beta^\mu B. \label{lag3}
 \end{eqnarray}
 Here, keeping the reducible nature of the theory, we have introduced extra auxiliary and ghost fields.  For instance, antisymmetric ghost    and   antighost fields  $c_{\mu\nu}$ and $\bar c_{\mu\nu}$  are   
Grassmannian and  the vector field 
 $\phi_\mu$, antisymmetric auxiliary fields  $B_{\mu\nu}, \bar B_{\mu\nu}$  and  auxiliary fields  $B, 
B_1, B_2$  are bosonic. The fields  $\beta_\mu$ and $\bar\beta_\mu$ are ghost of ghosts  and are bosonic in nature. However,  $c_2$ and $\bar{c}_2$  are ghost of ghost of ghosts  with
fermionic  nature. The rest of the Grassmannian fields, i.e., $c_1, \bar c_1, f_\mu$ and $\bar F_\mu$  are  auxiliary fields.  It has been shown in Ref. \cite{sud} that
the ghosts  $c_{\mu\nu}$ and $\bar c_{\mu\nu}$, ghost of ghosts  $\beta_\mu$ and $\bar\beta_\mu$ and ghost of ghost of ghosts  $c_2$ and $\bar{c}_2$, are massive.

To write the absolutely anticommuting BRST and anti-BRST invariant 
BV action for 3-form gauge theory in VSR,
we consider an  equivalent candidate to the above gauge-fixing Lagrangian density 
as following:
\begin{eqnarray}
{\cal L}^{\tilde B}_{gf}  
&=& -\partial_\mu B^{\mu\nu\eta}{\tilde B}_{\nu\eta} +\frac{1}{2}\frac{m^2}{n\cdot \partial}n^\mu B^{\mu\nu\eta}{\tilde B}_{\nu\eta} +\frac{1}{2}B_{\mu\nu}{\tilde B}^{\mu\nu}+ \partial_\mu \tilde c_{\nu\eta}\partial ^\mu 
c^{\nu\eta} + m^2\tilde c_{\nu\eta}c^{\nu\eta}\nonumber\\
&+& \partial_\nu \tilde c_{\eta\mu}\partial ^\mu 
c^{\nu\eta} -\partial_\nu \tilde c_{\eta\mu}\frac{1}{2}\frac{m^2}{n\cdot \partial}n^\mu c^{\nu\eta} -  \partial^\mu c^{\nu\eta}\frac{1}{2}\frac{m^2}{n\cdot \partial}n_\nu \tilde c^{\eta\mu}+ n_\nu \tilde c^{\eta\mu}\frac{1}{4}\frac{m^2}{n\cdot \partial} n^\mu c^{\nu\eta}\nonumber\\
&+& \partial_\mu \tilde c_{\mu\nu}\partial^\mu c^{\nu\eta}-\partial_\mu \tilde c_{\mu\nu}\frac{1}{2}\frac{m^2}{n\cdot \partial}n^\mu c^{\nu\eta}- \partial^\mu c^{\nu\eta}\frac{1}{2}\frac{m^2}{n\cdot \partial}n_\eta \tilde c_{\mu\nu} +n^\eta \tilde c^{\mu\nu}\frac{1}{4}\frac{m^2}{n\cdot \partial}n_\eta \tilde c_{\mu\nu}\nonumber\\
&-&\partial_\mu\tilde\beta_\nu \partial^\mu\beta^\nu - m^2\tilde\beta_\nu \beta^\nu +\partial_\nu\tilde\beta_\mu \partial^\mu\beta^\nu - \partial_\nu\tilde\beta_\mu\frac{1}{2}\frac{m^2}{n\cdot \partial}n^\mu \beta^\nu -\partial^\mu\beta^\nu\frac{1}{2}\frac{m^2}{n\cdot \partial}n_\nu \tilde\beta_\mu  \nonumber\\
&+& \frac{1}{4}\frac{m^2}{n\cdot \partial}n_\nu \tilde\beta_\mu n^\mu \beta^\nu -  BB_2 -\frac{1}{2} B_1^2+ \partial_\mu c^{\mu\nu}\tilde{f}_\nu - \partial_\mu\tilde{c}^{\mu\nu}F_\nu + \frac{1}{2}\frac{m^2}{n\cdot \partial}n_\mu\tilde c^{\mu\nu}F_\nu  \nonumber\\
&-&   \frac{1}{2}\frac{m^2}{n\cdot \partial}n_\mu c^{\mu\nu}\tilde{f}_\nu  + \partial_\mu \tilde c_2 \partial^\mu c_2 + m^2\tilde c_2 c_2+\tilde{f}_\mu f_\mu -\tilde F_\mu F^\mu+\partial_\mu \beta^\mu B_2 \nonumber\\
& -& \frac{1}{2}\frac{m^2}{n\cdot \partial}n_\mu \beta^\mu B_2+ \partial_\mu \phi^\mu B_1 -\partial_\mu\tilde\beta^\mu B-\frac{1}{2}\frac{m^2}{n\cdot \partial}n_\mu \phi^\mu B_1 + \frac{1}{2}\frac{m^2}{n\cdot \partial}n_\mu
\tilde\beta^\mu B.\label{lag32}
\end{eqnarray} 
These two Lagrangian densities  are equivalent as they describe  same dynamics of the theory
on the following  CF type restricted surface:
\begin{eqnarray}
 f_\mu +F_{\mu}& =& \partial_\mu c_1  -\frac{1}{2}\frac{m^2}{n\cdot \partial}n_\mu c_1 ,  \ \
\bar f_\mu +\bar F_{\mu} = \partial_\mu\bar c_1 - \frac{1}{2}\frac{m^2}{n\cdot \partial}n_\mu \bar c_1, \nonumber\\
B_{\mu\nu}+\bar B_{\mu\nu} &=& \partial_\mu \phi_\nu - \partial_\nu\phi_\mu - \frac{1}{2}\frac{m^2}{n\cdot \partial}n_\mu \phi_\nu +\frac{1}{2}\frac{m^2}{n\cdot \partial}n_\nu\phi_\mu.
\end{eqnarray}
The VSR modified BRST transformations which leave the effective Lagrangian density,
${\cal L}={\cal L}_0 + {\cal L}^{B}_{gf}$, 
invariant are given by
\begin{eqnarray}
s_b B_{\mu\nu\eta} 
&=&(\partial_\mu c_{\nu\eta}+\partial_\nu c_{\eta\mu} +\partial_\eta c_{\mu\nu}- \frac{1}{2}\frac{m^2}{n\cdot \partial}n_\mu  c_{\nu\eta} -\frac{1}{2}\frac{m^2}{n\cdot \partial}n_\nu c_{\eta\mu} - \frac{1}{2}\frac{m^2}{n\cdot \partial}n_\eta c_{\mu\nu}), \nonumber\\
 s_b\tilde c_{\mu\nu}&=& B_{\mu\nu},\ \ s_b B_{\mu\nu}=  {\partial}_\mu f_\nu - {\partial}_\nu f_\mu -\frac{1}{2}\frac{m^2}{n\cdot \partial}n_\mu f_\nu +\frac{1}{2}\frac{m^2}{n\cdot \partial}n_\nu  f_\mu  ,  \ \ \ 
s_b\tilde\beta_\mu = \tilde F_\mu, \nonumber\\
s_b\beta_\mu &=&
  (\partial_\mu - \frac{1}{2}\frac{m^2}{n\cdot \partial}n_\mu) c_2,   \ \  
s_b F_\mu  =-(\partial_\mu -\frac{1}{2}\frac{m^2}{n\cdot \partial}n_\mu ) B ,\nonumber\\
s_b\tilde f_\mu &=&  (\partial_\mu -\frac{1}{2}\frac{m^2}{n\cdot \partial}n_\mu )B_1 ,  \ \ s_b\tilde c_2 =B_2,\ s_b c_1 =-B,\ \ s_b \phi_\mu =f_\mu,\nonumber\\
 s_b \tilde c_1 &=&
 B_1,\ \
 s_b [c_2, f_\mu, \tilde F_\mu, B, B_1, B_2, B_{\mu\nu}] =0.\label{3brst}
 \end{eqnarray}
Since the gauge fixed Lagrangian density ${\cal L}^B_{gf}$ is BRST exact, so it can be written in terms of BRST variation of a gauge-fixing fermion $\Psi$ 
as  
\begin{eqnarray}
{\cal L}^B_{gf}=s_b \Psi ,
\end{eqnarray}
where the explicit form of $\Psi$ is   as following:
\begin{eqnarray}
\Psi &=& -\frac{1}{2}\tilde c_2 B+\frac{1}{2}B_2 c_1 -\frac{1}{2}\tilde c_1 B_1 -\frac{1}{2}(\partial_\mu\tilde \beta_\nu - \partial_\nu\tilde\beta_\mu -  \frac{1}{2}\frac{m^2}{n\cdot \partial}n_\mu \tilde\beta_\nu +  \frac{1}{2}\frac{m^2}{n\cdot \partial}n_\nu  \tilde\beta_\mu)c^{\mu\nu} \nonumber\\
&+& \frac{1}{2}\tilde c_{\mu\nu} \tilde B^{\mu\nu}-\partial_\mu \tilde c_2\beta^\mu + \frac{1}{2}\frac{m^2}{n\cdot \partial}(n\cdot \beta)\tilde c_2 - \tilde \beta_\mu F^\mu -\phi_\mu \tilde f^\mu - \frac{1}{3} B_{\mu\nu\eta} \left(\partial^\mu \tilde c^{\nu\eta} +\partial^\nu \tilde c^{\eta\mu}  \right. \nonumber\\
&+& \left. \partial^\eta \tilde c^{\mu\nu}-\frac{1}{2}\frac{m^2}{n\cdot \partial}n^\mu \tilde c^{\mu\eta} - \frac{1}{2}\frac{m^2}{n\cdot \partial} n^\nu  c^{\eta\mu} - \frac{1}{2}\frac{m^2}{n\cdot \partial}n^\eta \tilde c^{\mu\nu}\right ).
\label{gfff}
\end{eqnarray}
Here, one can see the VSR modification in  the gauge-fixing fermion clearly.
The anti-BRST symmetry transformations $(s_{ab})$ for  the above 
Lagrangian densities  are given by
 \begin{eqnarray}
s_{ab} B_{\mu\nu\eta} &=&    (\partial_\mu \tilde c_{\nu\eta}+\partial_\nu \tilde c_{\eta\mu} +\partial_\eta 
\tilde c_{\mu\nu}-\frac{1}{2}\frac{m^2}{n\cdot \partial}n_\mu \tilde c_{\nu\eta} -\frac{1}{2}\frac{m^2}{n\cdot \partial}n_\nu  \tilde c_{\eta\mu}
-\frac{1}{2}\frac{m^2}{n\cdot \partial}n_\eta\tilde c_{\mu\nu} )\nonumber\\
s_{ab}\tilde c_{\mu\nu}  &=&   \partial_\mu \tilde\beta_\nu -\partial_\nu \tilde\beta_\mu - \frac{1}{2}\frac{m^2}{n\cdot \partial}n_\mu \tilde\beta_\nu + \frac{1}{2}\frac{m^2}{n\cdot \partial}n_\nu\tilde\beta_\mu  ,  \ \ s_{ab}  c_{\mu\nu}=
\tilde B_{\mu\nu}, \nonumber\\
s_{ab} B_{\mu\nu}&=&  \partial_\mu \tilde{f}_\nu -\partial_\nu\tilde{ f_\mu} -\frac{1}{2}\frac{m^2}{n\cdot \partial}n_\mu \tilde{f}_\nu +\frac{1}{2}\frac{m^2}{n\cdot \partial}n_\nu \tilde{ f}_\mu  ,  \ \ \ 
s_{ab}\beta_\mu = F_\mu,\
\nonumber\\
s_{ab}\tilde{\beta}_\mu &=& 
=\partial_\mu\tilde c_2 - \frac{1}{2}\frac{m^2}{n\cdot \partial}n_\mu\tilde c_2 ,\ \  s_{ab}\tilde F_\mu = -\partial_\mu B_2 + \frac{1}{2}\frac{m^2}{n\cdot \partial}n_\mu B_2 ,\nonumber\\
s_{ab} f_\mu &=&   -\partial_\mu B_1 + \frac{1}{2}\frac{m^2}{n\cdot \partial}n_\mu B_1  ,  \ \ s_{ab}  c_2 =B,\ s_{ab} c_1 =-B_1, \ \  s_{ab} \phi_\mu = \tilde f_\mu,\nonumber\\
s_{ab} \tilde c_1 &=& - B_2,\ \ s_{ab} [\tilde c_2, \tilde f_\mu,  F_\mu, B, B_1, B_2, \tilde B_{\mu\nu}] =0. \label
{an}
\end{eqnarray} 
Since both the BRST and anti-BRST transformations are absolutely anticommuting in nature,
so both the BRST and anti-BRST exact  Lagrangian densities can be expressed as  
follows,
\begin{eqnarray}
{\cal L}^B_{gf}  &=& 
s_bs_{ab} \left[\frac{1}{2}\tilde c_2 c_2 -\frac{1}{2}\tilde c_1 c_1 -\frac{1}{2}\tilde c_{\mu\nu} c^{\mu\nu}-
\tilde\beta_\mu \beta^\mu -\frac{1}{2}\phi_\mu\phi^\mu -\frac{1}{6}B_{\mu\nu\eta}B^{\mu\nu\eta} \right].
\end{eqnarray}
Now, we will study the construction of VSR modified extended BRST transformation
for Abelian 3-form gauge theory in next subsection.
\subsection{VSR modified extended BRST invariant BV action }
Following the previous sections, we generalize the VSR modified extended BRST 
construction for the case of  Abelian 3-form gauge theory in VSR.
In this regard, 
we   shift all the fields from their original values as follows
\begin{eqnarray} 
&&B_{\mu\nu\eta} \rightarrow   B_{\mu\nu\eta}-\bar B_{\mu\nu\eta},\   c_{\mu\nu}\rightarrow  c_{\mu\nu} 
-\bar c_{\mu\nu}, \ 
\tilde c_{\mu\nu}\rightarrow   \tilde c_{\mu\nu}-\bar {\tilde c}_{\mu\nu},\ B_{\mu\nu} \rightarrow    
B_{\mu\nu} - \bar B_{\mu\nu},\nonumber\\ 
&&
\tilde B_{\mu\nu} \rightarrow \tilde B_{\mu\nu}-\bar{\tilde B}_{\mu\nu}, \ \
\beta_\mu\rightarrow  \beta_\mu - \bar\beta_\mu,\ \
\tilde \beta_\mu \rightarrow \tilde\beta_\mu-\bar{\tilde \beta}_\mu,  \ \
F_\mu\rightarrow  F_\mu - \bar{F}_\mu,\nonumber\\
&&\tilde F_\mu \rightarrow   \tilde F_\mu  -\bar{\tilde F}_\mu,\ \
f_\mu  \rightarrow   f_\mu - \bar{f}_\mu,\ \
\tilde f_\mu  \rightarrow   \tilde f_\mu  -\bar{\tilde f}_\mu,\ \
c_2\rightarrow  c_2 - \bar c_2,\nonumber\\
&&\tilde c_2  \rightarrow  \tilde c_2  -\bar{\tilde c}_2,\  
c_1\rightarrow  c_1- \bar c_1,\
\tilde c_1  \rightarrow  \tilde c_1  -\bar{\tilde c}_1,\
\phi_\mu \rightarrow   \phi_\mu- \bar \phi_\mu,\nonumber\\
&&B   \rightarrow  B  -\bar B, \ \
B_1\rightarrow  B_1- \bar B_1,\
B_2  \rightarrow   B_2  -\bar B_2.
\end{eqnarray} 
Following the above shifts in fields, 
the effective Lagrangian density of the theory is modified by
\begin{eqnarray}
\bar{\cal L} &=& {\cal L} (B_{\mu\nu\eta}-\bar B_{\mu\nu\eta}, c_{\mu\nu} 
-\bar c_{\mu\nu}, \tilde c_{\mu\nu}-\bar {\tilde c}_{\mu\nu}, B_{\mu\nu} - \bar B_{\mu\nu}, 
\tilde B_{\mu\nu}-\bar{\tilde B}_{\mu\nu}, \beta_\mu - \bar\beta_\mu, \tilde\beta_\mu-\bar{\tilde \beta}_
\mu,\nonumber\\
&& F_\mu - \bar{F}_\mu, \tilde F_\mu  -\bar{\tilde F}_\mu, f_\mu - \bar{f}_\mu, \tilde f_\mu  -\bar{\tilde f}_\mu,
 c_2 - \bar c_2, \tilde c_2  -\bar{\tilde c}_2,  c_1- \bar c_1, \tilde c_1  -\bar{\tilde c}_1, \phi_\mu- \bar 
\phi_\mu,\nonumber\\
&& B  -\bar B,  B_1- \bar B_1, B_2  -\bar B_2 ).
\end{eqnarray}
This Lagrangian density
remains invariant under the BRST transformation  (\ref{3brst}) corresponding to the 
shifted fields. In fact, it is   invariant  under the
following extended BRST transformations of fields:
\begin{eqnarray}
s_b \Phi (x)= \alpha (x),\ \ s_b \bar\Phi (x)&=& \alpha (x) -\beta (x),\label{sw}
\end{eqnarray}
where $\Phi (x)$   and $\bar\Phi (x)$  represent 
the collective original and shifted fields respectively.
Here $\beta (x)$ refers the original BRST transformation of with respect to the 
shifted fields,
whereas $\alpha (x)$ corresponds the ghost fields associated with shift
symmetry collectively.  The explicit form of the extended BRST transformation (\ref{sw}) is given explicitly in Eq. (\ref{bs1}) of
Appendix B.   
Now, in order to make theory unchanged, we need to remove the contribution of
these ghosts from the physical states. To do so,  we further
 introduce the following antifields (anti-ghosts): $B_{\mu\nu\eta}^\star,$ 
$c_{\mu\nu}^\star,$  $\tilde 
c_{\mu\nu}^\star,$ $B_{\mu\nu}^\star,$ 
$\tilde B_{\mu\nu}^\star,$ $\eta_\mu^\star,$   $\tilde \beta_\mu^\star,$  $F_\mu^\star,$ 
$\tilde F_\mu^\star,$ $f_\mu^\star,$ $\tilde f_\mu^\star,$ $c_2^\star,$ $\tilde c_2^\star,$ $c_1^\star,$ 
$\tilde c_1^\star,$ 
$\phi_\mu^\star,$ $B^\star,$ $B_1^\star,$ $B_2^\star$ to the theory.
The BRST variation of these antifields are given in  (\ref{bs2}).
Now, to remove the redundancies, we fix the gauge for shift symmetry, which is achieved 
by adding the following gauge fixing term to the VSR quantum
action
\begin{eqnarray}
\bar{\cal L}^B_{gf}  
&=& l_{\mu\nu\eta}{\bar B}^{\mu\nu\eta} -B_{\mu\nu\eta}^\star (L^{\mu\nu\eta}- \partial^\mu c^{\nu \eta} +
\partial^\mu \bar{c}^{\nu \eta} - \partial^\nu c^{\eta \mu} + \partial^\nu \bar{c}^{\eta \mu} - \partial^\eta c^{\mu \nu} + \partial^\eta \bar{c}^{\mu \nu}\nonumber\\
&+& \frac{1}{2}\frac{m^2}{n\cdot \partial}n^\mu c^{\nu \eta} - \frac{1}{2}\frac{m^2}{n\cdot \partial}n^\mu \bar{c}^{\nu \eta} + \frac{1}{2}\frac{m^2}{n\cdot \partial}n^\nu c^{\eta \mu} - \frac{1}{2}\frac{m^2}{n\cdot \partial}n^\nu \bar{c}^{\eta \mu} + \frac{1}{2}\frac{m^2}{n\cdot \partial}n^\eta c^{\mu \nu}\nonumber\\
& -& \frac{1}{2}\frac{m^2}{n\cdot \partial}n^\eta \bar{c}^{\mu \nu} )+ \tilde {c}_{\mu\nu}^\star (M^{\mu\nu}-\partial^\mu\beta^\nu +\partial^\mu\bar{\beta}^\nu + \partial^\nu\beta^\mu -\partial^\nu\bar{\beta}^\mu + \frac{1}{2}\frac{m^2}{n\cdot \partial}n^\mu \beta^\nu  \nonumber\\
&-&  \frac{1}{2}\frac{m^2}{n\cdot \partial}n^\mu \bar{\beta}^\nu - \frac{1}{2}\frac{m^2}{n\cdot \partial}n^\nu \beta^\mu +\frac{1}{2}\frac{m^2}{n\cdot \partial}n^\nu \bar{\beta}^\mu)+ m_{\mu\nu}\bar {\tilde c}^{\mu\nu} +{ c}_{\mu\nu}^\star (\tilde M^{\mu\nu}- 
B^{\mu\nu} +\bar B^{\mu\nu})\nonumber\\
&+&  n_{\mu\nu}\bar B^{\mu\nu} -B_{\mu\nu}^\star N^{\mu\nu} +\bar n_{\mu\nu}\bar {\tilde B}^{\mu\nu} -
\tilde B_{\mu\nu}^\star (\tilde N^{\mu\nu} - \partial^\mu F^\nu +\partial^\mu \bar{F}^\nu +\partial^\nu F^\mu  \nonumber\\
&-&  \partial^\nu \bar{F}^\mu + \frac{1}{2}\frac{m^2}{n\cdot \partial}n^\mu F^\nu- \frac{1}{2}\frac{m^2}{n\cdot \partial}n^\mu \bar{F}^\nu   -\frac{1}{2}\frac{m^2}{n\cdot \partial}n^\nu F^\mu + \frac{1}{2}\frac{m^2}{n\cdot \partial}n^\nu \bar{F}^\mu ) + o_\mu \bar \beta ^\mu  \nonumber\\
&-&  \beta_\mu^\star (O^\mu -\partial^\mu c_2+\partial^\mu \bar c_2 + \frac{1}{2}\frac{m^2}{n\cdot \partial}n^\mu c_2 - \frac{1}{2}\frac{m^2}{n\cdot \partial}n^\mu \bar{c}_2) ++\bar o_\mu \bar {\tilde 
\beta}^\mu -\tilde \beta_\mu^\star (\tilde O^\mu -\tilde F ^\mu \nonumber\\
&+&\bar {\tilde F}^\mu) 
+\bar p_\mu \bar F^\mu + \tilde F_\mu^\star (P^\mu +\partial^\mu B-  \partial^\mu \bar B -\frac{1}{2}\frac{m^2}{n\cdot \partial}n^\mu B +\frac{1}{2}\frac{m^2}{n\cdot \partial}n^\mu \bar B ) + p_\mu \bar {\tilde F} ^\mu 
\nonumber\\
&+&  F_\mu^\star \tilde P^\mu + \bar q_\mu \bar f^\mu + \tilde f_\mu^\star Q^\mu +  q_\mu \bar {\tilde f^\mu}+ f_\mu^\star (\tilde Q^\mu -\partial ^\mu B_1+    \partial ^\mu \bar B_1 +\frac{1}{2}\frac{m^2}{n\cdot \partial}n^\mu  B_1 \nonumber\\
&-& \frac{1}{2}\frac{m^2}{n\cdot \partial}n^\mu \bar B_1 )+\bar r\bar c_2 +\tilde c_2^\star \mathfrak{R} 
+ r \bar {\tilde c}_2 + c_2^\star (\tilde {\mathfrak{R}} -B_2 +\bar B_2 ) + \bar s \bar c_1  +  s \bar {\tilde c}_1+v\bar B_1 
\nonumber\\
&+& \tilde c_1^\star ({\mathfrak{S}}+B-\bar B) +   c_1^\star (\tilde {\mathfrak{S}} -B_1 +\bar B_1)+ t_\mu \bar 
\phi^\mu -\phi_\mu ^\star (T^\mu -f^\mu +\bar f^\mu ) + u\bar B  \nonumber\\
&-& B^\star \mathbb{U} +w\bar B_2 -B_2 ^\star \mathfrak{W} -B_1^\star \mathfrak{V}.\label{brl}
\end{eqnarray}
where the fields $L_{\mu\nu\eta},$  $M_{\mu\nu},$ $\tilde M_{\mu\nu},$ $N_{\mu\nu},$ $\tilde N_{\mu\nu},$ $O_\mu,$ 
$\tilde O_\mu,$ $P_\mu,$ $\tilde P_\mu,$ $Q_\mu,$ $\tilde Q_\mu,$ $\mathfrak{R},$ $\tilde {\mathfrak{R}},$ ${\mathfrak{S}},$ $\tilde {\mathfrak{S}},$ $T_\mu,$ $\mathbb{U},$ 
$\mathfrak{V} $ and $\mathfrak{W}$ are the ghost fields 
associated with the shift symmetries for fields 
$B_{\mu\nu\eta},$ $c_{\mu\nu},$ $\tilde c_{\mu\nu},$ $B_{\mu\nu},$ $\tilde B_{\mu\nu},$ $\beta_\mu,$ $\tilde \beta_\mu,$ $F_\mu,$ 
$\tilde F_\mu,$ $f_\mu,$ $\tilde f_\mu,$ $c_2,$ $\tilde c_2,$ $c_1,$ $\tilde c_1,$ $\phi_\mu,$ $B,$ $B_1,$ $B_2$ 
respectively and the fields $l_{\mu\nu\eta},$  $m_{\mu\nu},$ $\bar m_{\mu\nu},$ $n_{\mu\nu},$ $\bar 
n_{\mu\nu},$ $o_\mu,$ $\bar
o_\mu,$ $p_\mu,$ $\bar p_\mu,$ $q_\mu,$  $\bar q_\mu,$ $r,$ $\bar r,$ $s,$ $\bar s,$ $t_\mu,$ $u,$ $v,$ $w$  
are the Nakanishi-Lautrup type auxiliary fields corresponding to the antighost fields $B_{\mu\nu\eta}^\star,$ 
$c_{\mu\nu}^\star,$  $\tilde 
c_{\mu\nu}^\star,$ $B_{\mu\nu}^\star,$ 
$\tilde B_{\mu\nu}^\star,$ $\eta_\mu^\star,$   $\tilde \beta_\mu^\star,$  $F_\mu^\star,$ 
$\tilde F_\mu^\star,$ $f_\mu^\star,$ $\tilde f_\mu^\star,$ $c_2^\star,$ $\tilde c_2^\star,$ $c_1^\star,$ 
$\tilde c_1^\star,$ 
$\phi_\mu^\star,$ $B^\star,$ $B_1^\star,$ $B_2^\star$, respectively.

The gauge-fixed Lagrangian density corresponding to shift symmetry, 
$\bar{\cal L}^B_{gf}$, is invariant 
under the extended BRST symmetry transformations given in Eqs. (\ref{bs1}) and (\ref{bs2}) of Appendix B. If we perform the equations of motion for auxiliary fields, then all  bar fields 
disappear   and  
we left with the following  gauge-fixed Lagrangian density:
\begin{eqnarray}
\bar{\cal L}^B_{gf} &= &   -B_{\mu\nu\eta}^\star (L^{\mu\nu\eta}-
\partial^\mu c^{\nu \eta} - \partial^\nu c^{\eta\mu} -\partial^\eta c^{\mu\nu }  + \frac{1}{2}\frac{m^2}{n\cdot \partial}n^\mu c^{\nu \eta} + \frac{1}{2}\frac{m^2}{n\cdot \partial}n^\nu c^{  \eta\mu} \nonumber\\
&+&   \frac{1}{2}\frac{m^2}{n\cdot \partial}n^\eta c^{\mu\nu })+\tilde {c}_{\mu\nu}^\star (M^{\mu\nu}-\tilde{\partial}^\mu\beta^\nu  +\tilde{\partial}^\nu\beta^\mu + \frac{1}{2}\frac{m^2}{n\cdot \partial}n^\mu \beta^\nu -\frac{1}{2}\frac{m^2}{n\cdot \partial}n^\nu           \beta^\mu ) \nonumber\\
&+&  {c}_{\mu\nu}^\star (\tilde M^{\mu\nu}- B^{\mu\nu} ) -B_{\mu\nu}^\star N^{\mu\nu}+\tilde B_{\mu\nu}^\star (\tilde N^{\mu\nu} -\partial^\mu f^\nu   +\partial^\nu f^\mu +
\frac{1}{2}\frac{m^2}{n\cdot \partial}n^\mu f^\nu \nonumber\\
&-& \frac{1}{2}\frac{m^2}{n\cdot \partial}n^\nu f^\mu )-\beta_\mu^\star(O^\mu -\partial^\mu c_2 +\frac{1}{2}\frac{m^2}{n\cdot \partial}n^\mu c_2)+\tilde \beta_\mu^\star (\tilde O^\mu -\tilde F ^\mu )  \nonumber\\ 
&+&   \tilde F_\mu^\star (P^\mu +\partial^\mu B  - \frac{1}{2}\frac{m^2}{n\cdot \partial}n^\mu B)+ F_\mu^\star \tilde P^\mu  +\tilde  f_\mu^\star Q^\mu + f_\mu^\star (\tilde Q^\mu -\partial^\mu B_1  \nonumber\\
&+& \frac{1}{2}\frac{m^2}{n\cdot \partial}n^\mu B_1 ) + \tilde c_2^\star {\mathfrak{R}} + 
 c_2^\star (\tilde {\mathfrak{R}} -B_2 ) 
 + \tilde c_1^\star ({\mathfrak{S}}+B) + c_1^\star (\tilde {\mathfrak{S}} -B_1 ) 
 \nonumber\\
&-&  \phi_\mu ^\star (T^\mu -f^\mu ) -B^\star \mathbb{U}   -B_1^\star \mathfrak{V}   -B_2 ^\star \mathfrak{W}.\label{ki}
\end{eqnarray}
Furthermore, if the general gauge-fixing fermion of the Abelian 3-form gauge theory
in VSR,  $\Psi$, depends on the original fields only, then we can write
a general gauge-fixed Lagrangian density 
for this theory with the original BRST symmetry in terms of $\Psi$ as
\begin{equation}
{\cal L}^B_{gf}  =  s_b \Psi [\Phi]= \sum_\Phi  (s_b \Phi) \frac{\delta\Psi}{\delta \Phi},
\end{equation}
where $\Phi$ is the generic notation for all fields in the theory.
Keeping the  nature of  fields (i.e fermionic and bosonic) in mind the above 
gauge-fixed Lagrangian density in can be expressed as 
\begin{eqnarray}
{\cal L}^B_{gf} &= & -\frac{\delta\Psi}{\delta B_{\mu\nu\eta}}L_{\mu\nu\eta}+
 \frac{\delta\Psi}{\delta c_{\mu\nu}}M_{\mu\nu} + \frac{\delta\Psi}
{\delta\tilde{c}_{\mu\nu}}\tilde M_{\mu\nu} -
 \frac{\delta\Psi}{\delta B_{\mu\nu}}N_{\mu\nu}-\frac{\delta\Psi}{\delta \tilde B_{\mu\nu}}\tilde N_{\mu\nu}
\nonumber\\
&-& \frac{\delta\Psi}{\delta
\beta_\mu}O_\mu - \frac{\delta\Psi}{\delta
\tilde\beta_\mu}\tilde O_\mu +\frac{\delta\Psi}{\delta F_\mu}P_\mu +\frac{\delta\Psi}{\delta \tilde F_\mu}
\tilde P_\mu + \frac{\delta\Psi}{\delta f_\mu}Q_\mu +\frac{\delta\Psi}{\delta \tilde f_\mu}\tilde Q_\mu \nonumber\\
&+&
 \frac{\delta\Psi}{ \delta c_2}{\mathfrak{R}} + \frac{\delta\Psi}{ \delta \tilde c_2}\tilde {\mathfrak{R}}
 + \frac{\delta\Psi}{ \delta c_1}{\mathfrak{S}} + \frac{\delta\Psi}{ \delta \tilde c_1}\tilde {\mathfrak{S}} 
 -  \frac{\delta\Psi}{ \delta \phi_\mu}T_\mu - \frac{\delta\Psi}{ \delta B}\mathbb{U}
-\frac{\delta\Psi}{\delta B_1}\mathfrak{V}-\frac{\delta\Psi}{\delta B_2}\mathfrak{W}. \label{ci} 
\end{eqnarray}
Now, we are able to define the BV action for Abelian 3-form gauge theory in VSR by combining  Eqs. (\ref{lag3}), (\ref{ki}) and (\ref{ci}) as follows  
\begin{eqnarray}  
 {\cal L}_{eff}&=& {\cal L}_0(B_{\mu\nu\rho}-\bar B_{\mu\nu\rho})+ 
{\cal L}^B_{gf}+\bar{\cal L}^B_{gf}.\label{comp}
\end{eqnarray}
The explicit expression of ${\cal L}_{eff}$ is given in (\ref{b00}) (see in Appendix B). 

To obtain the identifications on the antifields of 3-form theory in VSR, we
 integrate   the ghosts associated
with the shift symmetry and utilize the gauge-fixing fermion (\ref{gfff}). 
We thus obtain the explicit values of antifields as following:
\begin{eqnarray}
&&B^{\mu\nu\eta\star}  =\frac{1}{3} ( {\partial^\mu} \tilde c^{\nu\eta}
+ {\partial}^\nu \tilde c^{\eta\mu} + {\partial}^\eta \tilde c^{\mu\nu} -\frac{1}{2}\frac{m^2}{n\cdot \partial}n^\mu\tilde c^{\nu\eta} -\frac{1}{2}\frac{m^2}{n\cdot \partial}n^\nu\tilde c^{\eta\mu}   -\frac{1}{2}\frac{m^2}{n\cdot \partial}n^\eta\tilde c^{\mu\nu}                                ),\nonumber\\  
&& c^{\mu\nu\star}  = -\frac{1}{2}B^{\mu\nu},
\ \ B^{\mu\nu\star} =-\frac{1}{2}\tilde c^{\mu\nu},\ \
\tilde\beta^{\mu \star}  =F^\mu +\partial_\nu c^{\mu\nu} -\frac{1}{2}\frac{m^2}{n\cdot \partial}n_\nu c^{\mu\nu}  ,\nonumber\\
&&\tilde c^{\mu\nu\star} =\frac{1}{2}(\partial^\mu\tilde \beta^\nu -\partial^\nu\tilde\beta^\mu -\frac{1}{2}\frac{m^2}{n\cdot \partial}n^\mu\tilde \beta^\nu +\frac{1}{2}\frac{m^2}{n\cdot \partial}n^\nu\tilde\beta^\mu  )-
\partial_\eta B^{\mu\nu\eta} +\frac{1}{2}\frac{m^2}{n\cdot \partial}n_\eta B^{\mu\nu\eta}  ,\nonumber\\
&&\beta^{\mu\star}  =-\frac{1}{2}\partial^\mu\tilde c_2 +\frac{1}{4}\frac{m^2}{n\cdot \partial}n^\mu\tilde c_2 ,\ 
c_2^{\star} = \frac{1}{2}B-( \partial_\mu -\frac{1}{2}\frac{m^2}{n\cdot \partial}n_\mu) \beta^\mu,
\  \tilde c_1^{\star} =-\frac{1}{2}B_2,\nonumber\\  
&&\tilde F^{\mu\star} =\tilde\beta^\mu,\ 
 f^{\mu\star}  = \phi^\mu,\ \
 c_1^{\star} =\frac{1}{2}B_1,\ \phi^{\mu\star} =\tilde f^\mu,\
B^{\star} =\frac{1}{2}\tilde c_2,\ 
  B_1^{\star} =\frac{1}{2}\tilde c_1,\
B_2^{\star} = -\frac{1}{2}c_1,\nonumber\\
&& [\tilde B^{\mu\nu\star}, F^{\mu\star}, f^{\mu \star}, \tilde c_2^{\star}]=0 .
\label{an3}
\end{eqnarray}
Now, we are able to express gauge-fixed  Lagrangian density corresponding to
extended BRST
transformations in terms of  generalized gauge-fixing fermion as follows, 
\begin{eqnarray}
 {\cal L}^B_{gf} +{\bar{\cal L}}^B_{gf}&=&  s_b \left(B_{\mu\nu\eta}^\star \bar B^{\mu\nu\eta} +
c_{\mu\nu}^\star {\bar {\tilde c}}^{\mu\nu} +\tilde c_{\mu\nu}^\star \bar {c}^{\mu\nu} +
B_{\mu\nu}^\star \bar B^{\mu\nu} +\beta_{\mu}^\star \bar \beta^{\mu} +\tilde\beta_{\mu}^\star 
{\bar{\tilde \beta}}^{\mu}+F_{\mu}^\star {\bar{\tilde F}}^{\mu}+ \tilde F_{\mu}^\star \bar{ F}^{\mu}\right.\nonumber\\
&+&\left.f_{\mu}^\star \bar{\tilde f}^{\mu}+ \tilde f_{\mu}^\star \bar{ f}^{\mu} 
 +
c_2^\star {\bar{\tilde c}}_2 + \tilde c_2^\star \bar{ c_2} 
 + c_1^\star {\bar{\tilde c}}_1 + \tilde c_1^\star \bar{ c_1} 
+\phi_\mu^\star\bar\phi^\mu +B^\star\bar B+B_1^\star\bar B_1+B_2^\star\bar B_2
\right),\nonumber\\ 
&=:&s_b \left(\Phi^\star\bar\Phi\right),
\end{eqnarray}
where the collective fields $\Phi$ and $\bar\Phi$ denote the
  original fields and corresponding shifted fields respectively. Here
the  ghost number  of the expression  $\Phi^\star\bar\Phi=-1$,  as we expect. Here 
we see a difference in the expression of generalized gauge-fixing fermion 
to that of the ordinary gauge-fixing fermion  
\begin{eqnarray}
\Psi &=& - [ B_{\mu\nu\eta}B^{\mu\nu\eta\star} +\tilde c_{\mu\nu} c^{\mu\nu\star} +\tilde\beta_\mu
\tilde\beta^{\mu\star}+\phi_\mu\phi^{\mu\star}+\tilde c_2c_2^\star +\tilde c_1 c_1^\star +B_2 B_2^\star].
\end{eqnarray} 
Plugging
the values of the  antifields (\ref{an3}), we can recover the original Lagrangian density of 3-form gauge theory in VSR.
\subsection{   VSR modified extended BRST invariant BV action in superspace } 
Furthermore, while the   BRST and the  anti-BRST symmetries of 3-form
theories can be given a geometrical meaning and have led to a superspace formulation of
such theories \cite{sudb}, a superspace description of the VSR modified BV action does not exist so far.
 Here, we develop  a superspace formalism of the VSR modified 3-form theory having extended
BRST invariance only. To study the  theory with extended
BRST invariance only, we need
  one extra fermionic parameter $\theta$ together with $x_\mu$. 
 The components of superfields, $\mathfrak{T}$,  in terms of generic fields $\Phi$
 is given by
\begin{eqnarray}
\mathfrak{T} (x, \theta) =\Phi (x) +\theta (s_b\Phi ).\label{brss}
\end{eqnarray}
With the help of extended BRST transformation (\ref{bs1}),
the explicit expressions for the superfield are listed in Eq. (\ref{supb}) 
of Appendix. 

The gauge-fixed Lagrangian density corresponding to shift symmetry
mentioned in  (\ref{brl}) in the superspace is  
 is simplified as
\begin{eqnarray}
\bar{\cal L}^B_{gf} &=& \frac{\delta}{\delta\theta}\left[\bar{\cal B}_{\mu\nu\eta}^\star \bar{\cal 
B}^{\mu\nu\eta} +\bar{\tilde{\cal C}}_{\mu\nu}^\star  \bar{\cal C}^{\mu\nu}
+\bar{\tilde{\cal C}}_{\mu\nu}  \bar{\cal C}^{\mu\nu\star}+\bar{\tilde{\cal B}}_{\mu\nu}^\star  
\bar{\tilde{\cal B}}^{\mu\nu}
+\bar{{\cal B}}_{\mu}^\star  \bar{\cal B}^{\mu}+ \bar{\tilde{\cal F}}_\mu^\star \bar {\cal F}^\mu +{\bar{\tilde{\mathfrak{F}}}}_\mu^\star {\bar {\mathfrak{F}}}^\mu 
\right.\nonumber\\
&+&\left. {\bar{\tilde{\cal F}}}_\mu  \bar {\cal F}^{\mu\star} +{\bar{\tilde{\mathfrak{F}}}}_\mu {\bar {\mathfrak{F}}}^{\mu\star}
 +{\bar{\tilde{\cal C}}}_1^\star  \bar{\cal 
C}_1  + {\bar{\tilde{\cal C}}}_1 \bar{\cal 
C}^\star _1+ {\bar{\tilde{\cal C}}}_2^\star  \bar{\cal 
C}_2  + {\bar{\tilde{\cal C}}}_2 \bar{\cal 
C}^\star _2+\bar{\cal B}^\star  \bar{\cal 
B} +\bar{\cal B}_1^\star  \bar{\cal 
B}_1  +\bar{\cal B}_2^\star  \bar{\cal 
B}_2\right],\nonumber\\
&=&\frac{\delta}{\delta\theta}\left[\bar{\mathfrak{T}} ^\star \bar{\mathfrak{T}}  \right].
\end{eqnarray}
Being the $\theta$ component of a superfields,  $\bar{\cal L}^B_{gf}$ 
 is invariant under the
extended BRST transformation. 
In the extended BRST invariant superspace, the gauge-fixed fermion
of the original VSR modified 3-form gauge theory in component form  translates as
\begin{equation}
{\Gamma  } (x,\theta) =\Psi (x) +\theta s_b \Psi. 
\end{equation}
If we assume a general gauge-fixed fermion $\Psi$ depending on all the original fields, 
then a ${\Gamma  } (x,\theta)$ reads,
\begin{eqnarray}
\Gamma (x,\theta) &=&\Psi (x)+\theta \left[-\frac{\delta\Psi}{\delta B_{\mu\nu\eta}}L_{\mu\nu\eta}+
 \frac{\delta\Psi}{\delta c_{\mu\nu}}M_{\mu\nu} + \frac{\delta\Psi}
{\delta\tilde{c}_{\mu\nu}}\tilde M_{\mu\nu} -
 \frac{\delta\Psi}{\delta B_{\mu\nu}}N_{\mu\nu}
-\frac{\delta\Psi}{\delta \tilde B_{\mu\nu}}\tilde N_{\mu\nu}\right.\nonumber\\
&-&\left. \frac{\delta\Psi}{\delta
\beta_\mu}O_\mu - \frac{\delta\Psi}{\delta
\tilde\beta_\mu}\tilde O_\mu +\frac{\delta\Psi}{\delta F_\mu}P_\mu +\frac{\delta\Psi}{\delta \tilde F_\mu}
\tilde P_\mu + \frac{\delta\Psi}{\delta f_\mu}Q_\mu +\frac{\delta\Psi}{\delta \tilde f_\mu}\tilde Q_\mu
 + \frac{\delta\Psi}{ \delta c_2}{\mathfrak{R}}
\right.\nonumber\\
&+&  \left. \frac{\delta\Psi}{ \delta \tilde c_2}\tilde {\mathfrak{R}}
 + \frac{\delta\Psi}{ \delta c_1}{\mathfrak{S}} + \frac{\delta\Psi}{ \delta \tilde c_1}\tilde {\mathfrak{S}}-
 \frac{\delta\Psi}{ \delta \phi_\mu}T_\mu - \frac{\delta\Psi}{ \delta B}\mathbb{U}
-\frac{\delta\Psi}{\delta B_1}\mathfrak{V}-\frac{\delta\Psi}{\delta B_2}\mathfrak{W} \right].
\end{eqnarray}
Therefore, the  VSR modified  gauge-fixing Lagrangian density (\ref{lag3}) 
in the superspace formalism takes very compact form as
\begin{equation}
{\cal L}^B_{gf} =\frac{\delta\Gamma (x,\theta)}{\delta\theta}.
\end{equation}
Here we note that  the invariance of this
Lagrangian density is obvious under the extended BRST transformation as 
it belongs to $\theta$ component.
 Now, we are able to write the extended BRST invariant BV action in the superspace as 
 follows
\begin{eqnarray}
{\cal L}_{eff}  = {\cal L}_0 (B_{\mu\nu\rho}-\bar{B}_{\mu\nu\rho})
+\frac{\delta\Gamma (x,\theta)}{\delta\theta}+ \frac{\delta}{\delta\theta}\left[\bar{\mathfrak{T}} ^\star \bar{\mathfrak{T}} \right].                                        
\end{eqnarray} 
Here we observe that eliminating  auxiliary fields and ghost fields of
the shift symmetry through using equations of motion, this effective Lagrangian density reduces to the original BRST invariant Lagrangian density of 3-form gauge theory  in VSR also.

\subsection{ VSR modified extended anti-BRST invariant BV action} 
 In this subsection, we discuss the extended 
anti-BRST symmetry for Abelian 3-form gauge theory in VSR and their 
superspace description. 

As the gauge-fixing  Lagrangian density is anti-BRST exact, so one can write this 
 in terms of corresponding gauge-fixing fermion $\bar\Psi$ as
\begin{eqnarray}
{\cal L}^{B}_{gf} = s_{ab} \bar\Psi,
\end{eqnarray}
where $\bar\Psi$ fhas the following expression:
\begin{eqnarray}
\bar\Psi &=&  -\frac{1}{2} B_2 c_2 +\frac{1}{2} B_1c_1 +\frac{1}{2} \tilde c_1 B +\frac{1}{2} B_{\mu\nu} c^{\mu\nu} -\frac{1}{2}\tilde c_{\mu\nu} (\partial^\mu\beta^\nu -\partial^\nu \beta^\mu - \frac{1}{2}\frac{m^2}{n\cdot \partial}n^\mu\beta^\nu \nonumber\\
&+& \frac{1}{2}\frac{m^2}{n\cdot \partial}n^\nu \beta^\mu )  +\tilde F_\mu \beta^\mu + \beta _\mu 
\partial^\mu c_2 - \beta _\mu \frac{1}{2}\frac{m^2}{n\cdot \partial}n^\mu c_2 + \frac{1}{2} \phi_\mu f^\mu +\frac{1}{3} B_{\mu\nu\eta} (\partial^\mu c^{\nu\eta}  \nonumber\\
&+& \partial^\nu c^{
\eta\mu} +\partial^\eta c^{\mu\nu} - \frac{1}{2}\frac{m^2}{n\cdot \partial}n^\mu                                           c^{\nu\eta} -\frac{1}{2}\frac{m^2}{n\cdot \partial}n^\nu c^{
\eta\mu} - \frac{1}{2}\frac{m^2}{n\cdot \partial}n^\eta c^{\mu\nu}  ).
\end{eqnarray}  
Keeping the  structure of  (\ref{bs1}) (given in Appendix B) in mind, 
we demand that anti-BRST transformation of shifted fields $(\Phi-\bar\Phi)$ 
coincides with the anti-BRST transformations of the original fields $(\Phi)$ 
if all the bar fields vanish. 
This
requirement leads to the following form of the
 (VSR modified)  extended anti-BRST transformations:
\begin{eqnarray}
s_{ab} \bar {B}_{\mu\nu\eta} &=& B_{\mu\nu\eta}^\star,\nonumber\\
s_{ab} {B}_{\mu\nu\eta}&=&  B_{\mu\nu\eta}^\star +
\partial_\mu \tilde c_{\nu\eta}-\partial_\mu \bar{\tilde c}_{\nu\eta}+\partial_\nu \tilde c_{\eta\mu}
-\partial_\nu \bar{\tilde c}_{\eta\mu}  +\partial_\eta 
\tilde c_{\mu\nu}-\partial_\eta \bar{\tilde c}_{\mu\nu} -\frac{1}{2}\frac{m^2}{n\cdot \partial}n_\mu\tilde c_{\nu\eta}\nonumber\\
&+&\frac{1}{2}\frac{m^2}{n\cdot \partial}n_\mu \bar{\tilde c}_{\nu\eta}- \frac{1}{2}\frac{m^2}{n\cdot \partial}n_\nu\tilde c_{\eta\mu} + \frac{1}{2}\frac{m^2}{n\cdot \partial}n_\nu \bar{\tilde c}_{\eta\mu}-\frac{1}{2}\frac{m^2}{n\cdot \partial}n_\eta \tilde c_{\mu\nu} + \frac{1}{2}\frac{m^2}{n\cdot \partial}n_\eta \bar{\tilde c}_{\mu\nu},\nonumber\\
s_{ab} \bar{\tilde c}_{\mu\nu} &=& \tilde c_{\mu\nu}^\star,\ \
s_{ab} {\tilde c}_{\mu\nu} =   \tilde c_{\mu\nu}^\star +\partial_\mu \tilde\beta_\nu -\partial_\mu 
\bar{\tilde\beta}_\nu -\partial_\nu \tilde\beta_\mu +\partial_\nu \bar{\tilde\beta}_\mu 
- \frac{1}{2}\frac{m^2}{n\cdot \partial}n_\mu\tilde\beta_\nu\nonumber\\
&+&\frac{1}{2}\frac{m^2}{n\cdot \partial}n_\mu\bar{\tilde\beta}_\nu +\frac{1}{2}\frac{m^2}{n\cdot \partial}n_\nu\tilde\beta_\mu  - \frac{1}{2}\frac{m^2}{n\cdot \partial}n_\nu \bar{\tilde\beta}_\mu,  \ \
s_{ab} \bar{ c}_{\mu\nu} =c_{\mu\nu}^\star,\nonumber\\
s_{ab} { c}_{\mu\nu} &=& c_{\mu\nu}^\star +\tilde B_{\mu\nu} -\bar{\tilde B}_{\mu\nu}, \  \
s_{ab} { B}_{\mu\nu} =  B_{\mu\nu}^\star + \partial_\mu\tilde f_\nu -\partial_\mu\bar{\tilde f}_\nu
-\partial_\nu\tilde f_\mu +\partial_\nu\bar{\tilde f}_\mu\nonumber\\
&-& \frac{1}{2}\frac{m^2}{n\cdot \partial}n_\mu\tilde f_\nu +\frac{1}{2}\frac{m^2}{n\cdot \partial}n_\mu\bar{\tilde f}_\nu +\frac{1}{2}\frac{m^2}{n\cdot \partial}n_\nu\tilde f_\mu -                     \frac{1}{2}\frac{m^2}{n\cdot \partial}n_\nu\bar{\tilde f}_\mu,\ s_{ab} \phi_\mu  = \phi_\mu^\star +\tilde f_\mu -\bar{\tilde f}_\mu, \nonumber\\
s_{ab} \bar{ B}_{\mu\nu} &=& B_{\mu\nu}^\star,\ \
s_{ab} \bar \beta_\mu =\beta_\mu^\star,\ \ s_{ab} \beta_\mu
=\beta_\mu^\star +F_\mu -\bar F_\mu, \ \
s_{ab} \bar {\tilde \beta}_\mu  = \tilde\beta_\mu^\star,\nonumber\\
s_{ab} \tilde \beta_\mu &=& \tilde\beta_\mu^\star +\partial_\mu \tilde c_2-\partial_\mu \bar{\tilde c}_2 - \frac{1}{2}\frac{m^2}{n\cdot \partial}n_\mu \tilde c_2+\frac{1}{2}\frac{m^2}{n\cdot \partial}n_\mu\bar{\tilde c}_2, \ \ 
s_{ab} \bar {\tilde F}_\mu  = \tilde F_\mu^\star,\nonumber\\
 s_{ab} \tilde F_\mu
&=&  \tilde F_\mu^\star -\tilde{\partial}_\mu B_2+\tilde{\partial}_\mu \bar B_2 + \frac{1}{2}\frac{m^2}{n\cdot \partial}n_\mu   B_2 - \frac{1}{2}\frac{m^2}{n\cdot \partial}n_\mu \bar B_2    
,\ s_{ab} \bar {f}_\mu  =f_\mu^\star, \nonumber\\
 s_{ab} f_\mu
  &=&f_\mu^\star -\partial_\mu B_1+\partial_\mu \bar B_1 +\frac{1}{2}\frac{m^2}{n\cdot \partial}n_\mu B_1 - \frac{1}{2}\frac{m^2}{n\cdot \partial}n_\mu \bar B_1, \ \ s_{ab} \bar {c}_2 =c_2^\star,  \nonumber\\
  s_{ab} c_2
 &=&c_2^\star +B -\bar B, \ \
s_{ab} \bar {c}_1 =c_1^\star,\ \
s_{ab} c_1 =c_1^\star -B_1 +\bar B_1,\ \ s_{ab} \bar {\phi}_\mu  = 
\phi_\mu^\star,\nonumber\\
s_{ab} \bar {\tilde c}_1  &=&\tilde c_1^\star,\ \ s_{ab} \tilde c_1
=\tilde c_1^\star -B_2 +\bar B_2,\ \ 
s_{ab} \bar {\tilde c}_2 =\tilde c_2^\star,\ \ s_{ab} \tilde c_2
=\tilde c_2^\star, \ \
s_{ab} \bar {\tilde f}_\mu  = \tilde f_\mu^\star,\nonumber\\
s_{ab} \tilde f_\mu
&=&\tilde f_\mu^\star,\ \ 
 s_{ab} \bar {F}_\mu   = F_\mu^\star,\   s_{ab} F_\mu
=F_\mu^\star, \  s_{ab} \bar B  = B^\star,\ \
s_{ab} B =B^\star, \ 
 s_{ab} \bar B_1  = B_1^\star,\nonumber\\
s_{ab} B_1
&=&B_1^\star,\ 
 s_{ab} \bar B_2 =B_2^\star,\  s_{ab} B_2
=B_2^\star,\ 
s_{ab} \bar {\tilde B}_{\mu\nu}  = \tilde B_{\mu\nu}^\star,    s_{ab} \tilde B_{\mu\nu}
=\tilde B_{\mu\nu}^\star. 
\end{eqnarray}
However, the following fields: $B_{\mu\nu\eta}^\star$, $\tilde c_{\mu\nu}^\star$,  $c_{\mu\nu}^\star$, $B_{\mu\nu}^\star$,
 $\beta_\mu^\star$, $\tilde \beta_\mu^\star$, $\tilde F_\mu^\star$, $f_\mu^\star$, $c_2^\star$, $c_1^\star$,
 $\phi_\mu^\star$, $\tilde c_1^\star$, $\tilde c_2^\star$, $\tilde f_\mu^\star$, $F_\mu^\star$, $B^\star$, $B_1^\star$, $B_2^\star$ and $\tilde B_{\mu\nu}^\star$,  do not change under extended anti-BRST transformation which 
 ensures the nilpotency of the transformation.
 The ghost fields associated with the shift symmetry change under the extended
anti-BRST transformations as
\begin{eqnarray}
s_{ab}L_{\mu\nu\eta}&=& l_{\mu\nu\eta},\ \ s_{ab}M_{\mu\nu} = m_{\mu\nu},\ \
 s_{ab}\tilde M_{\mu\nu} =\bar m_{\mu\nu},\ \
s_{ab}N_{\mu\nu} = n_{\mu\nu},\nonumber\\
 s_{ab}\tilde N_{\mu\nu} &=&\bar n_{\mu\nu},\ \
s_{ab}O_{\mu} = o_{\mu},\  
s_{ab}\tilde O_{\mu} = \bar o_{\mu},\  
 s_{ab}P_{\mu} =p_{\mu},\
s_{ab}\tilde P_{\mu} = \bar p_{\mu},\nonumber\\
 s_{ab}Q_{\mu} &=&q_{\mu},\ \
s_{ab}\tilde Q_{\mu} = \bar q_{\mu},\ \ 
s_{ab}{\mathfrak{R}}   = r,\ \
 s_{ab}\tilde {\mathfrak{R}} =\bar r,\ \
s_{ab}{\mathfrak{S}} = s,\nonumber\\
s_{ab}\tilde {\mathfrak{S}}  &=& \bar s,\ \
s_{ab}T_\mu =t_\mu,\ \
s_{ab}\mathbb{U} = u,\ \ 
s_{ab}\mathfrak{V}   = v,\ \ s_{ab}\mathfrak{W}   = w.
\end{eqnarray}
Rest of fields, whose transformation is written here,  do not change under the extended anti-BRST transformation. 
 To describe  the superspace formulation of the Abelian 3-form gauge theory in VSR
which has only extended anti-BRST invariance,  we need
one extra Grassmannian coordinate, say $\bar\theta$.
It is now straightforward to define the superfields in this superspace 
which involve the extended anti-BRST transformations along the $\bar{\theta}$ coordinates respectively.
Therefore, the superspace description of Abelian 3-form gauge theory 
in VSR also holds.
 
\subsection{VSR modified extended BRST and anti-BRST invariant superspace formulation}
In this section, we 
construct the superspace formulation for the 
Abelian 3-form gauge theory in VSR which is   manifestly invariant under both 
the extended
BRST transformations and the  extended anti-BRST transformations.
To describe such superspace, we introduce two extra  Grassmann
coordinates,   $\theta$ and $\bar\theta$
together with $x_\mu$.
Now, we compute the components of the superfields   by requiring  
  the field strength  to vanish along the directions of
$\theta$ and $\bar\theta$. This leads to the 
 superfields to have following form in this superspace generically:
\begin{equation}
\mathfrak{T} (x, \theta, \bar\theta )=\Phi (x) +\theta (s_b \Phi) +\bar\theta (s_{ab} \Phi)
+\theta\bar\theta (s_bs_{ab}\Phi ).
\end{equation} 
Here $\mathfrak{T}$ and $\Phi$ describe  all the superfields and the fields generically.
The explicit expression  for the individual superfields 
can be found in Eq. (\ref{superf}) of the Appendix B.   

Exploiting superfields (\ref{superf})
we compute the relations (\ref{lsr}). Interestingly, we establish a relation
 between the gauge-fixing Lagrangian density corresponding to shift symmetry, $\bar {\cal L}^B_{gf}$ 
 (\ref{lag3}),  and composite superfields as follows, 
\begin{eqnarray}
\bar {\cal L}^B_{gf} 
&=&\frac{1}{2}\frac{\delta}{\delta\bar\theta}\frac{\delta}{\delta\theta}\left[
\bar {\cal B}_{\mu\nu\eta} \bar {\cal B}^{\mu\nu\eta} +2\bar{\tilde {\cal C}}_{\mu\nu}\bar {\cal C}^{\mu\nu} 
+ {\bar{\tilde {\cal B}}}_{\mu\nu} {\bar{\tilde {\cal B}}^{\mu\nu}}+\bar{\cal B}_{\mu\nu} \bar{\cal B}^{\mu\nu}
 +\bar{\cal B}_{\mu } \bar{\cal B}^{\mu}+{\bar{\tilde {\cal B}}}_{\mu } 
 {\bar{\tilde{\cal B}}}^{\mu}\right.\nonumber\\
&+&\left. 2{\bar{\tilde {\cal F}}}_{\mu } \bar{\cal F}^{\mu}+ 2{\bar{\tilde {\mathfrak{F}}}}_{\mu } \bar{\mathfrak{F}}^{\mu} 
 + 2{\bar{\tilde {\cal C}}}_{2 } \bar
{\cal C}_2 +2{\bar{\tilde {\cal C}}}_{1} \bar{{\cal C}}_1+\bar{\Phi}_{\mu } \bar{ \Phi}^{\mu}+\bar {\cal B}\bar {\cal B}
+\bar {\cal B}_1\bar {\cal B}_1 +\bar {\cal 
B}_2\bar {\cal B}_2\right],
\end{eqnarray}
This relation manifests that the gauge-fixed Lagrangian density $\bar {\cal L}^B_{gf} $
is invariant under both the extended BRST and extended anti-BRST transformations.

The super gauge-fixing fermion for the theory having both the   extended BRST and extended anti-BRST invariance in superspace is given by
\begin{equation}
\Gamma  (x, \theta, \bar\theta )=\Psi (x) +\theta s_b\Psi +\bar\theta s_{ab}\Psi +\theta
\bar\theta s_b s_{ab}\Psi.\label{last}
\end{equation} 
In general, all four components of the super gauge-fixing fermion will be non-trivial, implying that if we
choose as ${\cal L}^B_{gf} = s_b\Psi$,
then it will not be invariant under generalized anti-BRST transformations.
This follows
from the fact that the  last component of the super gauge-fixing fermion (\ref{last}) is 
non-vanishing in
general.  However,  if the gauge-fixed Lagrangian density in VSR is both extended BRST and
anti-BRST invariant, then the $\theta
\bar\theta $ component of   super gauge-fixing fermion would vanish,
because when we use the equations of motion  the bar fields vanish and the theory
reduces to the original theory, where, by assumption, the gauge-fixed Lagrangian
density in VSR is both
extended BRST and anti-BRST invariant. 
Therefore, for an arbitrary super gauge-fixing fermion
  that leads to a BRST and anti-BRST invariant gauge-fixing
Lagrangian density, one can choose 
\begin{equation}
{\cal L}^B_{gf} =\frac{\delta}{\delta\theta}\left[\delta(\bar\theta)\Gamma  (x, \theta, 
\bar\theta) \right] = s_b\Psi.
\end{equation} 
Now, the effective Lagrangian density (\ref{comp}) possessing both  the extended BRST 
and anti-BRST  symmetries in superspace can be  expressed by
\begin{eqnarray}
{\cal L}_{eff} 
&=&{\cal L}_{0}(B_{\mu\nu\rho}-\bar B_{\mu\nu\rho} )+\frac{1}{2}\frac{\delta}{\delta\bar\theta}\frac{\delta}{\delta\theta}\left[
\bar {\cal B}_{\mu\nu\eta} \bar {\cal B}^{\mu\nu\eta} +2\bar{\tilde {\cal C}}_{\mu\nu}\bar {\cal C}^{\mu\nu} 
+ {\bar{\tilde {\cal B}}}_{\mu\nu} {\bar{\tilde {\cal B}}^{\mu\nu}}+\bar{\cal B}_{\mu\nu} \bar{\cal B}^{\mu\nu}
 +\bar{\cal B}_{\mu } \bar{\cal B}^{\mu}\right.\nonumber\\
&+&\left.  {\bar{\tilde {\cal B}}}_{\mu } 
 {\bar{\tilde{\cal B}}}^{\mu}+2{\bar{\tilde {\cal F}}}_{\mu } \bar{\cal F}^{\mu}+ 2{\bar{\tilde {\mathfrak{F}}}}_{\mu } \bar{\mathfrak{F}}^{\mu} 
 + 2{\bar{\tilde {\cal C}}}_{2 } \bar
{\cal C}_2 +2{\bar{\tilde {\cal C}}}_{1} \bar{{\cal C}}_1+\bar{\Phi}_{\mu } \bar{ \Phi}^{\mu}+\bar {\cal B}\bar {\cal B}
+\bar {\cal B}_1\bar {\cal B}_1 +\bar {\cal 
B}_2\bar {\cal B}_2\right]\nonumber\\
&+&\frac{\delta}{\delta\theta}\left[\delta(\bar\theta)
\Gamma (x, \theta, \bar\theta)\right].
\end{eqnarray}
Using the auxiliary fields equations of motion   the bar fields can be set zero. However,
  integration of the ghost fields of the shift symmetry leads to the explicit
expressions for the antifields, which, when
substituted into the VSR modified Lagrangian density,   yield the BV action.
 The superspace formulation of BV action for the VSR modified  3-form gauge theory
  has similar description as the Lorentz invariant case.
    
\section{Concluding Remarks}
In VSR,  the space–time translational symmetry is retained to preserve the
 energy-momentum and also the usual relativistic dispersion
relation. Keeping the significance of VSR in mind, 
in this paper, we have discussed the VSR description of the non-Abelian 1-form,
Abelian 2-form and Abelian 3-form gauge theories.
We have constructed the  extended BRST and anti-BRST transformation
  (which include a shift symmetry) for these theories. 
  To fix the shift symmetry, we need the antighost fields which  
  coincide with the antifields of the   BV formulation for each 
  gauge theories in VSR. Furthermore, we have formulated the
  these VSR modified theories in   superspace.
  First, we have found that 
 the extended BRST invariant  Lagrangian densities   of $p=1,2,3$-form gauge theories in VSR can be written   manifestly 
 in a superspace with one additional fermionic coordinate, i.e., $(x_\mu,\theta)$. 
 Similarly, extended anti-BRST invariant   Lagrangian densities  of these theories in VSR can be written   manifestly  in a superspace with  coordinates  $(x_\mu,\bar\theta)$  where $\bar\theta$ is 
 another fermionic coordinate. 
Finally, a superspace description of the (manifestly covariant) BV action of these theories in VSR  having both the extended BRST and extended anti-BRST invariance requires two additional Grassmann coordinates $(x_\mu,\theta, \bar\theta)$. 
In this context, we have noted that  if the gauge-fixed Lagrangian density in VSR is both extended BRST and anti-BRST invariant, then the $\theta \bar\theta $ component of   super gauge-fixing fermion would vanish,
because when we use the equations of motion  the bar fields would vanish and the theory
reduces to the original theory, where, by assumption, the gauge-fixed Lagrangian
density in VSR is both extended BRST and anti-BRST invariant. 
 The structure of the results we obtained here by studying BV action of  $p=1,2,3$-forms gauge theories with preferred direction  is   not very different to that of Lorentz invariant case.  Unlike  the Lorentz invariant case, the novel observation  is that in VSR scenario,  all the fields and superfields   acquire mass, which modifies the masses of the original dispersion relations. It will be interesting to extend this superspace formulation
   to some  regularization procedure at one-loop order, where we believe
 that the superfield associated
with the one-loop order term of the action  may have the
VSR modified anomalies and Wess-Zumino terms. Also,   the extension
of this superspace formulation for the more general cases in
which VSR modified anomalies and Wess-Zumino terms depend on the antifields will be interesting to
explore.

\acknowledgments 
We are thankful to Dr. Anton Ilderton for his careful reading  of the manuscript and helpful comments.
 
\appendix
\section{Mathematical details of VSR modified Abelian 2-form gauge theory in superspace}
The  explicit component form of superfields
for  VSR modified Abelian 2-form gauge theory in superspace
having both extended BRST and anti-BRST invariance are:
\begin{eqnarray}
 {\cal B}_{\mu\nu}(x, \theta, \bar\theta )  
 &=& B_{\mu\nu} (x) +\theta \psi_{\mu\nu}
+\bar\theta ( B_{\mu\nu}^\star + 
\partial_\mu\tilde\rho_\nu -\partial_\mu\bar{ \tilde\rho}_\nu-\partial_\nu\tilde\rho_\mu
+\partial_\nu\bar{\tilde\rho}_\mu -\frac{1}{2}\frac{m^2}{n\cdot \partial}n_\mu\tilde\rho_\nu  \nonumber\\
 &+ & \frac{1}{2}\frac{m^2}{n\cdot \partial}n_\mu\bar{ \tilde\rho}_\nu +\frac{1}{2}\frac{m^2}{n\cdot \partial}n_\nu\tilde\rho_\nu  -\frac{1}{2}\frac{m^2}{n\cdot \partial}n_\nu \bar{\tilde\rho}_\mu   )+\theta\bar\theta [L_{\mu\nu} +i(\partial_\mu\beta_\nu -\partial_\nu\beta_\mu
 \nonumber\\
&-& \partial_\mu\bar\beta_\nu +\partial_\nu\bar\beta_\mu -\frac{1}{2}\frac{m^2}{n\cdot \partial}n_\mu\beta_\nu +\frac{1}{2}\frac{m^2}{n\cdot \partial}n_\nu\beta_\mu + \frac{1}{2}\frac{m^2}{n\cdot \partial}n_\mu\bar\beta_\nu  - \frac{1}{2}\frac{m^2}{n\cdot \partial}n_\nu\bar\beta_\mu )],\nonumber\\
 \bar{{\cal B}}_{\mu\nu}(x, \theta, \bar\theta )  
  &=& \bar{B}_{\mu\nu} (x) 
 +\theta (
\psi_{\mu\nu}-\partial_\mu\rho_\nu + \partial_\mu\bar{\rho}_\nu +
\partial_\nu\rho_\mu -\partial_\nu\bar{\rho}_\mu + \frac{1}{2}\frac{m^2}{n\cdot \partial}n_\mu\rho_\nu\nonumber\\
&-&  \frac{1}{2}\frac{m^2}{n\cdot \partial}n_\mu\bar{\rho}_\nu    
- \frac{1}{2}\frac{m^2}{n\cdot \partial}n_\nu\rho_\mu +\frac{1}{2}\frac{m^2}{n\cdot \partial}n_\nu\bar{\rho}_\mu )+\bar\theta B_{\mu\nu}^\star +\theta\bar\theta L_{\mu\nu},
\nonumber\\
{\cal M}_{\mu}(x, \theta, \bar\theta ) 
&=&\rho_{\mu} (x) +\theta \epsilon_\mu +\bar\theta 
( \rho_\mu^\star  -i\beta_\mu +i\bar{\beta}_\mu ) +\theta\bar\theta M_{\mu},\nonumber\\
\bar{{\cal{M}}}_{\mu}(x, \theta, \bar\theta ) 
&=& \bar{\rho}_{\mu} (x) 
 +\theta (\epsilon_\mu -i\partial_\mu\sigma +i\partial_\mu\bar\sigma + \frac{i}{2}\frac{m^2}{n\cdot \partial}n_\mu \sigma - \frac{i}{2}\frac{m^2}{n\cdot \partial}n_\mu\bar\sigma  ) +\bar\theta 
\rho_\mu^\star +\theta\bar\theta M_{\mu},\nonumber\\
{\cal{N}}(x, \theta, \bar\theta )&=&\sigma (x) +\theta 
\varepsilon  
  +\bar\theta(\sigma^\star +\chi -\bar\chi) +\theta\bar\theta N,\nonumber\\
   \bar{{\cal{N}}}(x, \theta, \bar\theta ) &=& \bar{\sigma}(x) +\theta \varepsilon  
+\bar\theta \sigma^\star +\theta\bar\theta N,\nonumber\\
 \tilde{\cal M}_{\mu}(x, \theta, \bar\theta )  &=& \tilde\rho_{\mu} (x) +\theta \xi_\mu +
\bar\theta (\tilde\rho_\mu^\star -i\partial_\mu\tilde\sigma+\frac{i}{2}\frac{m^2}{n\cdot \partial}n_\mu\tilde\sigma +i\partial_\mu\bar{\tilde\sigma} -\frac{i}{2}\frac{m^2}{n\cdot \partial}n_\mu \bar{\tilde\sigma} ) \nonumber\\
&+& \theta\bar\theta (\bar M_\mu -i\partial_\mu\tilde\chi +\frac{i}{2}\frac{m^2}{n\cdot \partial}n_\mu\tilde\chi+ 
i\partial_\mu\bar{\tilde\chi}-\frac{i}{2}\frac{m^2}{n\cdot \partial}n_\mu\bar{\tilde\chi}),\nonumber\\ 
{\bar{\tilde{\cal M}}}_{\mu}(x, \theta, \bar\theta )&=& \bar{\tilde\rho}_{\mu} (x) +\theta (
\xi_\mu -i\beta_\mu +i\bar\beta_\mu ) +\bar\theta\tilde\rho_\mu^\star +\theta\bar\theta \bar 
M_\mu,\nonumber\\
{\cal S}_{\mu}(x, \theta, \bar\theta )  
&=&\beta_{\mu} (x) +\theta \eta_\mu +
\bar\theta\beta_\mu^\star +\theta\bar\theta S_\mu, \nonumber\\
 \bar{\cal S}_{\mu}(x, \theta, \bar\theta )&=& \bar\beta_{\mu} (x) +\theta \eta_\mu 
+\bar\theta\beta_\mu^\star +\theta\bar\theta S_\mu,\nonumber\\
 \tilde{\cal N}(x, \theta, \bar\theta )&=& \tilde\sigma (x) +\theta \psi
 +\bar\theta\tilde\sigma^\star +\theta\bar\theta \bar N,\nonumber\\
{\bar{\tilde{\cal N}}}(x, \theta, \bar\theta )&=&\bar{\tilde\sigma} (x) 
+\theta (\psi -\tilde
\chi +\bar{\tilde\psi}) 
 +\bar\theta\tilde\sigma^\star +\theta\bar\theta \bar N,\nonumber\\ 
 {\cal{O}}(x, \theta, \bar\theta )&=& \chi (x) +\theta 
\Sigma   +\bar\theta\chi^\star +\theta\bar\theta O,\nonumber\\
\bar{\cal{O}}(x, \theta, \bar\theta ) &=& \bar\chi (x) 
  +\theta 
\Sigma   +\bar\theta\chi^\star +\theta\bar\theta O,\nonumber\\ 
 \tilde{\cal{O}}(x, \theta, \bar\theta )&=& \tilde\chi (x) +\theta 
\eta +\bar\theta\tilde\chi^\star +\theta\bar\theta \bar O,\nonumber\\ 
{\bar{\tilde{\cal{O}}}}(x, \theta, \bar\theta ) 
 &=& \bar{\tilde\chi} (x) +\theta 
\eta +\bar\theta\tilde\chi^\star +\theta\bar\theta \bar O,\nonumber\\
 {\cal{T}}(x, \theta, \bar\theta ) &=& \varphi (x) +\theta 
\phi +\bar\theta (\varphi^\star -\tilde\chi +\bar{\tilde\chi}) +\theta\bar\theta T,
\nonumber\\
 \bar{\cal{T}}(x, \theta, \bar\theta ) &=&\bar\varphi (x) +\theta 
(\phi -\chi +\bar\chi ) +\bar\theta\varphi^\star +\theta\bar\theta T.\label{a1}
\end{eqnarray}

\section{ Mathematical details of VSR modified Abelian 3-form gauge theory} 
The explicit form of the BV action of VSR modified Abelian 3-form gauge theory is
calculated as 
\begin{eqnarray}  
 {\cal L}_{eff}  
&=& \frac{1}{24}H_{\mu \nu \eta\xi}H^{\mu \nu \eta\xi} +B_{\mu\nu\eta}^\star (
\partial^\mu c^{\nu \eta} + \partial^\nu c^{  \eta\mu} +
\partial^\eta c^{\mu\nu } - \frac{1}{2}\frac{m^2}{n\cdot \partial}n^\mu c^{\nu \eta} -  \frac{1}{2}\frac{m^2}{n\cdot \partial}n^\nu c^{\eta\mu} \nonumber\\
&-& 
\frac{1}{2}\frac{m^2}{n\cdot \partial}n^\eta c^{\mu \nu})- \tilde {c}_{\mu\nu}^\star ( \partial^\mu\beta^\nu  -\partial^\nu\beta^\mu -\frac{1}{2}\frac{m^2}{n\cdot \partial}n^\mu\beta^\nu +\frac{1}{2}\frac{m^2}{n\cdot \partial}n^\nu\beta^\mu  )-{c}_{\mu\nu}^\star 
 B^{\mu\nu}  \nonumber\\
 & +&\tilde B_{\mu\nu}^\star (\partial^\mu f^\nu   -\partial^\nu f^\mu -\frac{1}{2}\frac{m^2}{n\cdot \partial}n^\mu f^\nu +\frac{1}{2}\frac{m^2}{n\cdot \partial}n^\nu f^\mu ) 
+\beta_\mu^\star \partial^\mu c_2  +\tilde \beta_\mu^\star \tilde F ^\mu 
 + \tilde F_\mu^\star  \partial^\mu B   \nonumber\\
 & -&  f_\mu^\star  \partial ^\mu B_1  -\beta_\mu^\star \frac{1}{2}\frac{m^2}{n\cdot \partial}n^\mu c_2 - \tilde F_\mu^\star\frac{1}{2}\frac{m^2}{n\cdot \partial}n^\mu B+ f_\mu^\star \frac{1}{2}\frac{m^2}{n\cdot \partial}n^\mu B_1 - c_2^\star   B_2  
 +\tilde c_1^\star  B 
 \nonumber\\                                                
&-&    c_1^\star  B_1 
+\phi_\mu ^\star   f^\mu    
 - \left(B^{\mu\nu\eta\star}+\frac{\delta\Psi}{\delta B_{\mu\nu\eta}}\right)L_{\mu\nu\eta}+
\left( \tilde c^{\mu\nu\star} +
 \frac{\delta\Psi}{\delta c_{\mu\nu}}\right) M_{\mu\nu} \nonumber\\
&+&\left( c^{\mu\nu\star} + \frac{\delta\Psi}
{\delta\tilde{c}_{\mu\nu}}\right)\tilde M_{\mu\nu} -\left(B^{\mu\nu\star} +
 \frac{\delta\Psi}{\delta B_{\mu\nu}}\right) N_{\mu\nu}
-\left(\tilde B^{\mu\nu\star} +\frac{\delta\Psi}{\delta \tilde B_{\mu\nu}}\right) 
\tilde N_{\mu\nu}\nonumber\\
&-&\left(\beta^{\mu\star} +\frac{\delta\Psi}{\delta
\beta_\mu}\right)O_\mu - \left(\tilde\beta^{\mu\nu\star} +\frac{\delta\Psi}{\delta
\tilde\beta_\mu}\right)\tilde O_\mu +\left( \tilde F^{\mu\star} +\frac{\delta\Psi}{\delta F_\mu}\right)P_\mu 
\nonumber\\
&+&\left( F^{\mu\star} +\frac{\delta\Psi}{\delta \tilde F_\mu}\right)
\tilde P_\mu  
 +  \left( \tilde f^{\mu \star} +\frac{\delta\Psi}{\delta f_\mu}\right)Q_\mu +\left( f^{\mu\star} 
+\frac{\delta\Psi}{\delta \tilde f_\mu}\right)\tilde Q_\mu
 + \left( \tilde c_2^{\star} +\frac{\delta\Psi}{ \delta c_2}\right) {\mathfrak{R}} \nonumber\\
&+&\left( c_2^{\star} + \frac{\delta\Psi}{ \delta \tilde c_2}\right)\tilde {\mathfrak{R}}
 + \left( \tilde c_1^{\star} +\frac{\delta\Psi}{ \delta c_1}\right){\mathfrak{S}} 
+ \left(c_1^{\star} +\frac{\delta\Psi}{ \delta \tilde c_1}\right)\tilde {\mathfrak{S}}-
 \left(\phi^{\mu\star} +\frac{\delta\Psi}{ \delta \phi_\mu}\right) T_\mu\nonumber\\
 &-&  \left( B^{\star} +\frac{\delta\Psi}{ \delta B}\right) \mathbb{U}
-\left(B_1^{\star} +\frac{\delta\Psi}{\delta B_1}\right) \mathfrak{V}-\left(B_2^{\star} +\frac{\delta\Psi}{\delta B_2}
\right) \mathfrak{W}.\label{b00}
\end{eqnarray}
The explicit form of the VSR modified extended BRST transformations for the Abelian 3-form gauge theory is  
\begin{eqnarray}
s_b B_{\mu\nu\eta} &=& L_{\mu\nu\eta},\nonumber\\
s_b {\bar B}_{\mu\nu\eta}
&=& L_{\mu\nu\eta}-
\left(\partial_\mu c_{\nu\eta} -\partial_\mu \bar c_{\nu\eta}+\partial_\nu c_{\eta\mu} -\partial_\nu \bar c_{\eta\mu} +
\partial_\eta c_{\mu\nu}-\partial_\eta \bar c_{\mu\nu} -  \frac{1}{2}\frac{m^2}{n\cdot \partial}n_\mu       c_{\nu\eta}  \right.\nonumber\\
&+&\left. \frac{1}{2}\frac{m^2}{n\cdot \partial}n_\mu\bar c_{\nu\eta}       -\frac{1}{2}\frac{m^2}{n\cdot \partial}n_\nu c_{\eta\mu} +\frac{1}{2}\frac{m^2}{n\cdot \partial}n_\nu \bar c_{\eta\mu}-\frac{1}{2}\frac{m^2}{n\cdot \partial}n_\eta c_{\mu\nu} +\frac{1}{2}\frac{m^2}{n\cdot \partial}n_\eta \bar c_{\mu\nu} \right),\nonumber\\                                                                             
s_b c_{\mu\nu} &=&  M_{\mu\nu}, \ s_b \tilde c_{\mu\nu} = \tilde M_{\mu\nu},\ 
s_b\bar c_{\mu\nu} = M_{\mu\nu}-( {\partial}_\mu \beta_\nu - {\partial_\mu} 
\bar\beta_\nu -  {\partial}_\nu \beta_\mu + {\partial}_\nu \bar\beta_\mu \nonumber\\
&-&\frac{1}{2}\frac{m^2}{n\cdot \partial}n_\mu\beta_\nu +\frac{1}{2}\frac{m^2}{n\cdot \partial}n_\mu\bar\beta_\nu +\frac{1}{2}\frac{m^2}{n\cdot \partial}n_\nu \beta_\mu -\frac{1}{2}\frac{m^2}{n\cdot \partial}n_\nu \bar\beta_\mu 
 ) ,
\nonumber\\
s_b\bar {\tilde c}_{\mu\nu} &=&  \tilde M_{\mu\nu}-B_{\mu\nu} 
+\bar B_{\mu\nu},\ \
s_b B_{\mu\nu}  =  N_{\mu\nu},\ \ s_b\bar B_{\mu\nu} =N_{\mu\nu},\ \
s_b \tilde B_{\mu\nu} = \tilde N_{\mu\nu},\nonumber\\
s_b\bar {\tilde B}_{\mu\nu} &=&  \tilde N_{\mu\nu} -(\partial_\mu f_\nu
-\partial_\mu\bar f_\nu -\partial_\nu f_\mu +\partial_\nu \bar f_\mu \nonumber\\
&-&\frac{1}{2}\frac{m^2}{n\cdot \partial}n_\mu f_\nu +\frac{1}{2}\frac{m^2}{n\cdot \partial}n_\mu\bar f_\nu  +\frac{1}{2}\frac{m^2}{n\cdot \partial}n_\nu  f_\mu -\frac{1}{2}\frac{m^2}{n\cdot \partial}n_\nu \bar f_\mu 
),\nonumber\\
 s_b\beta_\mu  &=&  O_\mu,\   s_b\bar\beta_\mu  = O_\mu -\partial_\mu c_2 +\partial_\mu \bar c_2 +\frac{1}{2}\frac{m^2}{n\cdot \partial}n_\mu  c_2 - \frac{1}{2}\frac{m^2}{n\cdot \partial}n_\mu  \bar c_2      , 
\nonumber\\
s_b\tilde \beta_\mu &=& \tilde O_\mu,\ \ s_b\bar{\tilde\beta}_\mu = \tilde O_\mu -\tilde F_\mu +\bar {\tilde F}_\mu, 
\ 
s_bF_\mu  =  P_\mu,\nonumber\\
s_b\bar F_\mu &=&    P_\mu +\partial_\mu B -\partial_\mu \bar B -\frac{1}{2}\frac{m^2}{n\cdot \partial}n_\mu B +\frac{1}{2}\frac{m^2}{n\cdot \partial}n_\mu \bar B , 
\nonumber\\
s_b\tilde F_\mu &=& \tilde P_\mu,\ \ s_b\bar{\tilde F}_\mu = \tilde P_\mu, \ \
s_bf_\mu = Q_\mu,\ \ s_b\bar f_\mu = Q_\mu, \ s_b\tilde f_\mu = \tilde Q_\mu,
\nonumber\\
s_b\bar{\tilde f}_\mu &=&  \tilde Q_\mu -\partial_\mu B_1 +\partial_\mu \bar B_1 -\frac{1}{2}\frac{m^2}{n\cdot \partial}n_\mu  B_1 + \frac{1}{2}\frac{m^2}{n\cdot \partial}n_\mu \bar B_1 ,\nonumber\\ 
s_b c_2  &=& {\mathfrak{R}},\ \ s_b\bar c_2 ={\mathfrak{R}},\
 s_b \tilde c_2 =\tilde {\mathfrak{R}},\ s_b\bar {\tilde c}_2 =\tilde {\mathfrak{R}} -B_2 +\bar B_2, \nonumber\\
 s_b c_1 &=& {\mathfrak{S}},\ \ s_b\bar c_1 ={\mathfrak{S}} +B -\bar B, 
\ \
 s_b \tilde c_1=\tilde {\mathfrak{S}},\ \
s_b\phi_\mu  = T_\mu,\ \ s_b\bar{\phi}_\mu = T_\mu -f_\mu +\bar f_\mu,\nonumber\\
  s_b\bar {\tilde c}_1 & =&\tilde {\mathfrak{S}} -B_1 +\bar B_1, \
s_bB = \mathbb{U},\  
 s_bB_1 =\mathfrak{V},\ \ s_b\bar B_1 =\mathfrak{V},  \nonumber\\ 
s_b\bar B &=&\mathbb{U},\
s_bB_2  = \mathfrak{W},\ \ s_b\bar B_2 =\mathfrak{W},\ \
s_b \Omega =0, \label{bs1}
\end{eqnarray}
where $\Omega\equiv [L_{\mu\nu\eta}, M_{\mu\nu},  \tilde M_{\mu\nu}, N_{\mu\nu},  \tilde N_{\mu\nu}, O_\mu, \tilde 
O_\mu, P_\mu, \tilde P_\mu, Q_\mu, \tilde Q_\mu, {\mathfrak{R}}, \tilde {\mathfrak{R}},  {\mathfrak{S}}, \tilde {\mathfrak{S}}, T_\mu, \mathbb{U}, \mathfrak{V}, \mathfrak{W}]$
are the ghosts corresponding to the shift symmetry.

The VSR modified   BRST transformation of antifields are
\begin{eqnarray}
&&s_b B_{\mu\nu\eta}^\star =l_{\mu\nu\eta},\ \ 
s_b c_{\mu\nu}^\star = m_{\mu\nu},\ \
s_b \tilde{c}_{\mu\nu}^\star = \bar{m}_{\mu\nu},\ \
s_b B_{\mu\nu}^\star = n_{\mu\nu}, \nonumber\\
&&s_b \beta_\mu^\star = o_\mu, \ \
s_b \tilde {\beta}_\mu^\star = \bar o_\mu, \ \ 
s_b  {F_\mu}^\star = p_\mu, 
s_b \tilde{F_\mu}^\star = \bar p_\mu, \ \
s_b  {f_\mu}^\star = q_\mu,\nonumber\\
&&s_b \tilde{f_\mu}^\star = \bar q_\mu, \ \ 
s_b  {c_2}^\star = r,\ \
s_b \tilde{c_2}^\star = \bar r, \ \ 
s_b  {c_1}^\star = s,\ \
s_b \tilde{c_1}^\star = \bar s,\nonumber\\
&&s_b  {\phi_\mu}^\star = t_\mu,\  
s_b B^\star = u,\  
s_b  B_1^\star = v,\  
s_b B_2^\star  =  w,\ s_b \tilde{B}_{\mu\nu}^\star = \bar n_{\mu\nu},\ s_b \Lambda  =0.
\label{bs2}
\end{eqnarray}
where $\Lambda \equiv l_{\mu\nu\eta}, m_{\mu\nu}, \bar{m}_{\mu\nu}, n_{\mu\nu},  \bar n_{\mu\nu},   o_
\mu, \bar o_\mu,
p_\mu,  \bar p_\mu, q_\mu, \bar q_\mu, r, \bar r, s, \bar s,  t_\mu, u, v, w$
are the auxiliary fields.

The superfields and anti-superfields in component form for the VSR modified extended  BRST invariant
3-form theory
  are 
\begin{eqnarray} 
{\cal B}_{\mu\nu\eta}(x, \theta ) &=& B_{\mu\nu\eta} (x) +\theta L_{\mu\nu\eta},
 \ {\cal C}_{\mu\nu}(x, \theta )  = c_{\mu\nu} (x) +\theta M_{\mu\nu},\nonumber\\
\bar{{\cal B}}_{\mu\nu\eta}(x, \theta )  
&=&\bar{B}_{\mu\nu\eta} (x) +\theta \left[ L_{\mu\nu\eta}-
\left(\partial_\mu c_{\nu\eta} -\partial_\mu \bar c_{\nu\eta}+\partial_\nu c_{\eta\mu} -\partial_\nu \bar c_{\eta\mu} +
\partial_\eta c_{\mu\nu}-\partial_\eta \bar c_{\mu\nu} \right.\right.\nonumber\\
&-&\left.\left.  \frac{1}{2}\frac{m^2}{n\cdot \partial}n_\mu c_{\nu\eta}+\frac{1}{2}\frac{m^2}{n\cdot \partial}n_\mu \bar c_{\nu\eta} -\frac{1}{2}\frac{m^2}{n\cdot \partial}n_\nu c_{\eta\mu} +\frac{1}{2}\frac{m^2}{n\cdot \partial}n_\nu \bar c_{\eta\mu} -\frac{1}{2}\frac{m^2}{n\cdot \partial}n_\eta c_{\mu\nu} \right.\right.\nonumber\\
&+&\left.\left. \frac{1}{2}\frac{m^2}{n\cdot \partial}n_\eta \bar c_{\mu\nu} 
 \right)\right],\nonumber\\
\bar{\cal C}_{\mu\nu}(x, \theta ) &=&  \bar c_{\mu\nu} (x) +\theta \left[M_{\mu\nu}-\left(\partial_\mu \beta_\nu -\partial_\mu 
\bar\beta_\nu - \partial_\nu \beta_\mu +\partial_\nu \bar\beta_\mu -\frac{1}{2}\frac{m^2}{n\cdot \partial}n_\mu\beta_\nu \right.\right.\nonumber\\
&+&\left.\left.  \frac{1}{2}\frac{m^2}{n\cdot \partial}n_\mu\bar\beta_\nu +\frac{1}{2}\frac{m^2}{n\cdot \partial}n_\nu \beta_\mu -\frac{1}{2}\frac{m^2}{n\cdot \partial}n_\nu \bar\beta_\mu 
\right)\right] ,
\nonumber\\
\tilde{\cal C}_{\mu\nu}(x, \theta ) &=&\tilde c_{\mu\nu} (x) +\theta \tilde M_{\mu\nu}, \
{\bar{\tilde{\cal C}}}_{\mu\nu}(x, \theta )  =  {\bar {\tilde c}}_{\mu\nu} (x) +\theta ( \tilde M_{\mu\nu}-B_{\mu\nu} 
+\bar B_{\mu\nu}),\nonumber\\ 
{\cal B}_{\mu\nu}(x, \theta ) &=& B_{\mu\nu} (x) +\theta N_{\mu\nu}, \ \
\bar{\cal B}_{\mu\nu}(x, \theta )  =  \bar B_{\mu\nu} (x) +\theta N_{\mu\nu},\nonumber\\
\tilde{\cal B}_{\mu\nu}(x, \theta ) &=& \tilde B_{\mu\nu} (x) +\theta \tilde N_{\mu\nu},\ {\cal B}_{\mu}(x, \theta ) 
=  \beta_{\mu} (x) +\theta O_{\mu},\nonumber\\
{\bar{\tilde{\cal B}}}_{\mu\nu}(x, \theta ) 
&=& \bar {\tilde B}_{\mu\nu} (x) +\theta \left[\tilde N_{\mu\nu} 
-\left(\partial_\mu f_\nu -\partial_\mu\bar f_\nu -\partial_\nu f_\mu +\partial_\nu \bar f_\mu \right.\right. \nonumber\\
&-& \left.\left. \frac{1}{2}\frac{m^2}{n\cdot \partial}n_\mu  f_\nu + \frac{1}{2}\frac{m^2}{n\cdot \partial}n_\mu \bar f_\nu +\frac{1}{2}\frac{m^2}{n\cdot \partial}n_\nu  f_\mu -\frac{1}{2}\frac{m^2}{n\cdot \partial}n_\nu \bar f_\mu
 \right)\right],\nonumber\\
\bar{\cal B}_{\mu}(x, \theta ) &=&  \bar \beta_{\mu } (x) +\theta \left(O_\mu -\partial_\mu c_2 +\partial_\mu \bar c_2  +\frac{1}{2}\frac{m^2}{n\cdot \partial}n_\mu c_2 -\frac{1}{2}\frac{m^2}{n\cdot \partial}n_\mu   \bar c_2 \right),
\nonumber\\
\tilde{\cal B}_{\mu}(x, \theta ) &=&\tilde \beta_{\mu} (x) +\theta \tilde O_{\mu}, \
{\bar{\tilde{\cal B}}}_{\mu}(x, \theta ) = \bar {\tilde\beta}_{\mu } (x) +\theta (\tilde O_\mu -\tilde F_\mu +\bar 
{\tilde F}_\mu),\nonumber\\
\bar{\cal F}_{\mu}(x, \theta ) &=& \bar F_{\mu } (x) +\theta \left(P_\mu + {\partial}_\mu B - {\partial}_\mu \bar B -\frac{1}{2}\frac{m^2}{n\cdot \partial}n_\mu B
+\frac{1}{2}\frac{m^2}{n\cdot \partial}n_\mu B\right),
\nonumber\\
\tilde{\cal F}_{\mu}(x, \theta ) &=&\tilde F_{\mu} (x) +\theta \tilde P_{\mu},\ 
{\bar{\tilde{\cal F}}}_{\mu}(x, \theta )  =  {\bar {\tilde F}}_{\mu } (x) +\theta \tilde P_{\mu},\ \
{\cal F}_{\mu}(x, \theta ) = F_{\mu} (x) +\theta P_{\mu},\nonumber\\
{\mathfrak{F}}_{\mu}(x, \theta ) &=& f_{\mu} (x) +\theta Q_{\mu},\ \
\bar{\mathfrak{F}}_{\mu}(x, \theta )  =  \bar f_{\mu} (x) +\theta  Q_{\mu},\ \
\tilde{\mathfrak{F}}_{\mu}(x, \theta ) = \tilde f_{\mu} (x) +\theta \tilde Q_{\mu},\nonumber\\  
{\bar{\tilde{\mathfrak{F}}}}_{\mu}(x, \theta )  &=&    {\bar{\tilde f}}_{\mu} (x) +\theta \left(\tilde Q_\mu -\partial_\mu B_1 +\partial_\mu \bar B_1 +\frac{1}{2}\frac{m^2}{n\cdot \partial}n_\mu B_1 -\frac{1}{2}\frac{m^2}{n\cdot \partial}n_\mu \bar B_1   \right),\nonumber\\
{\cal C}_{2}(x, \theta ) &=& c_{2} (x) +\theta {\mathfrak{R}},\ \
\bar{\cal C}_{2}(x, \theta )  =  \bar c_{2} (x) +\theta {\mathfrak{R}},\ \ \tilde{\cal C}_{2}(x, \theta ) = \tilde c_{2} (x) +\theta \tilde {\mathfrak{R}},
\nonumber\\
{\bar{\tilde{\cal C}}}_{2}(x, \theta ) &=&  {\bar {\tilde c}}_{2} (x) +\theta (\tilde {\mathfrak{R}} -B_2 +\bar B_2), \
{\cal C}_{1}(x, \theta ) = c_{1} (x) +\theta {\mathfrak{S}},
\nonumber\\
{\tilde{\cal C}}_{1}(x, \theta ) &=& \tilde c_{1} (x) +\theta \tilde {\mathfrak{S}},\ 
{\bar{\tilde{\cal C}}}_{1}(x, \theta )  =  {\bar {\tilde c}}_{1} (x) +\theta (\tilde {\mathfrak{S}} -B_1 +\bar B_1),\nonumber\\
{ \Phi}_{\mu}(x, \theta )&=& \phi_\mu (x) +\theta T_\mu,\ 
\bar{  \Phi}_{\mu}(x, \theta )  =  \bar \phi_\mu (x) +\theta ( T_\mu -f_\mu +\bar f_\mu),
\nonumber\\
{ \cal B}(x, \theta ) &=&B(x) +\theta \mathbb{U},\ \
\bar{  \cal B}(x, \theta )  =  \bar B (x) +\theta \mathbb{U},\ \ { \cal B}_1(x, \theta ) =B_1(x) +\theta \mathfrak{V},
\nonumber\\
\bar{  \cal B}_1(x, \theta )  &=&  \bar B_1 (x) +\theta \mathfrak{V},
\ \
{ \cal B}_2(x, \theta )=B_2(x) +\theta \mathfrak{W},\ \
\bar{  \cal B}_2(x, \theta )  =  \bar B_2 (x) +\theta \mathfrak{W},\nonumber\\
\bar{\cal B}_{\mu\nu\eta}^\star(x, \theta ) &=& B_{\mu\nu\eta}^\star(x)  +\theta L_{\mu\nu\eta}, \ \
\bar{{\cal C}}_{\mu\nu}^\star (x, \theta )= c_{\mu\nu}^\star (x) +\theta M_{\mu\nu}, \ \
{\bar{\tilde{\cal C}}}_{\mu\nu}^\star(x, \theta )  =  \tilde c_{\mu\nu}^\star (x) +\theta \tilde M_{\mu\nu},\nonumber\\
\bar{{\cal B}}_{\mu\nu}^\star(x, \theta ) &=& B_{\mu\nu}^\star (x) +\theta N_{\mu\nu},\ \
{\bar{\tilde{\cal  B}}}_{\mu\nu}^\star (x, \theta ) =  \tilde B_{\mu\nu}^\star (x) +\theta \tilde N_{\mu\nu},\ \
\bar{{\cal B}}_{\mu}^\star (x, \theta )= \beta_{\mu}^\star (x) +\theta O_{\mu},\nonumber\\
{\bar{\tilde{\cal B}}}_{\mu}^\star(x, \theta ) &=&\tilde \beta_{\mu}^\star (x) +\theta\tilde O_{\mu},\ \ 
\bar{{\cal F}}_{\mu}^\star (x, \theta )= F_{\mu}^\star (x) +\theta P_{\mu},\ \
{\bar{\tilde{\cal F}}}_{\mu}^\star (x, \theta ) =  \tilde F_{\mu}^\star (x) +\theta \tilde P_{\mu},\nonumber \\
\bar{{\mathfrak{F}}}_{\mu}^\star (x, \theta )&=&f_{\mu}^\star (x) +\theta Q_{\mu},\ \
{\bar{\tilde{\mathfrak{F}}}}_{\mu}^\star  (x, \theta )= \tilde f_{\mu}^\star (x) +\theta \tilde Q_{\mu},\ \
\bar{{\cal C}}_{2}^\star (x, \theta ) = c_{2}^\star (x) +\theta {\mathfrak{R}},\nonumber\\ 
{\bar{\tilde{\cal C}}}_{2}^\star(x, \theta ) &=&\tilde c_{2}^\star (x) +\theta \tilde {\mathfrak{R}},\ \
\bar{{\cal C}}_{1}^\star (x, \theta )= c_{1}^\star (x) +\theta {\mathfrak{S}},\ \
{\bar{\tilde{\cal C}}}_{1}^\star  (x, \theta )= \tilde c_{1}^\star (x) +\theta \tilde {\mathfrak{S}},\nonumber\\
\bar{ {\cal B}}^\star  (x, \theta )&=&B^\star (x) +\theta \mathbb{U},\ \
\bar{{\cal B}}_{1}^\star (x, \theta )=B_{1}^\star (x) +\theta \mathfrak{V},\ \
\bar{{\cal B}}_{2}^\star (x, \theta )=B_{2}^\star (x) +\theta \mathfrak{W},\nonumber\\
{\bar{\cal C}}_{1}(x, \theta ) & =&  \bar c_{1} (x) +\theta ({\mathfrak{S}} +B -\bar B),\ \
\bar{  \Phi}_{\mu}^\star (x, \theta )=\phi_\mu^\star (x) +\theta T_\mu. \label{supb}
\end{eqnarray}
The superfields for both   extended BRST and anti-BRST invariant 3-form theory  are 
\begin{eqnarray}  
{\cal B}_{\mu\nu\eta}(x, \theta, \bar\theta )  
 &=& B_{\mu\nu\eta} (x) +\theta L_{\mu\nu\eta}
+\bar \theta(B_{\mu\nu\eta}^\star +\partial_\mu 
\tilde c_{\nu\eta}-\partial_\mu \bar{\tilde c}_{\nu\eta}+\partial_\nu \tilde c_{\eta\mu}
-\partial_\nu \bar{\tilde c}_{\eta\mu}  +\partial_\eta 
\tilde c_{\mu\nu}\nonumber\\
&-&\partial_\eta \bar{\tilde c}_{\mu\nu} -\frac{1}{2}\frac{m^2}{n\cdot \partial}n_\mu{\tilde c}_{\nu\eta} +\frac{1}{2}\frac{m^2}{n\cdot \partial}n_\mu \bar{\tilde c}_{\nu\eta} -\frac{1}{2}\frac{m^2}{n\cdot \partial}n_\nu\tilde c_{\eta\mu}+\frac{1}{2}\frac{m^2}{n\cdot \partial}n_\nu\bar{\tilde c}_{\eta\mu} \nonumber\\
&-&\frac{1}{2}\frac{m^2}{n\cdot \partial}n_\eta \tilde c_{\mu\nu} + \frac{1}{2}\frac{m^2}{n\cdot \partial}n_\eta \bar{\tilde c}_{\mu\nu} ) 
+\theta\bar\theta (l_{\mu\nu\eta} +{\partial}_\mu B_{\nu\eta} - {\partial}_\mu \bar B_{\nu\eta}
+ {\partial}_\nu B_{\eta\mu}  \nonumber\\
&-& {\partial}_\nu \bar B_{\eta\mu} +  {\partial}_\eta B_{\mu\nu} - {\partial}_\eta \bar B_{\mu\nu}-\frac{1}{2}\frac{m^2}{n\cdot \partial}n_\mu B_{\nu\eta} + \frac{1}{2}\frac{m^2}{n\cdot \partial}n_\mu \bar B_{\nu\eta} \nonumber\\ 
&-&\frac{1}{2}\frac{m^2}{n\cdot \partial}n_\nu  B_{\eta\mu} + \frac{1}{2}\frac{m^2}{n\cdot \partial}n_\nu\bar B_{\eta\mu} - \frac{1}{2}\frac{m^2}{n\cdot \partial}n_\eta B_{\mu\nu}
+\frac{1}{2}\frac{m^2}{n\cdot \partial}n_\eta  \bar B_{\mu\nu}      ),
\nonumber\\
  {\cal C}_{\mu\nu}(x, \theta, \bar\theta ) & =& c_{\mu\nu} (x) +\theta M_{\mu\nu} +\bar \theta (c_{\mu\nu}^\star +
\tilde B_{
\mu\nu} -\bar{\tilde B}_{\mu\nu} )+\theta\bar\theta (m_{\mu\nu} +\partial_\mu f_\nu -\partial_\mu \bar f_\nu -
\partial_\nu f_\mu \nonumber\\
&+&\partial_\nu \bar f_\mu -\frac{1}{2}\frac{m^2}{n\cdot \partial}n_\mu  f_\nu+\frac{1}{2}\frac{m^2}{n\cdot \partial}n_\mu\bar f_\nu+\frac{1}{2}\frac{m^2}{n\cdot \partial}n_\nu f_\mu -\frac{1}{2}\frac{m^2}{n\cdot \partial}n_\nu \bar f_\mu ),\nonumber\\
\bar{{\cal B}}_{\mu\nu\eta}(x, \theta, \bar\theta ) &=&   \bar{B}_{\mu\nu\eta} (x) +\theta (L_{\mu\nu\eta}-
(\partial_\mu c_{\nu\eta} -\partial_\mu \bar c_{\nu\eta}+\partial_\nu c_{\eta\mu} -\partial_\nu \bar c_{\eta\mu} +
\partial_\eta c_{\mu\nu}-\partial_\eta \bar c_{\mu\nu}\nonumber\\
&-& \frac{1}{2}\frac{m^2}{n\cdot \partial}n_\mu c_{\nu\eta} +\frac{1}{2}\frac{m^2}{n\cdot \partial}n_\mu \bar{c}_{\nu\eta} -\frac{1}{2}\frac{m^2}{n\cdot \partial}n_\nu c_{\eta\mu}+\frac{1}{2}\frac{m^2}{n\cdot \partial}n_\nu\bar{c}_{\eta\mu} -\frac{1}{2}\frac{m^2}{n\cdot \partial}n_\eta  c_{\mu\nu} \nonumber\\
&+&  \frac{1}{2}\frac{m^2}{n\cdot \partial}n_\eta \bar{ c}_{\mu\nu}) )+\bar\theta B_{\mu\nu\eta}^\star +\theta\bar\theta l_{\mu\nu\eta},\nonumber\\
\bar{\cal C}_{\mu\nu}(x, \theta, \bar\theta ) 
 &=& \bar c_{\mu\nu} (x) +\theta ( M_{\mu\nu}-(\partial_\mu \beta_\nu -
\partial_\mu 
\bar\beta_\nu - \partial_\nu \beta_\mu +\partial_\nu \bar\beta_\mu - \frac{1}{2}\frac{m^2}{n\cdot \partial}n_\mu \beta_\nu + \frac{1}{2}\frac{m^2}{n\cdot \partial}n_\mu \bar\beta_\nu \nonumber\\
&+& \frac{1}{2}\frac{m^2}{n\cdot \partial}n_\nu \beta_\mu -\frac{1}{2}\frac{m^2}{n\cdot \partial}n_\nu \bar\beta_\mu ))+\bar\theta c_{\mu\nu}^\star
 + \theta\bar\theta   m_{\mu\nu},\nonumber\\
\tilde{\cal C}_{\mu\nu}(x, \theta, \bar\theta )  
&=&\tilde c_{\mu\nu} (x) +\theta \tilde M_{\mu\nu}+\bar\theta(\tilde 
c_{\mu\nu}^
\star +\partial_\mu \tilde\beta_\nu -\partial_\mu 
\bar{\tilde\beta}_\nu -\partial_\nu \tilde\beta_\mu +\partial_\nu \bar{\tilde\beta}_\mu -\frac{1}{2}\frac{m^2}{n\cdot \partial}n_\mu\tilde\beta_\nu \nonumber\\
&+& \frac{1}{2}\frac{m^2}{n\cdot \partial}n_\mu \bar{\tilde\beta}_\nu +\frac{1}{2}\frac{m^2}{n\cdot \partial}n_\nu \tilde\beta_\mu - \frac{1}{2}\frac{m^2}{n\cdot \partial}n_\nu \bar{\tilde\beta}_\mu) 
+\theta\bar\theta ( \bar m_{\mu\nu}+\partial_\mu \tilde F_\nu -\partial_\mu\bar{\tilde F}_\nu  \nonumber\\
& -&  \partial_\nu \tilde F_\mu +\partial_\nu\bar{\tilde F}_\mu  - \frac{1}{2}\frac{m^2}{n\cdot \partial}n_\mu  \tilde F_\nu+\frac{1}{2}\frac{m^2}{n\cdot \partial}n_\mu \bar{\tilde F}_\nu + \frac{1}{2}\frac{m^2}{n\cdot \partial}n_\nu \tilde F_\mu -\frac{1}{2}\frac{m^2}{n\cdot \partial}n_\nu \bar{\tilde F}_\mu), 
\nonumber\\
{\bar{\tilde{\cal C}}}_{\mu\nu}(x, \theta, \bar\theta ) & =&  \bar {\tilde c}_{\mu\nu} (x) +\theta ( \tilde M_{\mu\nu
}-B_{\mu\nu} 
+\bar B_{\mu\nu}) +\bar\theta  {\tilde c}_{\mu\nu}^\star +\theta\bar\theta \bar m_{\mu\nu},\nonumber\\ 
{\cal B}_{\mu\nu}(x, \theta, \bar\theta )  
&=& B_{\mu\nu} (x) +\theta N_{\mu\nu}+\bar\theta (B_{\mu\nu}^\star + 
\partial_\mu\tilde f_\nu -\partial_\mu\bar{\tilde f}_\nu
-\partial_\nu\tilde f_\mu +\partial_\nu\bar{\tilde f}_\mu - \frac{1}{2}\frac{m^2}{n\cdot \partial}n_\mu \tilde f_\nu  \nonumber\\
&+& \frac{1}{2}\frac{m^2}{n\cdot \partial}n_\mu  \bar{\tilde f}_\nu +\frac{1}{2}\frac{m^2}{n\cdot \partial}n_\nu \tilde f_\mu -\frac{1}{2}\frac{m^2}{n\cdot \partial}n_\nu \bar{\tilde f}_\mu )+\theta\bar\theta n_{\mu\nu},\nonumber\\
\bar{\cal B}_{\mu\nu}(x, \theta, \bar\theta ) & =&  \bar B_{\mu\nu} (x) +\theta N_{\mu\nu}+\bar\theta B_{\mu\nu}^\star
+\theta\bar\theta n_{\mu\nu},\nonumber\\
\tilde{\cal B}_{\mu\nu}(x, \theta, \bar\theta ) &=& \tilde B_{\mu\nu} (x) +\theta \tilde N_{\mu\nu}+\bar\theta \tilde
 B_{\mu\nu}^\star
+\theta\bar\theta \bar n_{\mu\nu},
\nonumber\\
 {\cal B}_{\mu}(x, \theta, \bar\theta ) 
&=&     \beta_{\mu} (x) +\theta O_{\mu}+\bar\theta ( \beta_{\mu}^\star +F_\mu -\bar F_\mu )+\theta\bar\theta (
o_\mu -\partial_\mu B_2+\partial_\mu \bar B_2+\frac{1}{2}\frac{m^2}{n\cdot \partial}n_\mu B_2   \nonumber\\
&-&\frac{1}{2}\frac{m^2}{n\cdot \partial}n_\mu \bar B_2),
\nonumber\\
{\bar{\tilde{\cal B}}}_{\mu\nu}(x, \theta, \bar\theta )  
 &=& \bar {\tilde B}_{\mu\nu} (x) +\theta (\tilde N_{\mu\nu} 
-(\partial_\mu f_\nu -\partial_\mu\bar f_\nu -\partial_\nu f_\mu +\partial_\nu \bar f_\mu -\frac{1}{2}\frac{m^2}{n\cdot \partial}n_\mu f_\nu +\frac{1}{2}\frac{m^2}{n\cdot \partial}n_\mu \bar f_\nu \nonumber\\
&+&\frac{1}{2}\frac{m^2}{n\cdot \partial}n_\nu  f_\mu -\frac{1}{2}\frac{m^2}{n\cdot \partial}n_\nu  \bar f_\mu ))+\bar\theta {\tilde B}_{
\mu\nu}^\star
+\theta\bar\theta \bar n_{\mu\nu},\nonumber\\
\bar{\cal B}_{\mu}(x, \theta, \bar\theta )  
 &=& \bar \beta_{\mu } (x) +\theta (O_\mu -\partial_\mu c_2 +\partial_\mu 
\bar c_2 +\frac{1}{2}\frac{m^2}{n\cdot \partial}n_\mu c_2 -\frac{1}{2}\frac{m^2}{n\cdot \partial}n_\mu \bar c_2)+\bar\theta \beta_{\mu }^\star +\theta\bar\theta o_\mu,\nonumber
\end{eqnarray}
\begin{eqnarray}
\tilde{\cal B}_{\mu}(x, \theta, \bar\theta )  
&=& \tilde \beta_{\mu} (x) +\theta \tilde O_{\mu}+\bar\theta (
{\tilde\beta}_{\mu}^\star +\partial_\mu \tilde c_2 -\partial_\mu \bar{\tilde c}_2 -\frac{1}{2}\frac{m^2}{n\cdot \partial}n_\mu \tilde c_2 +\frac{1}{2}\frac{m^2}{n\cdot \partial}n_\mu  \bar{\tilde c}_2)\nonumber\\
&+&\theta\bar\theta ( \bar o_\mu +\partial_\mu B_2 -\partial_\mu \bar B_2 -\frac{1}{2}\frac{m^2}{n\cdot \partial}n_\mu B_2 +\frac{1}{2}\frac{m^2}{n\cdot \partial}n_\mu\bar B_2 ),
\nonumber\\
{\bar{\tilde{\cal B}}}_{\mu}(x, \theta, \bar\theta )& =& \bar {\tilde\beta}_{\mu} (x) +\theta (\tilde O_\mu -\tilde 
F_\mu +\bar {\tilde F}_\mu)+\bar\theta {\tilde\beta}_{\mu}^\star +\theta\bar\theta \bar o_\mu, \nonumber\\
{\cal F}_{\mu}(x, \theta, \bar\theta ) &=& F_{\mu} (x) +\theta P_{\mu}+\bar\theta F_{\mu}^\star 
+\theta\bar\theta p_\mu,\nonumber\\ 
\bar{\cal F}_{\mu}(x, \theta, \bar\theta )  
& =&  \bar F_{\mu} (x) +\theta (P_\mu +\partial_\mu B -\partial_\mu 
\bar B -\frac{1}{2}\frac{m^2}{n\cdot \partial}n_\mu B +\frac{1}{2}\frac{m^2}{n\cdot \partial}n_\mu \bar B)+\bar\theta F_{\mu}^\star +\theta\bar\theta p_\mu,\nonumber\\
\tilde{\cal F}_{\mu}(x, \theta, \bar\theta )  
 &=& \tilde F_{\mu} (x) +\theta \tilde P_{\mu}+\bar\theta (\tilde 
F_{\mu}^\star -\partial_\mu B_2 +
\partial_\mu\bar B_2 +\frac{1}{2}\frac{m^2}{n\cdot \partial}n_\mu B_2 -\frac{1}{2}\frac{m^2}{n\cdot \partial}n_\mu \bar B_2)+\theta\bar\theta \bar p_\mu,\nonumber\\ 
{\bar{\tilde{\cal F}}}_{\mu}(x, \theta, \bar\theta ) &= &\bar {\tilde F}_{\mu} (x) +\theta \tilde P_{\mu}
+\bar\theta {\tilde F}_{\mu}^\star +\theta\bar\theta \bar p_\mu,\nonumber\\
{\mathfrak{F}}_{\mu}(x, \theta, \bar\theta )  
 &=& f_{\mu} (x) +\theta Q_{\mu}+\bar\theta (f_{\mu}^\star -\partial_\mu B_1 +
\partial_\mu\bar B_1 +\frac{1}{2}\frac{m^2}{n\cdot \partial}n_\mu B_1 -\frac{1}{2}\frac{m^2}{n\cdot \partial}n_\mu \bar B_1 )+\theta\bar\theta  q_\mu,\nonumber\\
\bar{\mathfrak{F}}_{\mu}(x, \theta, \bar\theta )  &=&  \bar f_{\mu} (x) +\theta  Q_{\mu}+\bar\theta f_{\mu}^\star
+\theta\bar\theta q_\mu,\nonumber\\
\tilde{\mathfrak{F}}_{\mu}(x, \theta, \bar\theta ) &=& \tilde f_{\mu} (x) +\theta \tilde Q_{\mu}+\bar\theta
{\tilde f}_{\mu}^\star +
\theta\bar\theta \bar q_\mu,\nonumber\\  
\bar{\tilde{\mathfrak{F}}}_{\mu}(x, \theta, \bar\theta )  
&=& \bar {\tilde f}_{\mu} (x) +\theta (\tilde Q_\mu -
\partial_\mu B_1 +\partial_\mu \bar B_1 +\frac{1}{2}\frac{m^2}{n\cdot \partial}n_\mu B_1 -\frac{1}{2}\frac{m^2}{n\cdot \partial}n_\mu \bar B_1  )+\bar\theta {\tilde f}_{\mu}^\star +\theta\bar\theta \bar q_\mu,\nonumber\\
{\cal C}_{2}(x, \theta, \bar\theta ) &=& c_{2} (x) +\theta {\mathfrak{R}}+\bar\theta (c_{2}^\star +B-\bar B)+\theta\bar\theta r,
\nonumber\\
\bar{\cal C}_{2}(x, \theta, \bar\theta ) & = & \bar c_{2} (x) +\theta {\mathfrak{R}}+\bar\theta c_{2}^\star +\theta\bar\theta r,
\nonumber\\
\tilde{\cal C}_{2}(x, \theta, \bar\theta ) &=& \tilde c_{2} (x) +\theta \tilde {\mathfrak{R}}+\bar\theta \tilde c_{2}^\star
+\theta\bar\theta \bar r,\nonumber\\ 
{\bar{\tilde{\cal C}}}_{2}(x, \theta, \bar\theta )  &= & \bar {\tilde c}_{2} (x) +\theta (\tilde {\mathfrak{R}} -B_2 +\bar B_2)+
\bar\theta {\tilde c}_{2}^\star +\theta\bar\theta \bar r,\nonumber\\
{\cal C}_{1}(x, \theta, \bar\theta ) &=& c_{1} (x) +\theta {\mathfrak{S}}+\bar\theta (c_{1}^\star -B_1 +\bar B_1)+\theta\bar\theta
 s,\nonumber\\ 
\bar{\cal C}_{1}(x, \theta, \bar\theta )  &= & \bar c_{1} (x) +\theta ({\mathfrak{S}} +B -\bar B)+\bar\theta c_{1}^
\star +\theta\bar\theta s,\nonumber\\
\tilde{\cal C}_{1}(x, \theta, \bar\theta ) &=& \tilde c_{1} (x) +\theta \tilde {\mathfrak{S}}+\bar\theta (\tilde c_{1}^\star -B_2+
\bar B_2)+\theta\bar\theta \bar s,
\nonumber\\ 
{\bar{\tilde{\cal C}}}_{1}(x, \theta, \bar\theta ) &= & \bar {\tilde c}_{1} (x) +\theta (\tilde {\mathfrak{S}} -B_1 +\bar B_1)+
\bar\theta {\tilde c}_{1}^\star +\theta\bar\theta \bar s,\nonumber\\
{ \Phi}_{\mu}(x, \theta, \bar\theta ) &=& \phi_\mu (x)  +\theta T_\mu +\bar\theta (\phi_\mu^\star +\tilde f_\mu -
\bar{\tilde f}_\mu )+\theta\bar\theta (t_\mu + {\partial}_\mu B_1 - {\partial}_\mu \bar B_1 -\frac{1}{2}\frac{m^2}{n\cdot \partial}n_\mu B_1 \nonumber\\
&+&\frac{1}{2}\frac{m^2}{n\cdot \partial}n_\mu \bar B_1),\nonumber\\
\bar{  \Phi}_{\mu}(x, \theta, \bar\theta )  &= & \bar \phi_\mu (x) +\theta ( T_\mu -f_\mu +\bar f_\mu)+\bar\theta 
 \phi_\mu^\star
+\theta\bar\theta t_\mu,\nonumber\\
{ \cal B}(x, \theta, \bar\theta ) &=&B(x) +\theta \mathbb{U}+\bar\theta B^\star+\theta\bar\theta u,\nonumber\\
\bar{  \cal B}(x, \theta, \bar\theta ) & = & \bar B (x) +\theta \mathbb{U}+\bar\theta B^\star+\theta\bar\theta u,
\nonumber\\
{ \cal B}_1(x, \theta, \bar\theta ) &=&B_1(x) +\theta \mathfrak{V}+\bar\theta B_1^\star+\theta\bar\theta v,\nonumber\\
\bar{  \cal B}_1(x, \theta, \bar\theta )  &=&  \bar B_1 (x) +\theta \mathfrak{V}+\bar\theta B_1^\star+\theta\bar\theta v,
\nonumber\\
{ \cal B}_2(x, \theta, \bar\theta ) &=&B_2(x) +\theta \mathfrak{W}+\bar\theta B_2^\star+\theta\bar\theta w,\nonumber\\
\bar{  \cal B}_2(x, \theta, \bar\theta ) & =&  \bar B_2 (x) +\theta \mathfrak{W} +\bar\theta B_2^\star +\theta\bar\theta w. 
\label{superf}
\end{eqnarray}
Form the above relations, we calculate   
\begin{eqnarray}
 \frac{1}{2}\frac{\delta}{\delta\bar\theta}\frac{\delta}{\delta\theta}\bar {\cal B}_{\mu\nu\eta} \bar {\cal B}^{
\mu\nu\eta} 
&=&    l_{\mu\nu\eta}\bar B^{\mu\nu\eta} -B_{\mu\nu\eta}^\star (L^{\mu\nu\eta} -\partial^\mu c^{\nu\eta} +\partial^\mu 
\bar c^{\nu\eta} -\partial^\nu c^{\eta\mu}+\partial^\nu \bar c^{\eta\mu}-\partial^\eta c^{ \mu\nu}+\partial^\eta \bar c^{ \mu\nu} \nonumber\\
&+&\frac{1}{2}\frac{m^2}{n\cdot \partial}n^\mu c^{\nu\eta} - \frac{1}{2}\frac{m^2}{n\cdot \partial}n^\mu \bar c^{\nu\eta} + \frac{1}{2}\frac{m^2}{n\cdot \partial}n^\nu  c^{\eta\mu} +\frac{1}{2}\frac{m^2}{n\cdot \partial}n^\nu \bar c^{\eta\mu} + \frac{1}{2}\frac{m^2}{n\cdot \partial}n^\eta c^{ \mu\nu}  \nonumber\\
&-&\frac{1}{2}\frac{m^2}{n\cdot \partial}n^\eta  \bar 
c^{ \mu\nu}       ),             \nonumber\\
  \frac{\delta}{\delta\bar\theta}\frac{\delta}{\delta\theta}\bar{\tilde {\cal C}}_{\mu\nu}\bar {\cal C}^{
\mu\nu}
&=& \bar m_{\mu\nu}\bar c^{\mu\nu} +m_{\mu\nu}\bar {\tilde c}^{\mu\nu} +\tilde c_{\mu\nu}^\star (
M^{\mu\nu} -\partial^\mu \beta^\nu +\partial^\mu \bar \beta^\nu +\partial^\nu \beta^\mu -\partial^\nu \bar \beta^\mu +\frac{1}{2}\frac{m^2}{n\cdot \partial}n^\mu \beta^\nu  \nonumber\\
 &-& \frac{1}{2}\frac{m^2}{n\cdot \partial}n^\mu  \bar \beta^\nu- \frac{1}{2}\frac{m^2}{n\cdot \partial}n^\nu \beta^\mu + \frac{1}{2}\frac{m^2}{n\cdot \partial}n^\nu \bar \beta^\mu)+ c_{\mu\nu}^\star (\tilde M^{\mu\nu} -B^{\mu\nu} +\bar B^{\mu\nu}),
\nonumber\\
 \frac{1}{2}\frac{\delta}{\delta\bar\theta}\frac{\delta}{\delta\theta} \bar{\tilde {\cal B}}_{\mu\nu} \bar{\tilde {
\cal B}}^{\mu\nu}
&=&  n_{\mu\nu}\bar B^{\mu\nu}-B_{\mu\nu}^\star N^{\mu\nu}, 
\nonumber\\
 \frac{1}{2}\frac{\delta}{\delta\bar\theta}\frac{\delta}{\delta\theta}\bar{\cal B}_{\mu\nu} \bar{\cal B}^{\mu\nu}
&=&  \bar n_{\mu\nu}\bar {\tilde B}^{\mu\nu} -\tilde B_{\mu\nu}^\star (\tilde N^{\mu\nu} -\partial^\mu f^\nu +
\partial^\mu\bar f^\nu +\partial^\nu f^\mu -\partial^\nu \bar f^\mu +\frac{1}{2}\frac{m^2}{n\cdot \partial}n^\mu f^\nu\nonumber\\
& -& \frac{1}{2}\frac{m^2}{n\cdot \partial}n^\mu\bar f^\nu -\frac{1}{2}\frac{m^2}{n\cdot \partial}n^\nu f^\mu +\frac{1}{2}\frac{m^2}{n\cdot \partial}n^\nu \bar f^\mu ),
\nonumber\\
 \frac{1}{2}\frac{\delta}{\delta\bar\theta}\frac{\delta}{\delta\theta}\bar{\cal B}_{\mu } \bar{\cal B}^{\mu}
&=& o_\mu\bar\beta^\mu -\beta_\mu^\star (O^\mu -\partial^\mu c_2 +\partial^\mu \bar c_2 -\frac{1}{2}\frac{m^2}{n\cdot \partial}n^\mu c_2 -\frac{1}{2}\frac{m^2}{n\cdot \partial}n^\mu \bar c_2  ) ,
\nonumber\\
 \frac{1}{2}\frac{\delta}{\delta\bar\theta}\frac{\delta}{\delta\theta}\bar{\tilde {\cal B}}_{\mu } 
\bar{\tilde{\cal B}}^{\mu}
&=& \bar o_\mu\bar{\tilde \beta}^\mu -\tilde\beta_\mu^\star (\tilde O^\mu -\tilde F^\mu +\bar{\tilde F}^\mu),
\nonumber\\
  \frac{\delta}{\delta\bar\theta}\frac{\delta}{\delta\theta}\bar{\tilde {\cal F}}_{\mu } \bar{\cal F}^{\mu} 
&=&  \bar p_\mu\bar F^\mu +p_\mu \bar {\tilde F}^\mu +\tilde F_\mu^\star (P^\mu +\partial^\mu B -\partial ^\mu \bar 
B -\frac{1}{2}\frac{m^2}{n\cdot \partial}n^\mu B +\frac{1}{2}\frac{m^2}{n\cdot \partial}n^\mu \bar 
B  ) +F_\mu^\star \tilde P^\mu,\nonumber\\
  \frac{\delta}{\delta\bar\theta}\frac{\delta}{\delta\theta}\bar{\tilde {\mathfrak{F}}}_{\mu } \bar{\mathfrak{F}}^{\mu} 
&=& \bar q_\mu\bar f^\mu +q_\mu\bar{\tilde f}^\mu +\tilde f_\mu^\star Q^\mu +f_\mu^\star (
\tilde Q^\mu -\partial^\mu B_1+\partial^\mu\bar B_1 +\frac{1}{2}\frac{m^2}{n\cdot \partial}n^\mu B_1 -  \frac{1}{2}\frac{m^2}{n\cdot \partial}n^\mu\bar B_1),\nonumber\\
 \frac{\delta}{\delta\bar\theta}\frac{\delta}{\delta\theta} {\bar{\tilde {\cal C}}}_{2 } \bar
{\cal C}_2
&=&  \bar r\bar c_2 +r\bar {\tilde c}_2 +\tilde c_2^\star {\mathfrak{R}} +c_2^\star (\tilde {\mathfrak{R}} -B_2 +\bar B_2),\nonumber\\
  \frac{\delta}{\delta\bar\theta}\frac{\delta}{\delta\theta}{\bar{\tilde {\cal C}}}_{1} \bar{{\cal C}}_1
&=& \bar s \bar{ c}_1 +\tilde c_1^\star ( {\mathfrak{S}} +B -\bar B ) +s\bar{\tilde c}_1 +c_1^\star
(\tilde {\mathfrak{S}}-B_1 +\bar B_1),\nonumber\\
 \frac{1}{2}\frac{\delta}{\delta\bar\theta}\frac{\delta}{\delta\theta}\bar{\Phi}_{\mu } \bar{ \Phi}^{\mu}
&=& t_\mu\bar \phi^\mu -\phi_\mu^\star (T^\mu -f^\mu +\bar f^\mu ),\nonumber\\
 \frac{1}{2}\frac{\delta}{\delta\bar\theta}\frac{\delta}{\delta\theta}\bar {\cal B}\bar {\cal B}
&=&u\bar B -B^\star \mathbb{U},\nonumber\\
 \frac{1}{2}\frac{\delta}{\delta\bar\theta}\frac{\delta}{\delta\theta}\bar {\cal B}_1\bar {\cal B}_1 
&=&  v\bar B_1 -B_1^\star \mathfrak{V},\nonumber\\
 \frac{1}{2}\frac{\delta}{\delta\bar\theta}\frac{\delta}{\delta\theta}\bar {\cal B}_2\bar {\cal B}_2
&=&w\bar B_2 -B_2^\star \mathfrak{W}. \label{lsr}
\end{eqnarray}

\end{document}